\documentclass[11pt,a4paper]{article}

\usepackage[
top    = 2.50cm,
bottom = 2.50cm,
left   = 2.50cm,
right  = 2.50cm]{geometry}
\usepackage{jheppub}
\usepackage[sort&compress]{natbib}

\usepackage{amssymb}
\usepackage{amsfonts,dsfont}
\usepackage{graphicx}
\usepackage{slashed}
\usepackage{amsmath}
\usepackage{hyperref}
\usepackage{verbatim}

\usepackage{caption}
\usepackage{subcaption}

\usepackage{tikz-cd}

\usepackage{tikz}
\usetikzlibrary{mindmap,trees}
\usepackage{verbatim}
\usetikzlibrary{arrows,positioning}
\usepackage{smartdiagram}
\usetikzlibrary{calc,trees,positioning,arrows,chains,shapes.geometric,%
	decorations.pathreplacing,decorations.pathmorphing,shapes,%
	matrix,shapes.symbols}

\usepackage{lmodern}
\tikzset{snake it/.style={decorate, decoration=snake}}

\tikzset{
	>=stealth',
	punkt/.style={
		rectangle,
		rounded corners,
		draw=black, very thick,
		text width=6.5em,
		minimum height=2em,
		text centered},
	pil/.style={
		->,
		thick,
		shorten <=2pt,
		shorten >=2pt,}
}

\tikzset{
	>=stealth',
	punktchain/.style={
		rectangle, 
		rounded corners, 
		draw=black, very thick,
		text width=10em, 
		minimum height=3em, 
		text centered, 
		on chain},
	line/.style={draw, thick, <-},
	element/.style={
		tape,
		top color=white,
		bottom color=blue!50!black!60!,
		minimum width=8em,
		draw=blue!40!black!90, very thick,
		text width=10em, 
		minimum height=3.5em, 
		text centered, 
		on chain},
	every join/.style={->, thick,shorten >=1pt},
	decoration={brace},
	tuborg/.style={decorate},
	tubnode/.style={midway, right=2pt},
}

\newcommand{\yslant}{0.5}
\newcommand{\xslant}{-0.6}

\usetikzlibrary{decorations.pathmorphing}

\tikzset{zigzag/.style={decorate, decoration=zigzag}}
\def \L {2.}

\makeatletter
\def\@hex@@Hex#1%
{\if a#1A\else \if b#1B\else \if c#1C\else \if d#1D\else
	\if e#1E\else \if f#1F\else #1\fi\fi\fi\fi\fi\fi \@hex@Hex}
\makeatother

\definecolor{darkgreen}{HTML}{006622}

\newcommand{\be}{\begin{equation}}
\newcommand{\ee}{\end{equation}}
\newcommand{\bea}{\begin{eqnarray}}
\newcommand{\eea}{\end{eqnarray}}
\newcommand{\bse}{\begin{subequations}}
\newcommand{\ese}{\end{subequations}}
\newcommand{\beqa}{\begin{eqnarray}}
\newcommand{\eeqa}{\end{eqnarray}}
\newcommand{\beqar}{\begin{eqnarray*}}
\newcommand{\eeqar}{\end{eqnarray*}}
\newcommand{\bi}{\begin{itemize}}
\newcommand{\ei}{\end{itemize}}
\newcommand{\bn}{\begin{enumerate}}
\newcommand{\en}{\end{enumerate}}

\newcommand{\ba}{\begin{array}}
\newcommand{\ea}{\end{array}}
\newcommand{\bc}{\begin{center}}
\newcommand{\ec}{\end{center}}

\newcommand{\SU}{\mathrm{SU}}
\newcommand{\U}{\mathrm{U}}
\newcommand{\SO}{\mathrm{SO}}
\newcommand{\RR}{\mathbb{R}}

\title{Correlations vs connectivity in R-charge}

\author{Joan Sim\'on}
\affiliation{
	\it{School of Mathematics and Maxwell Institute for Mathematical Sciences,\\
		University of Edinburgh, Peter Guthrie Tait road,
		Edinburgh EH9 3FD, UK}}
\emailAdd{j.simon@ed.ac.uk}
\abstract{The holographic relation between quantum correlations and connectivity of spacetime is explored for single R-charged AdS$_5$ black holes and their half-BPS limits (superstars). In a two boundary set-up, the wormhole between both universes reduces to a designable and computable quantum mechanical correlation between the dual microscopic degrees of freedom in the BPS limit. This quantum connectivity is seen as a naked singularity by a single sided observer. In a single boundary set-up, as a small step towards the description of entangled black holes, we describe quantum teleportation between two labs in different locations of the transverse 5-sphere using entangled gravitons in a reference state that provides a classical channel between both labs.}

\begin{document}

\maketitle

\newpage

\section{Introduction}

AdS/CFT \cite{Maldacena:1997re,Witten:1998qj,Ryu:2006bv} supports the idea that connectivity in spacetime is due to the existence of quantum correlations and entanglement \cite{VanRaamsdonk:2010pw}. A generalisation of this notion leads to the EPR=ER conjecture \cite{Maldacena:2013xja}.

The possibility that quantum effects violating the averaged null energy condition (ANEC) could make the Einstein-Rosen wormhole \cite{erosen} traversable was analysed in \cite{Gao:2016bin}, where this was shown to occur by turning on some double trace deformation between the two boundaries in an eternal AdS black hole set-up.

Probing the validity of EPR=ER\footnote{There is a large literature building on the original ideas in \cite{VanRaamsdonk:2010pw,Maldacena:2013xja,Gao:2016bin}. EPR=ER was studied on a string worldsheet describing a quark-antiquark EPR pair in \cite{Jensen:2013ora,Sonner:2013mba}. Multiboundary wormholes were studied in \cite{Balasubramanian:2014hda}, whereas \cite{Numasawa:2016emc} applied relevant information theoretic protocols for EPR=ER in 2d CFTs. More recently, traversability of time shifted eternal black holes was achieved in \cite{vanBreukelen:2017dul} by turning on an appropriate coupling between two CFTs. In fact, turning on some specific interaction in a single boundary set-up allowed the authors in \cite{deBoer:2018ibj} to argue that particles can escape the interior of a single sided black hole. Double trace deformations in 3d $N=4$ gauge theories based on large linear quivers were studied in \cite{Bachas:2017rch} to realise bridges between large 4d AdS spaces.} to generate time machines by studying entangled black holes in a single boundary theory \cite{Susskind:2014yaa} and analysing whether by turning some non-local interaction between their degrees of freedom makes the associated wormhole traversable is an important question.

In this work, single AdS$_5$ R-charged black holes and their singular BPS limits (superstars) are considered. The Kruskal extension of these black holes contains classical bridges whose length increases as the non-extremal parameter decreases in the near-BPS limit. This behaviour is expected of near-extremal systems, where a long throat typically develops to capture the decrease in the boundary correlations\footnote{See \cite{Andrade:2013rra} for a thorough discussion of the near-extremal behaviour of correlations and entanglement in the Reissner-Nordstr\"om AdS$_5$ black hole.}. In our set-up, the throat is replaced by a naked singularity in the BPS limit because the quantum correlations and entropy responsible for the microscopic degrees of freedom sourcing the singularity are not large enough to support a classical spacetime.

The naked singularity is located at the origin of AdS and extends over the transverse 5-sphere. It corresponds to a distribution of giant gravitons wrapping 3-spheres in this 5-sphere while rotating at the speed of light. The dual description in N=4 SYM involves N free fermions in a 1d harmonic oscillator \cite{Corley:2001zk,Berenstein:2004kk}. The semiclassical limit of this quantum mechanical system emerges in the deep infra-red of an alternative bulk (LLM) description, only available in the BPS limit, where the phase space density is realised as the charged sources for a Laplace equation uniquely determining the solution to the supergravity equations of motion \cite{Lin:2004nb}.

Despite the lack of classical bridges in the BPS limit, the quantum control over the microscopic degrees of freedom and its dual realisation in the deep infra-red of the geometry in terms of some coarse-grained version of its phase space will allow us to make some statements :
\begin{itemize}
	\item Consider two non-interacting N=4 SYM theories in their half-BPS subspaces. Any maximally correlated state in this two boundary set-up gives rise to a singular spacetime from the perspective of a one-sided observer. One can design any correlation between both universes. In the semiclassical limit, the only remnant of the quantum connectivity between them is through the regions in phase space where quantum correlations have support, i.e. the naked singularity. This BPS behaviour is an example of  "quantum bridges" \cite{Maldacena:2013xja} and is compatible with the bridge length behaviour of the near-extremal black holes.
	\item Consider a single N=4 SYM theory. The effective factorisation of the Hilbert space formulated in \cite{Berenstein:2017rrx} describing excitations on top of some reference state, allows to describe an entangled gas of gravitons localised in different regions of the transverse 5-sphere. This allows to study quantum mechanical effects in a curved background, the gravity dual to the reference state, such as teleportation between different localised regions in a single boundary theory, with the reference state providing a classical channel between both observers.
	\item The natural notion of entanglement in the half-BPS sector of N=4 SYM is in R-charge, since fermions are delocalised on the boundary 3-sphere. But in the phase space bulk description, the emergence of locality in the 1d where fermions are trapped allows to use the methods developed in condensed matter physics to compute entanglement in real space for ultra-cold atoms. Using these results, we learn the amount of such entanglement in some "real space" region is related to the variance in the number of fermions in that region. The existence of some effective 2d CFT reproducing the entanglement for the Fermi sea (global AdS) with central charge equal to one is consistent with the lack of a classical spacetime picture advocated in our EPR=ER discussion.
\end{itemize}

The organisation of this work is as follows. Section \ref{sec:motivation} gives further motivation to study the problem analysed in these notes appealing to general string theory and holographic arguments. Section \ref{sec:5d} computes the bridge length for the non-extremal 5d black holes under consideration, based on the Kruskal extension presented in appendix \ref{kruskal}. Section \ref{sec:review} reviews some of the holographic background material, rederives the relation between ensembles and bulk naked singularities in subsection \ref{sec:supers} and extends this construction to localised superstars in subsection \ref{sec:localised}. Section \ref{sec:2-boundary} discusses EPR=ER in the two boundary set-up, whereas section \ref{sboundary} does so for a single boundary. Section \ref{sec:discussion} summarises the results obtained in this note together with the difficulties and expectations to describe entangled superstars. In appendices \ref{shock}-\ref{sec:matching}, a shock-wave analysis of our 5d black holes, their thermodynamical stability and their matching to the LLM configurations in their BPS limit is presented. Finally, appendix \ref{cond-mat} attempts to summarise some of the results in the condensed matter literature on entanglement entropy for N free fermions in a 1d harmonic potential.

\section{On the naturalness of EPR=ER}
\label{sec:motivation}

The original Strominger-Vafa black hole microstate entropy counting \cite{Strominger:1996sh}, based on Polchinski's notion of D-branes \cite{Polchinski:1995mt}, relies on non-renormalisation theorems ensuring the number of microstates remains invariant as the string coupling $g_s$ varies. Typically, the macroscopic black hole description helps to identify the microscopic components of the hole. Varying the string coupling, the classical gravitational description is eventually not reliable anymore, but one analyses the same degrees of freedom in a perturbative worldsheet description or in some relevant dual CFT. Thus, while the number of states does not change with $g_s$, the size of the states does.

The same feature appears in the Horowitz-Polchinski correspondence principle \cite{Horowitz:1996nw}, based on earlier ideas by Susskind on black holes as single string states \cite{Susskind:1993ws}. Here, macroscopic non-extremal black holes are compared with a gas of strings and/or D-branes (depending on the charges of the hole). Despite the generic lack of technical control, it is realised that the entropy of both systems match, up to coefficients of order one, when they are computed at the value of the string coupling where the curvature invariants of the hole evaluated at the horizon are of order the string scale and the masses of both systems are of the same order of magnitude. For example, for a (d+1)-dimensional Schwarzschild black hole, the Riemann squared evaluated at the horizon scale $r_0$
\begin{equation}
  \left. R_{\mu\nu\alpha\beta}R^{\mu\nu\alpha\beta}\right|_{r_0} \sim \frac{1}{r_0^4} \sim \frac{1}{(\alpha^\prime)^2} \quad \Rightarrow \quad r_0\sim \ell_s
\end{equation}
determines the horizon to be string scale. Comparing the mass of the hole $(M_{\text{BH}})$ with the mass of the string excitation spectrum at level $N$ $(M_{\text{string}})$, determines the string coupling
\begin{equation}
  M^2_{\text{BH}} \sim \frac{r_0^{2(d-2)}}{G^2} \sim M^2_{\text{string}} \sim \frac{N}{\ell_s^2} \quad \Rightarrow \quad g_s \sim N^{-1/4}\,,
\end{equation}
a result that is compatible with the string worldsheet perturbative description. At this stage, the black hole entropy formula is fixed
\begin{equation}
  S_{\text{BH}} \sim \frac{r_0^{d-1}}{G} \sim \sqrt{N} \sim S_{\text{string}}
\end{equation}
and matches the one derived from the asymptotic density of states in perturbative string theory. Variations of this argument for more general black holes, with or without D-brane charges were presented in \cite{Horowitz:1996nw}.

Both scenarios manage to assign a Hilbert space $\mathcal{H}_{\text{string}}$ to a macroscopic black hole by moving in the string coupling parameter space. The latter allows to perform some explicit microstate counting, or the use of statistical mechanics arguments suggesting that a rather accurate description for this gas of strings and/or D-branes is given by
\begin{equation}
	\rho_{\text{string}} = \sum_{E_s\in \mathcal{H}_{\text{string}}} \frac{e^{-\beta\,E_s}}{Z(\beta)}\,|E_s\rangle\langle E_s|\,,
\end{equation}
for an appropriate choice of the temperature.

The description of multiple non-interacting black holes in disconnected universes would involve the tensor product $\otimes \mathcal{H}_{\text{string}}$. The existence of entanglement in quantum mechanics raises the question as to whether by tuning the string coupling $g_s$ and maximising the amount of correlation between different subsystems gives rise to any classical gravitational effect.
Maldacena-Susskind's EPR=ER \cite{Maldacena:2013xja}, extending the ideas relating connectivity of space with entanglement \cite{VanRaamsdonk:2010pw}, claims the existence of non-traversable wormholes connecting these subsystems.

Consider maximally correlated states in a pair $\mathcal{H}^L_{\text{string}}\otimes \mathcal{H}^R_{\text{string}}$, for simplicity 
\begin{equation}
|\Psi\rangle = \sum_{i=1}^K a_i\,|i\rangle_{\text{L}}\otimes |i\rangle_{\text{R}}\,.
\end{equation} 
Choosing the coefficients adequately, and tracing out one of the Hilbert space copies, one recovers the relevant ensembles in statistical mechanics. In black hole physics, it is natural to consider the purification of the thermal ensemble 
\begin{equation}
|\Psi \rangle =  \sum_{E_s\in \mathcal{H}_{\text{string}}} \frac{e^{-\beta\,E_s/2}}{\sqrt{Z(\beta)}}\,|E^L_s\rangle\otimes |E^R_s\rangle\,,
\label{eq:maxi-pure}
\end{equation}
since this state describes the maximal Kruskal extension of the black hole \cite{Israel:1976ur,Maldacena:2001kr}. This is precisely the set-up that applies to eternal AdS black holes in AdS/CFT \cite{Maldacena:2001kr}, providing some arena where to check the EPR=ER conjecture\footnote{It is amusing, and well known, that the same mathematical structure is responsible for Unruh's effect in Rindler physics \cite{Unruh:1976db,Bisognano:1975ih}, or Hawking's effect in black holes \cite{Hawking:1974sw} and in cosmological horizons \cite{Gibbons:1977mu}.}. But, the arguments above are quite general, though they may appear formal and speculative to some. In particular, notice that any future development describing holography in some different asymptotics, for example, would still need to account for the extra entanglement and correlation between the two Hilbert spaces. This is part of the naturalness and universality of the physics predicted by EPR=ER \cite{Maldacena:2013xja}.

In fact, some of the recent work in the subject applies this logic to the SYK model \cite{Sachdev:1992fk,Kitaev-talks,Polchinski:2016xgd,Maldacena:2016hyu} and its gravitational dual description at large N and low energies \cite{Maldacena:2017axo,Maldacena:2018lmt}. In this case, one uses the Hilbert space of the Majorana fermions, together with its nearly-AdS$_2$ dual \cite{Maldacena:2016upp}, and adds correlations between two such systems, as in \eqref{eq:maxi-pure}. But one can equally well apply these ideas to the D1-D5 system \cite{David:2002wn} or any effective Hilbert space description capturing the microscopics of black holes in specific regimes, such as Kerr-CFT \cite{Guica:2008mu}, dual infra-red descriptions of R-charged black holes \cite{deBoer:2011zt} or extremal vanishing horizon set-ups \cite{SheikhJabbaria:2011gc,Johnstone:2013eg}. 

In this work, we follow this philosophy for near-extremal single R-charged AdS black holes, whose BPS limit reduces to the half-BPS sector of N=4 SYM.

\section{Single R-charged AdS$_5$ black holes}
\label{sec:5d}

Consider the 5-sphere reduction of type IIB supergravity \cite{Kim:1985ez,Gunaydin:1984fk} truncated to its ${\cal N}=2$ sector with $\U(1)^3$ gauge symmetry \cite{Gunaydin:1984ak}. The bosonic matter content includes two scalar fields, which are usually described in terms of three scalar fields $X_i$ $i=1,2,3$ satisfying the constraint $X_1X_2X_3=1$, and three $\U(1)$ gauge fields $A_i$ $i=1,2,3$. 5d black holes carrying three electric charges were found in \cite{Behrndt:1998ns,Behrndt:1998jd}\footnote{The three equal charge case corresponds to the Reissner-Nordstrom AdS black hole \cite{Romans:1991nq, London:1995ib} having $X_i=1\,\,\forall\,i$.}. Single non-extremal R-charged AdS black holes turn off two of the gauge fields. They are described by \cite{Behrndt:1998ns,Behrndt:1998jd}
\begin{equation}
\begin{aligned}
ds^2 &= H^{-2/3}\,f \left(-dt^2 + \frac{H}{f^2}dr^2\right) + H^{1/3}\,r^2\,d\Omega_3^2\,, \\
A &= \frac{\sqrt{q(q+\mu)}}{q}\left(H^{-1}-1\right)\,dt\,, \\
X_1 &=H^{-2/3}\,, \quad X_2=X_3=H^{1/3}
\end{aligned}
\label{eq:5dBH}
\end{equation}
where
\begin{equation}
  H=1+\frac{q}{r^2}\,, \quad f=1-\frac{\mu}{r^2} + \frac{r^2}{R_{\text{AdS}}^2}\,H\,.
\end{equation}
These configurations are determined by three parameters $R_{\text{AdS}},\,q,\,\mu$. $R_{\text{AdS}}$ is the radius of AdS$_5$ and sets the scale for the cosmological constant. The mass $M$ and electric charge $Q$ are determined by $q$ and $\mu$ \cite{Behrndt:1998jd,Batrachenko:2004fd} 
\begin{equation}
  M=\frac{\omega_3}{8\pi\,G_5}\left(\frac{3}{2}\mu + q\right)\,, \quad \quad Q=\frac{\omega_3}{8\pi\,G_5}\tilde{q}  \equiv \frac{\omega_3}{8\pi\,G_5}\sqrt{q(q+\mu)}\,,
\end{equation}
in terms of the volume of the transverse 3-sphere $\omega_3$ and 5d Newton's constant $G_5$.

Black holes \eqref{eq:5dBH} have a curvature singularity at the origin $r=0$ of AdS$_5$ cloaked by an event horizon located at $r=r_+$
\begin{equation}
 2r^2_+= -(q+R_{\text{AdS}}^2) + \sqrt{(q+R_{\text{AdS}}^2)^2+4\mu\,R_{\text{AdS}}^2}\,,
\label{eq:horizon}
\end{equation}
whenever $\mu$, the non-extremal parameter, is different from zero. Their thermodynamical properties are captured by the first law \cite{Batrachenko:2004fd}
\begin{equation}
TdS = dM - \Phi dQ\,,
\label{eq:1stlaw}
\end{equation}
where the temperature, chemical potential $\Phi$\footnote{The value of the chemical potential $\Phi$ equals the difference between the boundary and the horizon values of the gauge potential, i.e. $\Phi = A(r\to\infty) - A(r_+)$. In the AdS/CFT literature, the natural boundary conditions choose the gauge field to vanish at the horizon, so that $\Phi$ equals the gauge field at infinity, matching the expectation value of the dual density charge. These are the boundary conditions used in \cite{Chamblin:1999hg}, for example. These differ from the ones in \eqref{eq:5dBH}. Both choices are related through a large gauge transformation.} and entropy $S$ are given by 
\begin{eqnarray}
  T &=& \frac{\kappa}{2\pi}= \frac{q+R_{\text{AdS}}^2+2r_+^2}{2\pi\,R_{\text{AdS}}^2\,\sqrt{q+r_+^2}}\,\,, \quad \quad \Phi =\frac{\tilde{q}}{r_+^2 + q}\,, \label{eq:Htemp} \\
  S &=&\frac{\omega_3}{4G_5} r_+^2\,\sqrt{r_+^2+q}\,. \label{eq:GRentropy}
\end{eqnarray}

\begin{figure}
\centering

\begin{tikzpicture}

\draw[thick,red,zigzag] (-\L,\L) coordinate(stl) -- (\L,\L) coordinate (str);
\draw[thick,red,zigzag] (-\L,-\L) coordinate(stl) -- (\L,-\L) coordinate (str);
\draw[thick,black] (-\L,\L) -- (-\L,-\L);
\draw[thick,black] (\L,\L) -- (\L,-\L);
\draw[dashed,black] (\L,\L) -- (-\L,-\L);
\draw[dashed,black] (-\L,\L) -- (\L,-\L);

\draw[black] (0.2*\L,-0.6*\L) node[right] (scrip) {$\mathcal{H}^-$}
(0.5*\L,0.85*\L) node[right] (scrip) {$\mathcal{H}^+$};

\draw[thick,blue] (-1,1) -- (1,1);
\draw[blue] (-1,1) node[left] (scrip) {$u_L$};
\draw[blue] (1,1) node[right] (scrip) {$u_R$};

\end{tikzpicture}
\caption{Penrose diagram for a 5d single R-charged AdS black hole. The blue line represents the bridge between two equal time horizon crossing points at constant Kruskal time $v_0$.}

\label{fig:Penrose}	
\end{figure}
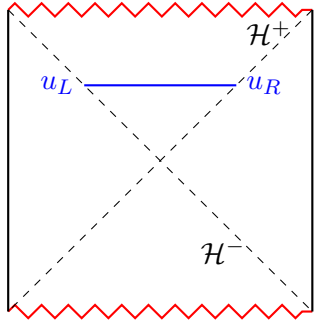

The maximal Kruskal extension \cite{Kruskal:1959vx} of these black holes is worked out in appendix \ref{kruskal}. It is convenient to present the extended metric as
\begin{equation}
\begin{aligned}
  ds^2 &= w^2(u_s,v_s)(du_s^2-dv_s^2) + H(u_s,v_s)\,r^2(u_s,v_s)\,d\Omega_3^2\,,\\
  w^2(u_s,v_s) &= \frac{r^2+|r_-|^2}{\kappa^2\,R_{\text{AdS}}^2r^2\,(H(r))^{2/3}}\,(r^2-r_+^2)\,e^{-2\kappa r_\star}\,,
\end{aligned}  
\end{equation}
where the new coordinates $(u_s,v_s)$ are related to the ones in the 5d black hole \eqref{eq:5dBH} by
\begin{equation}
  u_s=e^{\kappa r_\star}\cosh \kappa t\,,  \quad v_s=e^{\kappa r_\star} \sinh\kappa t\,,
\end{equation}
  where $r_\star$ is the tortoise coordinate described in \eqref{eq:tortoise}. Due to the existence of a single event horizon, the global structure of these black holes is similar to 5d Schwarzschild black holes, as indicated in the Penrose diagram \ref{fig:Penrose}. 

Consider the distance between two horizon crossing points, $u_L$ and $u_R$, at the same point on the 3-sphere and at constant Kruskal time $v_0$, corresponding to the bridge indicated by the blue line in figure \ref{fig:Penrose}. Since 
\begin{equation}
u_s^2-v_s^2 = e^{\kappa r_\star}\,,
\end{equation}
and $e^{\kappa r_\star(r_+)}$ vanishes at the horizon (see \eqref{eq:small-y}), one infers $u_L=-u_R = v_0$. Hence, their distance equals
\begin{equation}
\begin{aligned}
\text{Length}(u_L, u_R) &= \int^{u_R}_{u_L} \left.w(u_s,v_0)\right|_{r=r_+} du_s = 2|v_0|\,\left.w(u,v_0)\right|_{r=r_+} \\
&= |v_0|\,\frac{4R_{\text{AdS}}}{r_+^{1/3}\sqrt{r_+^2+ |r_-|^2}} (r_+^2+q)^{2/3}\,e^{-\kappa D(r_+)}\,,
\end{aligned}
\label{eq:bridgeL}
\end{equation}
where the near horizon analysis of the tortoise coordinate \eqref{eq:tortoise-exp} was used in the last equality.

For very massive black holes, i.e. $\mu \gg R_{\text{AdS}}^2 \geq q$ so that $r_+^2 \approx \sqrt{\mu\,R_{\text{AdS}}^2}$, one expects the bridge length to grow as $\mu$ decreases, indicating the decrease in quantum correlations between degrees of freedom in the two asymptotic regions.
This can be checked by expanding \eqref{eq:bridgeL} in this regime 
\begin{equation}
  \text{Length}(u_L, u_R) \sim  2\sqrt{2}\,|v_0|\,R_{\text{AdS}}\,e^{-\pi/4}\left(1 - \frac{1}{24r_+^2}\left(3\pi\,R_{\text{AdS}}^2 + (3\pi-16)q\right)\right)
\end{equation}
The bridge length indeed grows for $\frac{q}{R_{\text{AdS}}^2} \leq \frac{3\pi}{16\pi-3}$. 

The main focus in later sections is the BPS limit $\mu=0$. To interpolate with this regime, let us analyse the bridge length \eqref{eq:bridgeL} in the near-extremal regime $\mu\ll q$. Approximating the horizon $r_+^2\approx \frac{\mu\,R_{\text{AdS}}^2}{q+R_{\text{AdS}}^2}$ and $D\approx\frac{R_{\text{AdS}}^2}{q+R_{\text{AdS}}^2}\,R_{\text{AdS}}\,\arctan \frac{\sqrt{q}}{R_{\text{AdS}}}$ in this regime, one is finally left with
\begin{equation}
   \text{Length}(u_L, u_R) \sim 4|v_0|\,R_{\text{AdS}}\,\frac{\sqrt{q}}{\left[R_{\text{AdS}}\,(q+R_{\text{AdS}}^2)\right]^{1/3}}\,e^{-\frac{R_{\text{AdS}}}{\sqrt{q}} \arctan \frac{\sqrt{q}}{R_{\text{AdS}}}}\,\left(\frac{q}{\mu}\right)^{1/6}
\label{eq:bridge-div}
\end{equation}
Notice that as $\frac{\mu}{q}$ decreases, the bridge distance grows like $(\mu/q)^{-1/6}$ while the area of the horizon decreases as $\mu/q$. This is an explicit realisation of the pinching mechanism described by van Raamsdonk \cite{VanRaamsdonk:2010pw} when arguing connectivity in space is due to quantum entanglement (correlations). As reviewed in appendix \ref{sec:stability}, black holes \eqref{eq:5dBH} are thermodynamically unstable in this regime. Hence, our conclusion is not robust. Nevertheless, it is compatible with the behaviour found in the BPS limit, as it will be discussed in sections \ref{sec:supers} and \ref{sec:2-boundary}. In this respect, it is important to stress this behaviour is not happening because the geometry develops a throat, but because the quantum correlations among the microscopic degrees of freedom sourcing the naked singularity are not large enough to support any classical geometry.

Recent developments in holography have deepened the relation between black hole physics and quantum chaos \cite{Sekino:2008he,Maldacena:2015waa}, by studying the effect of a small perturbation in the entanglement structure described by the black hole. In the conformal field theory side, this can be analysed by computing 4-pt functions in the thermal field double (see \eqref{eq:dual-ensemble} for our specific setp-up) in the large central charge limit \cite{Roberts:2014ifa,Caputa:2015waa,Asplund:2015eha}. In the gravity side, this entanglement disruption is captured by a shock-wave \cite{Shenker:2013pqa}, following the original work in \cite{Dray:1984ha}. In appendix \ref{shock}, the shock-wave geometry due to a non-charged perturbation of mass $\delta M$ turned on very far in the past so that it follows a trajectory very close to the horizon in the 5d black hole \eqref{eq:5dBH} is computed following the general discussion in \cite{Leichenauer:2014nxa}. The scrambling time derived from this analysis behaves like
\begin{equation}
  t_\star \approx \frac{\beta}{2\pi} \log \frac{aM}{\delta M}\,,
\end{equation}
where the parameter $a$ depends on the regime of parameters describing the 5d black hole \eqref{eq:5dBH}. This is in agreement with the results presented in  \cite{Leichenauer:2014nxa}.

\paragraph{Type IIB embedding.} The 5d black hole \eqref{eq:5dBH} can be embedded into type IIB supergravity using the general embedding described in \cite{Cvetic:1999xp}. These are constant dilaton configurations with metric 
\begin{equation}
\begin{aligned}
  ds^2 &=\sqrt{\gamma}\left[-H^{-1}\,f\,dt^2 + \frac{dr^2}{f} + r^2\,d\Omega_3^2 + R_{\text{AdS}}^2\,d\theta^2\right] + \frac{R_{\text{AdS}}^2}{\sqrt{\gamma}}\sin^2\theta\,d\tilde{\Omega}_3^2 \\
  &+ \frac{H}{\sqrt{\gamma}}\cos^2\theta \left(R_{\text{AdS}}\,d\phi + A\right)^2\,,
\end{aligned}
\label{eq:10dBH}
\end{equation}
where $\gamma \equiv 1 + \frac{q}{r^2}\sin^2\theta$, and self-dual 5-form RR field strength (see \cite{Cvetic:1999xp} for details).

The 5d electric charge $Q$ is reinterpreted as angular momentum on the transverse 5-sphere in the 10d geometry \eqref{eq:10dBH}. In the dual $\mathcal{N}=4$ SYM gauge theory, it corresponds to R-charge $J=QR_{\text{AdS}}$ whereas the mass is encoded in the conformal dimension $\Delta = MR_{\text{AdS}}$\footnote{To derive these expressions one uses the Kaluza-Klein relation $G_5= \frac{G_{10}}{\omega_5 R_{\text{AdS}}^5}$ and the geometry facts $\omega_5=\pi^3$ and $\omega_3=2\pi^2$ together with the microscopic relations $R_{\text{AdS}}^4=4\pi g_sN\,\ell_s^4$ and $G_{10}=8\pi^6g_2^2\ell_s^8$ (see for example 
\cite{juanthesis}).}
\begin{equation}
  \Delta = \frac{N^2}{2}\left(\frac{3}{2}\frac{\mu}{R_{\text{AdS}}^2} + \frac{q}{R_{\text{AdS}}^2}\right)\,, \quad \quad J = \frac{N^2}{2}\,\sqrt{\frac{q}{R_{\text{AdS}}^2}\left(\frac{q}{R_{\text{AdS}}^2} + \frac{\mu}{R_{\text{AdS}}^2}\right)}
\end{equation}

The standard dual description for the non-extremal black hole \eqref{eq:10dBH} is in terms of the ensemble
\begin{equation}
  \rho(\tilde{\beta},\,\Phi) = \frac{1}{Z(\tilde{\beta},\,\Phi)}\sum_{\alpha\in\mathcal{H}} e^{-\tilde{\beta}(\Delta_\alpha-\Phi J_\alpha)}|\Delta_\alpha,\,J_\alpha\rangle\langle \Delta_\alpha,\,J_\alpha|\,,
\label{eq:dual-ensemble}
\end{equation}
where the sum is over the entire Hilbert space $\mathcal{H}$ of the $\mathcal{N}=4$ SYM gauge theory and $\tilde{\beta}=\frac{\beta}{R_{\text{AdS}}}$ is the dimensionless temperature in AdS radius units.

The 5d curvature singularity at $r=0$ in \eqref{eq:5dBH} has a non-trivial cone structure in its ten dimensional embedding \eqref{eq:10dBH}, whose microscopic interpretation in the BPS limit $\mu=0$ will be discussed in section \ref{sec:supers}. Given the discussion on the bridge length in the near-extremal limit $\mu\ll q$ \eqref{eq:bridge-div}, it is important to make sure \eqref{eq:10dBH} does not acquire quantum gravity and string effects, in this region of parameter space. Evaluating the Ricci and Riemann squared invariants at the horizon scale, one derives the condition \cite{Balasubramanian:2007bs}
\begin{equation}
\frac{1}{g_s N} \ll  \frac{\mu}{R_{\text{AdS}}^2} \ll \frac{q}{R_{\text{AdS}}^2} \sim \mathcal{O}(1)
\end{equation}
for the absence of quantum corrections.

\section{Half-BPS $\SO(4)$ invariant states in AdS/CFT}
\label{sec:review}

The BPS limit $(\mu=0)$ of the 5d R-charged black holes \eqref{eq:5dBH}, or their type IIB uplifts \eqref{eq:10dBH}, corresponds to half-BPS configurations  \cite{Behrndt:1998ns,Behrndt:1998jd} having a naked singularity at the origin of AdS \cite{Romans:1991nq} since the horizon disappears $(r_+=0)$. 

The main features of the field theory and gravity dual descriptions for the half-BPS $\SO(4)$ invariant sector of $N=4$ SYM that is relevant to understand this limit are reviewed below. The microscopics allows to interpret the singularity as a source, making it physical, and describable as an ensemble in the dual theory, as reviewed in subsection \ref{sec:supers}. The extension to localised singularities on the transverse 5-sphere is discussed in subsection \ref{sec:localised}.

\paragraph{Field theory.} Half-BPS $\SO(4)$ invariant states saturate the BPS bound $\Delta=J$, with $\Delta$ (conformal dimension) and $J$ (R-charge) associated with an $\SO(2)$ subgroup of the $\SO(2,4)$ conformal and $\SU(4)$ R-symmetry groups, respectively. 

Their physical interpretation depends on the scaling of the conformal dimension $\Delta$ with the rank $N$ of the gauge group $\SU(N)$ :  
\begin{itemize}
	\item $\Delta \sim \mathcal{O}(1)$ correspond to {\it gravitons}, pointlike particles rotating on the 5-sphere. Through the state-operator correspondence, these states are described by multitrace operators
	\begin{equation}
	  \prod_i \left(\text{tr}\,\Phi^{n_i}\right)^{m_i}\,, \quad \sum_i n_im_i=\Delta
	\end{equation}
	with $\Phi = X_1+iX_2$ and $X_i$ $i=1,\dots 6$ the six hermitian scalars in $N=4$ SYM.
	
	\item $\Delta =J \sim \mathcal{O}(N)$ correspond to {\it giant gravitons} \cite{McGreevy:2000cw} or {\it dual giants} \cite{Hashimoto:2000zp}, i.e. gravitons that expanded into spinning D3-branes, due to the Myers' effect \cite{Myers:1999ps}. In the probe approximation, giant gravitons correspond to D3-branes sitting at the origin of AdS$_5$, wrapping an S$^3$ in the transverse S$^5$ of size $\sin\theta$ while spinning along $\phi$. The relation between R-charge and 3-cycle size is 
	\begin{equation}
		\sin^2\theta = \frac{J}{N}\,, \quad \text{with} \quad ds^2(\Omega^5) = d\theta^2 + \cos^2\theta d\phi + \sin^2\theta\,ds^2(\Omega^3)\,.
	\label{giant-size}
	\end{equation}
	Their field theory description involves subdeterminant operators \cite{Balasubramanian:2001nh}
	\begin{equation}
	\text{det}_J \Phi = \frac{1}{J!}\varepsilon_{i_1\dots i_Ja_1\dots a_{N-J}}\varepsilon^{j_1\dots j_J a_1\dots a_{N-J}} \Phi^{i_1}_{j_1}\dots \Phi^{i_J}_{j_J}\,.
	\end{equation}
    In the same probe approximation, dual giants correspond to D3-branes wrapping an S$^3$ in AdS$_5$ at $\theta=0$ whose AdS radial size is determined by the R-charge
    \begin{equation}
      \frac{r}{R_{\text{AdS}}}=\sqrt{\frac{J}{N}}\,.
      \label{eq:dual-size}
    \end{equation}
    Notice that in this probe limit, the 3-sphere in AdS$_5$ has vanishing size for a giant graviton, whereas the 3-sphere in the transverse 5-sphere has vanishing size for the dual giants.
	
	\item $\Delta \sim \mathcal{O}(N^2)$ correspond to either {\it solitons} or {\it superstars}, i.e. bound states of giant gravitons. Superstars \cite{Myers:2001aq} are BPS limits of R-charged black holes \cite{Behrndt:1998ns,Behrndt:1998jd} with a naked singularity \cite{Romans:1991nq} that was interpreted as the source for a distribution of giant gravitons \cite{Myers:2001aq}
	\begin{equation}
	\frac{dn}{d\theta} = N_c\,\sin 2\theta\,, \quad N_c\sim N
	\label{gdist}
	\end{equation}		
	where $dn/d\theta$ stands for the number of giant gravitons per unit of $\theta$ angle in the 5-sphere \eqref{giant-size} and $N_c$ is the total number of giants. Solitons are BPS states with a smooth gravity dual.
\end{itemize}	

The $N=4$ SYM partition function for half-BPS $\SO(4)$ invariant states \cite{Berenstein:2005aa,Kinney:2005ej}
\begin{equation}
  Z(\nu,q) = \prod_{n=0}^\infty \frac{1}{1-\nu\,q^n}
\end{equation}
is not modified by quantum corrections. Hence it can be computed in free field theory and extrapolated to strong coupling. Furthermore, there is no phase transition at large N in this sector of the theory \cite{Kinney:2005ej}. The chemical potential for the number of D-branes is $\nu$, whereas $q=e^{-\hat\beta}$ is the chemical potential dual to R-charge $n=J$. The entropy of states with conformal dimension $\Delta= J \sim N^2$ is \cite{Berenstein:2005aa,Kinney:2005ej}
\begin{equation}
  S_{\text{1/2-BPS}} \propto N\log N\,.
\label{eq:fentropy} 
\end{equation}
This is consistent with the existence of a naked singularity in BPS superstars. Indeed, since $S_{\text{1/2-BPS}}\ll N^2$, the degeneracy of states is not enough to generate a macroscopic horizon. The addition of higher order supergravity corrections can not modify this conclusion \cite{Simon:2009mf}.

\begin{figure}
\centering
	\begin{tikzpicture}[scale=.7]
	\fill[black]
	(2.4,1.7) node[scale=1.5] {\{5,3,3,1\}};
	\draw (0,0) rectangle (1,1);
	\draw (1,0) rectangle (2,1);
	\draw (2,0) rectangle (3,1);
	\draw (3,0) rectangle (4,1);
	\draw (4,0) rectangle (5,1);
	\draw (0,-1) rectangle (1,0);
	\draw (1,-1) rectangle (2,0);
	\draw (2,-1) rectangle (3,0);
	\draw (0,-2) rectangle (1,-1);
	\draw (1,-2) rectangle (2,-1);
	\draw (2,-2) rectangle (3,-1);
	\draw (0,-3) rectangle (1,-2);
	\fill[black]
		(9,1.7) node[scale=1.5] {\{4,4,2\}};
		\draw (7,0) rectangle (8,1);
		\draw (8,0) rectangle (9,1);
		\draw (9,0) rectangle (10,1);
		\draw (10,0) rectangle (11,1);
		\draw (7,-1) rectangle (8,0);
		\draw (8,-1) rectangle (9,0);
		\draw (9,-1) rectangle (10,0);
		\draw (10,-1) rectangle (11,0);
		\draw (7,-2) rectangle (8,-1);
		\draw (8,-2) rectangle (9,-1);
	
	\end{tikzpicture}
\caption{Representation of excited states in terms of the integers $r_1 \leq \lambda_2 \leq \dots \leq \lambda_N$ defined in \eqref{eq:exc} using Young tableau.}
\label{ytableau}
\end{figure}
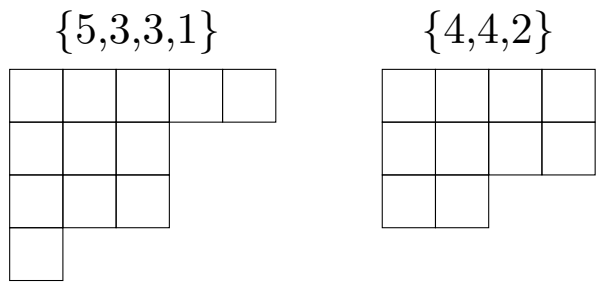

Half-BPS $\SO(4)$ invariant states can be described by $N$ free fermions in a 1d harmonic potential \cite{Corley:2001zk,Berenstein:2004kk}. Eigenstates of the hamiltonian are then labelled by an increasing set of $N$ integers $n_1 < n_2 < \dots < n_N$. These can be mapped to a Young tableau (YT) by recording the set of excitations above the Fermi sea
\begin{equation}
  r_i = n_i - i + 1\,, \quad \quad i=1,\dots , N
\label{eq:exc}
\end{equation}
as the number of boxes $r_i$ in the i-th row of the Young tableau, as indicated in figure \ref{ytableau}. In this description, the number of giant gravitons $N_c$ is approximately given by the number of excited columns in the tableau \cite{Corley:2001zk}\footnote{Further CFT evidence to support this picture includes \cite{deMelloKoch:2012ck}, together with \cite{Lin:2010sba,Lin:2012ey,Bissi:2011dc}. There is an extensive  literature on the subject. Readers interested in learning on the open string description of giant gravitons and its relation to N=4 SYM may consult \cite{Balasubramanian:2004nb,deMelloKoch:2007rqf,deMelloKoch:2007nbd,Koch:2008cm,deMelloKoch:2009jc} and references therein.}.

Let $\mathcal{H}_{\text{LLM}}$ be the Hilbert space spanned by these states. Eigenstates of the Hamiltonian will be denoted either by $|\Psi_{\vec{n}}\rangle$. The wavefunction of an individual fermion with excitation $n$ is given by 
\begin{equation}
  \psi_n(x) = \left[\frac{\alpha}{\sqrt{\pi}\,2^n\,n!}\right]^{1/2}\,e^{-\alpha^2 x^2/2}\,H_n(\alpha\,x)\quad \text{with}\quad \alpha = \sqrt{\frac{m\omega}{\hbar}}\,,
\end{equation}
with $H_n(\alpha x)$ the n-th Hermite polynomial of degree $n$. The N-particle wave function is given by the Slater determinant
\begin{equation}
  \Psi_{\vec{n}}(x_1,\dots x_N) = \frac{1}{\sqrt{N!}}\,\text{det}\left[\psi_{n_i}(x_j)\right]_{1\leq i,j\leq N}
\label{eq:slater}
\end{equation}
In particular, the wave function of the ground state is
\begin{equation}
  \Psi_{\text{vac}}(x_1,\dots x_N) \propto e^{-\frac{\alpha^2}{2}\sum_{i=1}
  	^N x_i^2}\,\text{det}\left[H_i(\alpha x_j)\right]\,,
\end{equation}
whereas its quantum probability density satisfies
\begin{equation}
  |\Psi_{\text{vac}}(x_1,\dots x_N)|^2 \propto \prod_{i<j} (x_i-x_j)^2\,e^{-\alpha^2\sum_{i=1}^N x_i^2}\,.
\label{eq:q-prob}
\end{equation}
Noting that the joint distribution of eigenvalues $\{\lambda_1,\dots \lambda_N\}$ of a random $N\times N$ complex Hermitian matrix $X$ with independent gaussian entries is given by \cite{mehta}
\begin{equation}
  P(\lambda_1,\dots \lambda_N) \propto \prod_{i<j} (\lambda_i-\lambda_j)^2\,e^{-\sum_{i=1}^N \lambda_i^2}\,,
\label{eq:m-theory}
\end{equation}
one concludes the quantum statistics of the fermion positions $x_i$ in a 1d harmonic potential at vanishing temperature is given, up to a trivial rescaling by $\alpha$, by the statistics of eigenvalues of a gaussian unitary ensemble\footnote{The Vandermonde term $\prod_{i<j} (\lambda_i-\lambda_j)^2$ has its origin in the Jacobian of the change of variables $X\to U^\dagger\, X\,U$ diagonalising the joint probability distribution $\text{Prob}[X]\,dX \propto e^{-\text{Tr}[X^2]}\,dX$, whereas the term $\prod_{i<j} (x_i-x_j)^2$ in the harmonic oscillator quantum probability originates from $\text{det}\left[H_i(\alpha x_j)\right]\propto \text{det}\,[x_i^{j-1}]\propto \prod_{i<j}(x_i-x_j)^2$.}.

\paragraph{Gravity.} The classical moduli space of gravity duals to half-BPS $\SO(4)$ invariant states was worked out in \cite{Lin:2004nb}. They are referred to as LLM configurations. These are half-supersymmetric type IIB supergravity on-shell solutions with $\mathbb{R}\times\SO(4)\times\SO(4)$ isometry group, metric
\begin{equation}
\begin{aligned}
	ds^2 &= - \frac{y}{\sqrt{\frac{1}{4}-z^2}}(dt + V_i dx^i)^2 + \frac{\sqrt{\frac{1}{4}-z^2}}{y} (dy^2 + dx^idx^i) \\
	&+ y\, \sqrt{\frac{\frac{1}{2}+z}{\frac{1}{2}-z}} d\Omega_3^2
	+ y\, \sqrt{\frac{\frac{1}{2}-z}{\frac{1}{2}+z}} d \tilde \Omega_3^2\,, \quad i=1,2
\end{aligned}
\label{eq:llm-metric}
\end{equation}
constant dilaton, RR 5-form $F_{(5)} = F\wedge d\Omega_3 + \tilde{F}\wedge d\tilde{\Omega}_3$ given in terms of the volume 3-forms on the 3-spheres $d\Omega_3$ and $d\tilde{\Omega}_3$ (see \cite{Lin:2004nb} for further details on the 2-forms $F$ and $\tilde{F}$) and all remaining type IIB bosonic fields vanishing.

Time translations generate $\mathbb{R}$. The bulk Killing vector field $k=\partial_t$ gives rise to a conserved charge, the mass $M$, which is related to the conformal dimension $\Delta=R_{\text{AdS}}\,M$, the eigenvalue of the dilatation operator in the field theory dual. By supersymmetry, $\Delta=J$, where $J$ is the eigenvalue associated with an $\SO(2)$ subgroup of the $\SO(6)$ R-symmetry group. The geometrical action of this $\SO(2)$ corresponds to rotations in the plane spanned by $x^1$ and $x^2$. Introducing polar coordinates $r$ and $\phi$, this action is adjusted to the Killing vector $k=\partial_\phi$. Hence, these transformations correspond to rotations in the $\phi$ direction of the transverse 5-sphere in \eqref{giant-size}. The $\SO(4)$ groups are realised as the isometries of two 3-spheres, the first at the boundary of AdS$_5$ $(d\Omega_3)$ and the second as a submanifold of the asymptotic transverse 5-sphere $(d\tilde\Omega_3)$. Hence, restricting our attention to the subset of configurations having $\mathbb{R}\times\SO(2)\times\SO(4)\times\SO(4)$ isometry group, these geometries depend on two coordinates : the radial coordinate $r$ in the $x^1$ and $x^2$ plane, together with $y$. As reviewed below, these encode the information on the standard global radial coordinate in AdS$_5$ together with the $\theta$ angle in the 5-sphere, as in \eqref{giant-size}.

LLM solutions depend on a single scalar function
\begin{equation}
z(y;\,x_1,\,x_2) = \frac{y^2}{\pi}\int dx_1'\,dx_2'\,
\frac{z(0;\,x_1',\,x_2')}{[(x-x')^2 + y^2]^2}~.
\label{eq:LLMscalar}
\end{equation}
which is uniquely determined by its boundary condition $z(0;\,x_1',\,x_2')$ on the $y=0$ plane (LLM plane from now on). The subset of smooth configurations corresponds to the subset of boundary conditions satisfying $z(0;x_1,x_2) = \pm \frac{1}{2}$ in a compact region of the LLM plane (droplets from now on) \cite{Lin:2004nb}.

Smoothness requires one of the 3-spheres to shrink to zero size while the second remains finite. LLM solutions may have a rich topological structure . Associating the colour black to $z(0;x_1,x_2) = - \frac{1}{2}$ and the colour white to $z(0;x_1,x_2) = \frac{1}{2}$, one can construct topologically non-contractible 5-spheres as follows. Consider a surface $\Sigma_2$ on the $(y,x_1,x_2)$ space ending on the $y=0$ plane on a closed, non-intersecting curve in a black (white) region. Fibering the finite size 3-sphere over $\Sigma_2$ gives rise to such a smooth 5-sphere \cite{Lin:2004nb}. Choosing different surfaces $\Sigma_2$ can give rise to different 5-manifolds. Hence, the larger the number of black and white regions, the richer its topological structure \cite{Lin:2004nb}. 

\begin{figure}
	\centering
	\begin{tikzpicture}[scale=0.7,every node/.style={minimum size=1cm},on grid]
	
	\begin{scope}[
	yshift=0,
	every node/.append style={yslant=\yslant,xslant=\xslant},
	yslant=\yslant,xslant=\xslant
	]
	\fill[white,fill opacity=.75] (0,0) rectangle (7,7); 
	\draw[black, dashed, thin] (0,0) rectangle (7,7); 
	
	\draw[fill]  (3.5,3.5) circle [radius=1];
	\fill[black]
	(0.5,6.5) node[right, scale=1] {$y=0$-plane};
	
	\filldraw[ball color=red] (3.5,2.5) circle (1.5);
	\draw[black, dashed, thin]  (3.5,3.5) circle [radius=1];
	\draw[black, dashed, thin]  (3,2.63) -- (3,4.3) ;
	\draw[black, dashed, thin]  (3.5,2.5) -- (3.5,4) ;
	\draw[black, dashed, thin]  (3.6,3.99) -- (3.6,2.5) ;
	\draw[black, dashed, thin]  (3.7,3.99) -- (3.7,2.52) ;
	\draw[black, dashed, thin]  (3.8,3.97) -- (3.8,2.55) ;
	\draw[black, dashed, thin]  (3.9,3.94) -- (3.9,2.58) ;
	\draw[black, dashed, thin]  (4,4.3) -- (4,2.63) ;
	\draw[black, dashed, thin]  (4.1,3.87) -- (4.1,2.7) ;
	\draw[black, dashed, thin]  (4.2,3.83) -- (4.2,2.79) ;
	\draw[black, dashed, thin]  (4.3,3.77) -- (4.3,2.9) ;
	\draw[black, dashed, thin]  (2.9,3.87) -- (2.9,2.7) ;
	\draw[black, dashed, thin]  (2.8,3.83) -- (2.8,2.78) ;
	\draw[black, dashed, thin]  (2.7,3.77) -- (2.7,2.86) ;
	\draw[black, dashed, thin]  (4.36,3) -- (2.63,3) ;
	\draw[black, dashed, thin]  (4.3,2.9) -- (2.7,2.9) ;
	\draw[black, dashed, thin]  (4.21,2.8) -- (2.79,2.8) ;
	\draw[black, dashed, thin]  (4.1,2.7) -- (2.9,2.7) ;
	\draw[black, dashed, thin]  (3.9,2.6) -- (3.06,2.6) ;
	\fill[red] (4.5,1.5) node[right, scale=1.5] {$\Sigma_2$};
	\end{scope} 
	
	\end{tikzpicture}
	
	\caption{2-dimensional surface $\Sigma$ where to evaluate the number of fermions in phase space insight $\Sigma$ by computing the flux of the RR 5-form over a 5-dimensional surface including a finite size 3-sphere.}
	\label{LLMcharge}	
\end{figure}
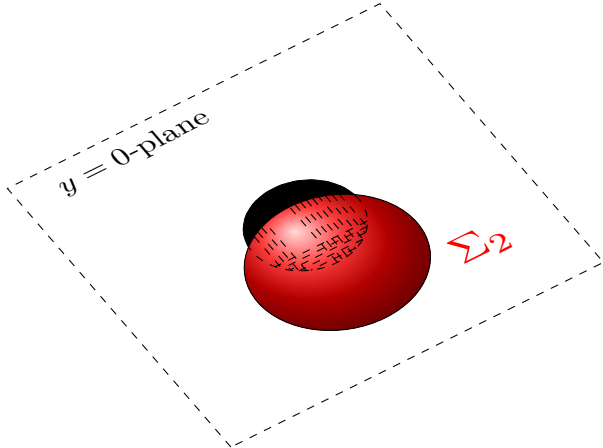

As explained in \cite{Lin:2004nb}, the RR 5-form flux across these 5-manifolds equals the area of its intersection with $\Sigma_2$ (as sketched in figure \ref{LLMcharge}). This is a quantised charge in the full theory
\begin{equation}
N_{\Sigma_2} = \frac{\text{Area}_{z(0;x_1,x_2)=-\frac{1}{2}}}{4\pi^2\ell_p^4}\,.
\label{LLM-charge}
\end{equation}
measuring the number of fermions in $\Sigma_2$.

\subsection{Holographic dictionary}
\label{sec:dictionary} 

The conformal dimension $\Delta = MR_{\text{AdS}}$ of these solutions equals \cite{Lin:2004nb}
\begin{equation}
\Delta = J = \int_{\RR^2}
\frac{d^2x}{2\pi\hbar}\,\frac{1}{2}\frac{x_1^2+x_2^2}{\hbar}\,
u(0;\,x_1,\,x_2)-\frac{1}{2}\left(\int_{\RR^2}
\frac{d^2x}{2\pi\hbar}\, u(0;\,x_1,\,x_2)\right)^2\,,
\end{equation}
where $u(0;\,x_1,\,x_2) = \frac{1}{2} - z(0;\,x_1,\,x_2)$. Hence, for smooth geometries, it only receives contributions from the regions in the LLM plane where the droplet is black, i.e. where $u(0;\,x_1,\,x_2)=1$. The normalisation $\hbar = 2\pi\ell_p^4$ is fixed by the RR 5-form flux quantisation condition
\begin{equation}
N =  \int_{\RR^2} \frac{d^2x}{2\pi\hbar}\,u(0;\,x_1,\,x_2)\,,
\end{equation}
ensuring the total number of fermions is $N$. 

These conserved charges allow to reinterpret the $y=0$ LLM plane as the phase space of a single fermion, with $u(0;\,x_1,\,x_2)$ being the semiclassical reduced phase space density of the full state\footnote{This also justifies why the normalisation $2\pi \ell_p^4$ is labelled as $\hbar$, since it plays the role of Planck's constant in the quantum mechanics defined on this phase space.}. It can be shown that the difference between phase space distributions can not be distinguished in the semiclassical limit $N\to \infty$ keeping $\hbar N$ fixed \cite{Balasubramanian:2005mg,Skenderis:2007yb}.

Consider the metric of global $AdS_5\times S^5$ 
\begin{equation}
ds^2 = -\left(r^2 +R_{\text{AdS}}^2\right)\,dt^2 + \frac{dr^2}{1+\frac{r^2}{R_{\text{AdS}}^2}} + r^2\,d\Omega_3^2 + R_{\text{AdS}}^2\left[d\theta^2 + \cos^2\theta\,d\tilde{\phi}^2 + \sin^2\theta\,d\tilde{\Omega}_3^2\right]\,.
\label{globalAdS}
\end{equation}
This is mapped to its LLM description by the diffeomorphism \cite{Lin:2004nb} 
\begin{equation}
\begin{aligned}
y = r\,R_{\text{AdS}}\,\sin\theta\,, & \quad r = R_{\text{AdS}}\,\sqrt{r^2+R_{\text{AdS}}^2}\,\cos\theta\,, \\
\tilde{\phi}&=\phi + t 
\end{aligned}
\label{vac-map}
\end{equation}
This corresponds to a rotationally invariant droplet with boundary conditions
\begin{equation}
  u(0; r,) = \Theta(R_{\text{AdS}}^2 - r) = \left\{
  \begin{array}{cc}
  1 & r\leq R_{\text{AdS}}^2 \\
  0 & r > R_{\text{AdS}}^2
 \end{array}
\right.
\label{eq:vac-dict}
\end{equation}
where the size of the droplet $r_0$ was already replaced by its relation to the AdS radius $r_0^2 = R_{\text{AdS}}^4 = 2\hbar\,N$.
Integration over momentum\footnote{The cartesian coordinates $(x_1,x_2)$ in LLM correspond to the position $x$ and momentum $p$ of a single harmonic oscillator phase space. Hence, $u(0; r,\,\phi)=u(0; x,\,p)$.} equals Wigner's semicircle distribution \cite{mehta}
\begin{equation}
\rho_1(x) = \int u(0; x,\,p)\,\frac{dp}{2\pi\hbar} = 2\int_0^{\sqrt{R_{\text{AdS}}^4-x^2}} \frac{dp}{2\pi\hbar} = \frac{1}{\pi\hbar}\sqrt{R_{\text{AdS}}^4-x^2}\,,
\end{equation}
using the normalisation $\text{tr}\,\rho_1 = N$. This bulk statement is consistent with the quantum mechanical equivalence between \eqref{eq:q-prob} and \eqref{eq:m-theory}.

The connection between classical gravity and the quantum fermion description requires the semiclassical limit $\hbar\to 0$, keeping $\hbar\,N$ fixed. Using the WKB approximation, the individual fermion wave functions $\psi_k(\alpha x)$ for large $k$ (and large $x$) become \cite{2016arXiv160904366D} 
\begin{equation}
\psi_k (\alpha x) \sim \left(\frac{2}{k}\right)^{1/4}\,\sqrt{\frac{\alpha}{\pi}}\,\frac{1}{(1-X^2)^{1/4}}\,g_k(X) \quad \text{with} \quad X = \frac{\alpha x}{\sqrt{2k}}\in(-1,\,1)\,,
\label{wf-approx}
\end{equation}
where $g_k(X) = \cos\left(MX\sqrt{1-X^2} +(M+1/2)\arcsin X - M\pi/2\right)$. Using this expression in the Slater determinant \eqref{eq:slater}, the phase space density reduces to
\begin{equation}
  \rho_1(x,p) = \rho_1(r,\phi) = \Theta (R_{\text{AdS}}^2 - r)\,,
\label{eq:phase-check}
\end{equation}
matching the gravity LLM boundary condition \eqref{eq:vac-dict}. Hence, the boundary conditions giving rise to smooth gravity configurations capture the regions of the single particle phase space where quantum fermions are excited (black droplet or $z_0=-\frac{1}{2}$) or unexcited (white droplet or $z_0=\frac{1}{2}$)), in the semiclassical limit.

For completeness, though it will not play a role in this work, Wigner's semicircle distribution does not capture the behaviour close to the edges of the droplet, where the density matrix for finite but large $N$ can be approximated by \cite{2016arXiv160904366D}
\begin{equation}
  \rho(x) \approx \frac{1}{\omega_N}\,F_1\left[\frac{x-\sqrt{2N}/\alpha}{\omega_N}\right] \quad \text{with} \quad \omega_N = \frac{1}{\alpha\,\sqrt{2}}\,N^{-1/6}\,,
\end{equation}
with $F_1(z)=[\text{Ai}^\prime(z)]^2 - z[\text{Ai}(z)]^2$ and $\text{Ai}(z)$ being the Airy function and $\text{Ai}^\prime(z)$ its derivative. 

This picture extends to rotationally invariant excitations of order N fermions giving rise to concentric rings in phase space \cite{Lin:2004nb,Ghodsi:2005ks}. For example, excited states with energy 
\begin{equation}
\Delta = \sum_{k=2}^p N_kM_k
\end{equation}
describing $N_1$ fermions in the Fermi sea, together with $N_k$ fermions carrying $M_k$ quanta, with $M_{k+1}> M_k$, have a one particle phase space density consisting of an inner black disk of radius $r_1$, followed by a collection of white and black annula with radia $r_{2k}$ and $r_{2k+1}$, respectively, given by
\begin{equation}
  \frac{r_{2s}^2}{R_{\text{AdS}}^4} = \frac{M_{s+1}}{N} + \sum_{a=1}^s \frac{N_a}{N}\,, \quad \frac{r_{2s+1}^2-r_{2s}^2}{R_{\text{AdS}}^4}= \frac{N_{s+1}}{N}\,, \quad s=1,\dots , p-1
\end{equation}
The second condition ensures all black annula encode the right number $N_{s+1}$ of excited fermions, whereas the first one matches the right excitation energy compatible with $N=\sum_{i=1}^{p}N_i$. These relations make explicit the necessity of both sets of data $\{N_1,\,N_k\}$ and $\{M_k\}$ to scale with $N$ in order to have a geometric description.

To ease the comparison with the holographic description of these states, it is convenient to introduce density matrices $\rho_N(\vec{n})=|\Psi_{\vec{n}}\rangle\langle \Psi_{\vec{n}}|$ satisfying the standard quantum mechanics normalisation $\text{Tr}\rho_N(\vec{n}) = 1$. Given a single Young tableau with quantum numbers $n_i$, its one particle reduced density matrix equals
\begin{equation}
\rho_1 = \frac{1}{N}\sum_{n_j}|n_j\rangle\langle n_j|\,.
\end{equation}
For the ground state, 
\begin{equation}
\rho_1 = \frac{1}{N}\sum_{i=0}^{N-1} |i\rangle\langle i|\,,
\end{equation}
whereas for a smooth soliton 
\begin{equation}
  \rho_1 = \frac{1}{N}\sum_{i=0}^{N_1-1} |i\rangle\langle i| + \frac{1}{N}\sum_{k=1}^{p-1}\sum_{i=N_k}^{N_{k+1}-1} |i+M_{k+1}\rangle\langle i+M_{k+1}|
\label{eq:rho-soliton}  
\end{equation}
Because of the scaling with $N$, the same WKB approximations used to derive \eqref{eq:phase-check} allow us to derive the semiclassical phase space correspondence
\begin{equation}
  u_{\text{soliton}}(0;r) = \sum_{k=1}^{p-1} (-1)^{k+1}\Theta (r_{k}-r)\,.
\label{eq:soliton-class}
\end{equation}

The supergravity description of these configurations was already worked out in \cite{Lin:2004nb}. Introducing polar coordinates so that $V_i dx^i\equiv V_\phi\,d\phi$, to reflect the rotational invariance of these states, a single black droplet of radius $r_i$ is described by \cite{Lin:2004nb}
\begin{equation}
\begin{aligned}
  z(y;r;r_i) &= f_-(r_i)\,, \quad \quad V_\phi (y;r;r_i) = \frac{1}{2}- f_+(r_i)\,, \\
  f_\pm (r_i) &= \frac{r^2+y^2\pm r_i^2}{2\sqrt{(r^2+y^2+r_i^2)^2-4r^2r_i^2}}\,.
\end{aligned}
\end{equation}
Given the linearity of the Laplace equation, solutions corresponding to different black and white annulus are given by \cite{Lin:2004nb}
\begin{equation}
z(y;r)= \sum_i (-1)^{i+1}\,z(r,y;r_i)\,, \quad \quad V_\phi(y;r) = \sum_i (-1)^{i+1}\,V_\phi(r,y;r_i)
\label{eq:smoothring}
\end{equation}
Hence, the gravity boundary condition $z(0;r)$ reproduces the semiclassical limit of the single particle quantum wave functions in  \eqref{eq:soliton-class}. 

The smoothness of the LLM geometry inherited from the phase space boundary condition stems from the semiclassical limit of the quantum mechanical wave functions, but also from the lack of uncertainty in the excitation of any of the fermions. To stress this last point, consider the smooth solitons dual to \eqref{eq:rho-soliton}. They are described by as many excitation levels as fermions, so that the reduced density matrix assigns the same probability, i.e. $\frac{1}{N}$, to each of them. The further condition of having a semiclassical limit, requires these excitations to be composed of order $N$ fermions to give rise to a smooth droplet \eqref{eq:soliton-class} when using the WKB approximation \eqref{wf-approx}. As soon as the number of excitations is larger than the number of fermions, the reduced phase space density will be of the form
\begin{equation}
  \rho_1 = \sum_j \frac{\alpha_j}{N} \sum_{i=N_j}^{N_{j+1}-1} |i\rangle\langle i|\,\,.
\label{eq:singrho1}
\end{equation}
This gives rise to a singular LLM geometry whenever there exists any $\alpha_j\neq 0,1$ compatible with a phase space density describing N fermions while surviving the semiclassical limit $\hbar\to 0$ with $\hbar\,N$ fixed. The emergence of these ensembles as effective descriptions of the singular BPS limits of non-extremal R-charged black holes is discussed next \cite{Balasubramanian:2005mg}.

\subsection{Superstars as ensembles and typical states} 
\label{sec:supers}

In holographic discussions, it is always important to understand the bulk description of thermal states \cite{Witten:1998zw}. In the half-BPS $\SO(4)$ invariant sector of $N=4$ SYM, this was achieved in \cite{Balasubramanian:2005mg}, where an interpretation for the BPS limit of the ensemble \eqref{eq:dual-ensemble} was put forward using the statistical mechanics of the free fermions and reproducing the giant graviton distribution \eqref{gdist} advocated in \cite{Myers:2001aq}.

\paragraph{Gravity description.} The naked singularity \cite{Romans:1991nq} emerging in the BPS limit of R-charged black holes \cite{Behrndt:1998ns,Behrndt:1998jd} at the origin of AdS $(r=0)$ must correspond to a {\it singular} LLM droplet boundary condition, i.e. $z(y=0;r_{\text{LLM}})\neq \pm \frac{1}{2}$\footnote{In this subsection, we introduced the subindex LLM to avoid any confusion with the radial coordinate $r$ used to describe previous 5d black holes \eqref{eq:5dBH} and their type IIB uplifts \eqref{eq:10dBH}.}, because it preserves the same symmetries LLM configurations do and the classification in \cite{Lin:2004nb} is complete. In the following, the relation between the description of this singularity in both coordinate systems is established.

The proper discussion of the matching between both descriptions is presented in appendix \ref{sec:matching}. The radial LLM coordinates $y,\,r_{\text{LLM}}$ are mapped to the superstar radial coordinate $r$ and the azimutal angle in the transverse 5-sphere $\theta$ in \eqref{eq:10dBH} by
\begin{equation}
y = rR_{\text{AdS}}\sin\theta\,, \quad \quad r^2_{\text{LLM}}=R_{\text{AdS}}^2\cos^2\theta\,\left(r^2+q+R_{\text{AdS}}^2\right)\,,
\label{eq:sing-map}
\end{equation}
while the scalar function determining the LLM geometry is identified as
\begin{equation}
  z(y;r_{\text{LLM}}) = \frac{1}{2}\frac{r^2 + \sin^2\theta (q-R_{\text{AdS}}^2)}{r^2+ \sin^2\theta (q+R_{\text{AdS}}^2)}~.
\label{eq:z-superstar}
\end{equation}
Since the LLM boundary condition corresponds to $z(0,r_{\text{LLM}})$, i.e. evaluating $z(y,r_{\text{LLM}})$ in the LLM plane $y=0$, let us examine the values of the scalar function \eqref{eq:z-superstar} on the latter. According to \eqref{eq:sing-map}, this corresponds to
\begin{enumerate}
	\item either $r=0$. The radial LLM coordinate equals $r^2_{\text{LLM}}= R_{\text{AdS}}^2\cos^2\theta\,\left(q+R_{\text{AdS}}^2\right)\in [0,\,r^2_{\text{sup}})$. Hence, the AdS origin $r=0$ corresponds to the interior of a finite LLM droplet of size $r_{\text{sup}}=R_{\text{AdS}}\,\sqrt{q+R_{\text{AdS}}^2}$. The LLM boundary condition reduces to
	\begin{equation}
	z(0;r_{\text{LLM}}) = \frac{1}{2}\frac{\omega -1}{\omega +1} \quad \Leftrightarrow \quad u(0;r) = \frac{1}{1 + \omega}\,, \quad\text{with} \quad \omega = \frac{N_c}{N}=\frac{q}{R_{AdS}^2}\,.
	\label{sstar}
	\end{equation}
	Hence, this is a {\it singular} droplet. Motion in the radial LLM coordinate inside the finite droplet is equivalent to motion in the $\theta$ direction in the transverse 5-sphere, whereas rotational motion inside the droplet is motion along the $\U(1)$ direction in the same 5-sphere. 
	\item or $\sin\theta=0$. The radial LLM coordinate $r^2_{\text{LLM}}= R_{\text{AdS}}^2\,\left(r^2 + r^2_{\text{sup}}\right) \in [r^2_{\text{sup}},\,\infty)$ explores the interior of the AdS$_5$ space, away from the origin. When this holds, $z(0;r_{\text{LLM}})=\frac{1}{2}$, so that the phase space distribution $u(0;r_{\text{LLM}})$ vanishes. Hence, this region corresponds to the outside of the finite singular droplet and it contains no singularity.
\end{enumerate}
Altogether, the LLM boundary condition for the superstar background is summarised by
\begin{equation}
u(0; r_{\text{LLM}}) = \frac{1}{1+\omega}\,\Theta(r^2_{\text{sup}} - r_{\text{LLM}}) = \left\{
\begin{array}{cc}
\frac{1}{1 + \omega} & r_{\text{LLM}} \leq r^2_{\text{sup}} \\
0 &  r_{\text{LLM}} > r^2_{\text{sup}}
\end{array}
\right.
\label{eq:supers-dict}
\end{equation}
In the microscopic fermion picture, there are {\it no} fermion excitations above the energy scale set by $r^2_{\text{sup}}$. Below this scale, there is a constant probability of finding an individual fermion inside the droplet. The existence of the singularity is captured by this constant probability not being equal to zero or one, i.e. by being compatible with the ensemble \eqref{eq:singrho1}.  This suggests the proper description is in terms of an ensemble, which matches the AdS/CFT intuition, since non-extremal black holes are believed to be described by ensembles and the superstar configurations correspond to the BPS limit of these. 

In the discussion below, we derive both the LLM singular boundary condition and the distribution of giant gravitons \eqref{gdist} supporting it by analysing the statistical mechanics of the N free fermions with energy of order $N\,N_c$ describing, at most, $N_c$ giant gravitons \cite{D'Errico:2007jm,Balasubramanian:2005mg}. This gives further evidence for the microscopic interpretation of the bulk singularity.

\paragraph{Quantum statistical mechanics matching.} The ensemble of $\SO(4)$ invariant half-supersymmetric states in $N=4$ SYM corresponds to the limit $\tilde{\beta}\to \infty$ keeping $\tilde{\beta}(1-\Phi)=\hat{\beta}$ fixed in \eqref{eq:dual-ensemble}. This truncates the sum over the Hilbert space of $\SO(4)$ symmetric states to the half-BPS sector $\Delta_\alpha=J_\alpha$. The corresponding density matrix matrix simplifies to\footnote{Regarding footnote 5, notice the diffeomorphism mapping the superstar and LLM descriptions discussed in appendix \ref{sec:matching} implements the large gauge diffeomorphism $\varphi_{\text{LLM}} + t_{\text{LLM}}=\varphi$ (see \eqref{eq:large-gauge}) relating the two gauge field boundary conditions.}
\begin{equation}
\rho(\beta,\,\mu) \to \rho(\hat{\beta})= \frac{1}{Z(\hat{\beta})}\sum_{\alpha\in \mathcal{H}_{\text{LLM}}} e^{-\hat{\beta}\Delta_\alpha}|\Delta_\alpha,\,\Delta_\alpha\rangle\langle \Delta_\alpha,\,\Delta_\alpha|\,.
\label{BPSrho}
\end{equation}

This ensemble can reproduce the average energy of the superstar, but it does so including tails of arbitrary large number of giant gravitons \cite{Buchel:2004mc,Balasubramanian:2005mg}. To take into account the physical constraint on such number of giant gravitons, reference \cite{Balasubramanian:2005mg} modified the above ensemble by introducing a Lagrange multilplier and derived the density \eqref{sstar} in the infinite effective temperature $\hat{\beta}\to 0$ limit. Later, it was realised \cite{D'Errico:2007jm} that an ensemble with smoother fluctuation behaviour is one summing over states having at most $N_c$ giant gravitons. Let us denote this Hilbert space by $\mathcal{H}^\prime$. Then, the superstar corresponds to the $\hat{\beta}\to 0$ limit, i.e. to the maximally entangled state in $\mathcal{H}^\prime$
\begin{equation}
\rho_{\text{superstar}} = \frac{1}{Z} \sum_{\vec{n}\in \mathcal{H}^\prime} |\Psi_{\vec{n}}\rangle\langle \Psi_{\vec{n}}|\,, \quad \quad Z = \binom{N+N_c}{N}
\label{eq:uniform}
\end{equation}
Notice how the resulting ensemble is effectively cutting off the energy in the standard canonical ensemble in \eqref{BPSrho} and maximises the entropy in the Hilbert space $\mathcal{H}^\prime$ \cite{D'Errico:2007jm,Balasubramanian:2005mg}
\begin{equation}
  S_{\text{superstar}} = \log Z \approx -N\log \frac{\omega^\omega}{(1+\omega)^{1+\omega}} \quad \text{with} \quad \omega = \frac{N_c}{N}\,. 
\label{eq:sstarentropy}
\end{equation}
As in \eqref{eq:fentropy}, the scaling in $N$ is not enough to generate a macroscopic horizon, which is again consistent with the existence of a naked singularity in the superstar geometry. 

This density matrix \eqref{eq:uniform} reproduces the singular boundary condition \eqref{sstar}. Indeed, its single particle reduced density matrix equals
\begin{equation}
\rho_1 = \frac{1}{N\,Z} \binom{N+N_c-1}{N-1} \sum_{i=1}^{N+N_c} |i\rangle\langle i|= \frac{1}{N+N_c}\sum_{i=1}^{N+N_c}|i\rangle\langle i| =\frac{1}{N}\frac{1}{1+\omega} \sum_{i=1}^{N+N_c} |i\rangle\langle i| \,,
\end{equation}
where the index $i$ labels single particle excitations. Notice that as soon as the ensemble describes a non-vanishing number of giant gravitons $(N_c\neq 0)$, the coefficients of $\rho_1$ differ from $\{0,N^{-1}\}$. Hence, its gravity dual will be singular, corresponding to a gray disk
\begin{equation}
  u_{\text{superstar}}(0;r) = \frac{1}{1+\omega} \Theta (r^2_{\text{sup}} - r^2) \quad \text{with} \quad r^2_{\text{sup}}= R^4_{\text{AdS}}\,(1+w)\,,
\label{eq:r-super}
\end{equation} 
explicitly matching the bulk boundary condition \eqref{eq:supers-dict}.

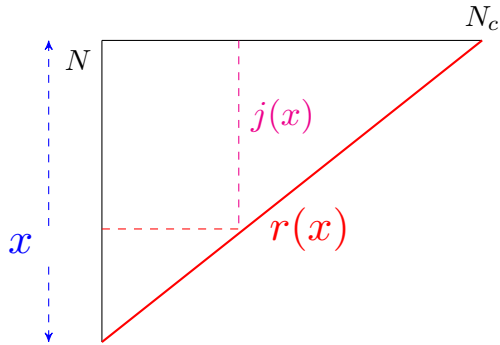
\begin{figure}
\centering
\begin{tikzpicture}
\draw (0,0) to (5,0);
\node [below left] at (0,0) {$N$};
\draw[<-, blue, dashed, thin] (-0.7,0) to (-0.7,-2.5);
\fill[blue] (-0.7,-2.7) node[left, scale=1.5] {$x$};
\draw[->, blue, dashed, thin] (-0.7,-3.0) to (-0.7,-4);
\draw (0,0) to (0,-4);
\draw[red, thick] (0,-4) --  (5,0);
\node [above] at (5,0) {$N_c$};
\draw[red, dashed, thin] (0,-2.5) to (1.8,-2.5);
\draw[magenta, dashed, thin] (1.8,0) to (1.8,-2.5);
\fill[magenta] (1.8,-1) node[right, scale=1.2] {$j(x)$};
\fill[red] (2,-2.5) node[right, scale=1.5] {$r(x)$};
\end{tikzpicture}
\caption{Triangular Young tableau describing the ensemble \eqref{eq:uniform}. It is characterised by the limit curve $r(x)$ describing the excitation above the Fermi sea of the fermion $x$. The length of the column $j(x)$ describes the R-charge carried by the giant graviton in that column.}
\label{fig:lcurve}
\end{figure}

The same conclusion is reached using the notion of {\it limit curve} capturing the shape of {\it typical} YTs \cite{Balasubramanian:2005mg}. In the large $N$ limit, one can approximate the discrete nature of the YT by a continuous YT described in terms of the curve $r(x)$ measuring the excitation of the x-th fermion. Combining this general large $N$ feature with the expectations values derived from the previous ensembles, one learns the typical shape is a triangular YT, as illustrated in figure \ref{fig:lcurve}. Combining this general large $N$ feature with the expectations values derived from any ensemble of YTs, one learns the shape of typical pure states. When applying these ideas to the previous ensembles, one derives
\begin{equation}
 r(x) = \int^N_{N-x} di\,\langle c_i \rangle = \frac{N_c}{N}\,x\,,
\end{equation}
where we used the result $\langle c_j \rangle = \frac{N_c}{N}$ \cite{D'Errico:2007jm,Balasubramanian:2005mg}. 

Using the phase space interpretation of the droplet boundary condition $u(0;r_{\text{LLM}})$ characterising the classical gravity solution in terms of the semiclassical limit of the single particle reduced density matrix, the two different perspectives described above can be matched \cite{Balasubramanian:2005mg}. Indeed,
\begin{itemize}
	\item the number of fermions $dx$ in the continuous Young tableau picture equals the phase space measure
	\begin{equation}
	  \frac{u(0;r_{\text{LLM}})}{2\hbar}\,dr_{\text{LLM}}^2 = dx\,. 
	\end{equation}
	\item the energy of the x-th fermion must equal the hamiltonian energy
	\begin{equation}
	  \frac{r_{\text{LLM}}^2}{2\hbar}= r(x)+x\,,
	\label{eq:babel1}  
	\end{equation}
	where $x$ describes the Fermi sea energy in the semiclassical limit.
\end{itemize}
Combining these requirements, a relation between the slope of the limit curve and the phase space density is derived \cite{Balasubramanian:2005mg}
\begin{equation}
  u(0;r_{\text{LLM}}) \equiv \frac{1}{1+r^\prime} = \frac{1}{1+\omega}=  u_{\text{superstar}}(0;r_{\text{LLM}})\,.
\end{equation}

\paragraph{Matching the giant graviton distribution.} The typical YT analysis also reproduces the distribution of giant gravitons \eqref{gdist}. 
Before rederiving this relation, it is important to further stress the holographic relation between the radial coordinate in the LLM plane $(r_{\text{LLM}}|_{r=0})$ and the azimutal angle $\theta$ in the transverse 5-sphere. Remember that giant gravitons are located at $r=0$ in global AdS \eqref{globalAdS}. Evaluating the LLM map \eqref{vac-map} in this location
\begin{equation}
r_{\text{LLM}}|_{r=0}= R_{\text{AdS}}^2\cos\theta\,, 
\label{eq:main-holo}
\end{equation} 
relates the two notions. Furthermore, since global AdS has no excitations, \eqref{eq:babel1} tells us how the continuous fermion label $x$ is related to the geometry of the 5-sphere
\begin{equation}
  x = N\,\cos^2\theta\,.
\label{eq:x-global}
\end{equation}

Next, consider the typical YT. Since the number of columns of length $j$, $\langle c_j\rangle$, is the averaged number of giant gravitons carrying charge $j$, one infers the phase space density of giants equals
\begin{equation}
u_{\text{giant}}=\omega=\frac{N_c}{N}\,.
\label{giant-dens}
\end{equation}
This allows to write the number of giants (columns) $(dn)$ having R-charge (lengths) in the interval $[j,j+dj]$ as 
\begin{equation}
dn = u_{\text{giant}}\,dj\,.
\end{equation}
As illustrated in figure \ref{fig:lcurve}, given some excitation $r(x)$ for the x-th fermion, the length of the typical YT column $j(x)$ equals
\begin{equation}
j(x) = N - x\,.
\label{eq:g-length}  
\end{equation}
Inserting \eqref{eq:x-global} in \eqref{eq:g-length}, one reproduces the relation \eqref{giant-size} between the R-charge of a single giant graviton with the geometry of the 3-cycles it wraps \cite{McGreevy:2000cw}
\begin{equation}
j(x(\theta)) = N\,\sin^2\theta\,.
\end{equation}
Hence, the number of giants wrapping 3-cycles with sizes between $\theta$ and $\theta+d\theta$ is given by
\begin{equation}
dn = u_{\text{giant}}\,dj = N_c\,\sin2\theta\,d\theta\,,
\end{equation}
in agreement with the distribution \eqref{gdist}. 

As mentioned below \eqref{eq:10dBH}, the superstar bulk singularity is a cone with non-trivial topological structure. This may be easier to see in the LLM description \eqref{eq:llm-metric}, where both 3-spheres shrink to zero size at the cone's apex $(y=0)$ at the same time whenever $z(0,r_{\text{LLM}}) \neq \pm\frac{1}{2}$. We would like to understand this fact microscopically.

In the probe approximation, giant gravitons correspond to adding a small white disk inside the black droplet describing the vacuum. Hence,  the 3-sphere in AdS$_5$ shrinks to zero size smoothly in the LLM description. Similarly, dual giants correspond to small black disks outside the vacuum black droplet. In this case, it is the 3-sphere in S$^5$ that shrinks to zero size smoothly. Notice both statements match the description given in section \ref{sec:review}. 

When interpreting singular LLM geometries described in terms of gray droplets, two facts need to be accounted for : the backreaction of many D-branes and the quantum uncertainty in the fermion excitations observed in \eqref{eq:singrho1} and further developed in this subsection. We discuss these separately below.

First, the backreaction of the wrapping D3-branes can change the geometry. In particular, there can exist geometric transitions in which the wrapped 3-sphere becomes contractible while the transverse 3-sphere becomes non-contractible as the D3-brane gets dissolved into flux in the process. These transitions give rise to new topologies \cite{Lin:2004nb}. This means that, for example, as soon as the backreacted geometry carries some amount of R-charge by adding a macroscopic white disk in the LLM plane, one could add dual giant excitations in that region without violating the exclusion principle. 

Second, whenever there is an uncertainty in the amount of excitation of a collection of fermions, this translates into a singular geometry in which both 3-spheres shrink at the same time, in agreement with the cone structure in the metric. This is consistent with the fact that each
half-BPS state can be given a description either in terms of giant gravitons or dual giants, since counting both separately would be a double count \cite{Suryanarayana:2004ig}. This is equivalent to the particle/hole duality in the fermion picture\footnote{This corresponds to the exchange symmetry $z\to -z$ in the LLM geometries describing the superstar \cite{Lin:2004nb}. A similar symmetry was observed in \cite{Balasubramanian:2001dx} when comparing the IR geometries in near-extremal single R-charged black holes.} and it boils down to either using the column $\{c_j\}$ or the row $\{r_i\}$ excitation quantum numbers describing the same Young tableau. 

Altogether, the information encoding this uncertainty in the density matrix is responsible for the simultaneous shrinking of both 3-spheres in the entire droplet.

\paragraph{Singular LLM configurations revisited.} It is convenient to derive \eqref{eq:z-superstar} from the solution to the Laplace equation \eqref{eq:LLMscalar}. This is used in appendix \ref{sec:matching}. The scalar function $z(y;r_{\text{LLM}})$ describing a smooth droplet disk centered at the LLM origin of radius $r_i$ can be derived from the
identity \cite{Lin:2004nb}
\begin{equation}
\text{Disk}(r_i)\equiv -\frac{y^2}{\pi}\int_{\text{Disk}(r_i)} r'dr'd\phi' \,
\frac{1}{[r^2 + r'^2 - 2rr'\cos\phi' + y^2]^2} = f_-(r_i) - \frac{1}{2}~\,.
\end{equation}
The smooth droplet involves the contribution from a black droplet, $\frac{1}{2}\text{Disk}(r_i)$, where the factor $\frac{1}{2}$ takes care of the smooth LLM boundary conditions, i.e. it equals $-z(0;r_{\text{LLM}})$, together with the contribution from a white annulus with inner radius $r_i$ and outer radius at infinity, which equals
\begin{equation}
-\frac{1}{2}\text{Disk}(r_\infty) + \frac{1}{2}\text{Disk}(r_i)\,.
\end{equation}
Altogether, the scalar function describing a smooth LLM droplet of radius $r_i$ equals
\begin{equation}
z(y;r_{\text{LLM}}) = \frac{1}{2}\text{Disk}(r_i) - \frac{1}{2}\text{Disk}(r_\infty) + \frac{1}{2}\text{Disk}(r_i) = \text{Disk}(r_i) - \frac{1}{2}\text{Disk}(r_\infty) = f_-(r_i)\,,
\end{equation}
where in the last step the identity $\text{Disk}(r_\infty)=-1$ was used. The generalisation to a singular droplet of radius $r_i$ with
boundary condition $z_i$ is
\begin{equation}
  z(y;r_{\text{LLM}}) = -z_i\text{Disk}(r_i) - \frac{1}{2}\text{Disk}(r_\infty) + \frac{1}{2}\text{Disk}(r_i) = \left(\frac{1}{2} -z_i\right)\,\left(f_-(r_i)-\frac{1}{2}\right) + \frac{1}{2}\,.
\label{eq:grayscalar}
\end{equation}
For a collection of $n$ rings with arbitrary boundary conditions, this reads
\begin{equation}
\begin{split}
  z(y;r_{\text{LLM}}) &= -z_1\text{Disk}(r_1) + \sum_{i=2}^{n-1} \left(-z_i\text{Disk}(r_i) + z_i\text{Disk}(r_{i-1})\right) - \frac{1}{2}\text{Disk}(r_\infty) + \frac{1}{2}\text{Disk}(r_n) \\
  &= \frac{1}{2}\left(\frac{1}{2}+z_1\right) + \sum_{i=1}^{n-1}(z_{i+1}-z_i)\,f_-(r_i) + \left(\frac{1}{2}-z_n\right)\,f_-(r_n)\,.
\end{split}
\end{equation}
Notice that for a collection of smooth rings, this reproduces \eqref{eq:smoothring}.

\subsubsection{Near-extremal R-charged black holes revisited}
\label{sec:near-ext-rev}

The identification of the microscopic degrees of freedom responsible for the naked singularity in the BPS limit allows us to revisit the entropy of near-extremal R-charged black holes \cite{Balasubramanian:2007bs} along the lines of the Horowitz-Polchinski correspondence principle in section \ref{sec:motivation}.

The appearance of an effective Planck constant $\hbar\propto\ell_p^4$ in the LLM plane suggests the transition between near-extremal black holes to an open string description\footnote{Interpreting the naked singularity as a distribution of giant gravitons makes the open string description very natural, at least, in the near-extremal regime.} occurs when the curvature invariants of the hole evaluated at the horizon are of order $\hbar^{-1}$. In the regime $q\sim R_{\text{AdS}}^2\gg r_+^2$, one finds
\begin{equation}
R_{\mu\nu}R^{\mu\nu}|_{r_+},\,R_{\mu\nu\alpha\beta}R^{\mu\nu\alpha\beta}|_{r_+}\sim \hbar^{-1} \quad \Rightarrow \quad q\,r_+^2\sim \hbar \quad \Rightarrow \quad \frac{r_+^2}{R_{\text{AdS}}^2} \sim\frac{1}{N}
\end{equation} 
It follows, the entropy \eqref{eq:GRentropy}
\begin{equation}
S \sim N^2\,\sqrt{\frac{q}{R_{\text{AdS}}^2}}\,\frac{r_+^2}{R_{\text{AdS}}^2}\sim N
\label{eq:near-extreme}
\end{equation}
scales linearly with $N$, as in the BPS microscopic discussion. 

Alternatively, one can also interpret the previous conclusion as coming from an stretched horizon perspective \cite{Susskind:1993if}, as already discussed in \cite{D'Errico:2007jm}. In the BPS limit, the LLM map \eqref{eq:sing-map} tells us the LLM scale at which the above curvature scales occur is at $y_{\text{LLM}}\sim \sqrt{\hbar}$, away from $\sin\theta=0$, since curvature invariants vanish there due to the absence of giant gravitons. 

This discussion may further suggest to study the inclusion of quantum gravity effects on the BPS limits of R-charged black holes since these may describe bulk horizons following the ideas in \cite{Dabholkar:2004dq}\footnote{The author would like to thank Alex Maloney and Sameer Murthy for stressing this point.}.

\paragraph{Decoupling limits.} The idea that excitations of open strings stretched between giant gravitons are responsible for the entropy of near-extremal R-charged black holes is further supported by the existence of decoupling limits in which $\alpha^\prime\to 0$, while keeping the mass of these excitations fixed \cite{Balasubramanian:2007bs}. In this reference, the limit $\epsilon\to0$
\begin{equation}
  r \to \epsilon^2 \, r \,, \qquad \mu \to \epsilon^4 \mu \,, \qquad \theta \to \theta_0 - \epsilon^2 \, \theta \,, \qquad d(\phi-t/R_{\text{AdS}}) \to \epsilon^2 d(\phi-t/R_{\text{AdS}}) \,,
\label{eq:dec-scaling}
\end{equation}
keeping $R_{\text{AdS}},\,g_s,\,q$ fixed, is motivated, resulting in a decoupled metric
\begin{equation}
ds^2 = \sin\theta_0 \left\{ z\left[ -f \, dt^2 + q \, ds^2_{S^3} + R_{\text{AdS}}^2 \, ds^2_{\tilde{S}^3} \right]
+ \frac{1}{z} \left[ \frac{q}{f} \, dz^2 + L^2 \, d\theta^2 + \frac{R_{\text{AdS}}^2}{\tan^2\theta_0} \, d\chi^2 \right]
\right\} \,,
\label{eq:decouple}
\end{equation}
written in terms of the new coordinate $z=\frac{r}{\sqrt{q}}$, with
\begin{equation}
  f = 1 +\frac{q}{R_{\text{AdS}}^2} - \frac{\mu}{q\,z^2} \,. 
\label{eq:fdec1}
\end{equation}
Hence, the decoupled metric \eqref{eq:decouple} keeps the information on the existence of horizons, but it does so only in a neightbourhood of the original $\theta_0$ azimutal location. Equivalently, the limit \eqref{eq:dec-scaling} focuses on a small annulus in the LLM plane around a ring whose size is controlled by $\theta_0$.

A full understanding on the meaning of these decoupled metrics remains an open question, though the possibility of the emergence of some $\U(K)$ gauge theory supported by the number $K$ of giants gravitons in this annulus was put forward in \cite{Balasubramanian:2007bs}.

\subsection{Localised superstars with and without solitons}
\label{sec:localised}

\begin{figure}
	\centering
	\begin{tikzpicture}[xscale=8, yscale=8]
	\draw[<->, help lines] (0.5,1.5) -- (0.5,1)  -- (1,1);
	\draw[red, ultra thick] (0.8,1) to [out=90 ,in=0] (0.5,1.3);
	\draw[->, dashed] (0.65,1) to [out=90 ,in=0] (0.5,1.15);
	\node [above right] at (0.51,1) {$\theta$};
	\node [above right] at (0.7,1.21) {$\frac{dn}{d\theta}=N_c\sin2\theta$};
	\node[above left] at (0.5,1.15) {$\frac{\pi}{2}$};
	\node[above right] at (0.65,1) {$0$};
	\draw[<->, help lines] (1.5,1.5) -- (1.5,1)  -- (2,1);
	\draw[->, dashed] (1.65,1) to [out=90 ,in=0] (1.5,1.15);
	\node [above right] at (1.51,1) {$\theta$};
	\draw (1.8,1) to [out=90 ,in=0] (1.5,1.3);
	\draw[red, ultra thick] (1.6,1.28) arc (70:90:0.30);
	\draw[red, dashed, thick] (1.5,1) to (1.6,1.28);
	\node [red, above right] at (1.6,1.27) {$\theta_0$};
	\node[above left] at (1.5,1.15) {$\frac{\pi}{2}$};
	\node[above right] at (1.65,1) {$0$};
	\end{tikzpicture}
	\caption{Comparison of a distribution of giant gravitons (superstar) wrapping all 3-cycles of the 5-sphere (left) with a localised superstar wrapping 3-cycles in the interval $\theta \in [\theta_0,\,\frac{\pi}{2}]$ (right).}
	\label{fig:loc-bulk}
\end{figure}
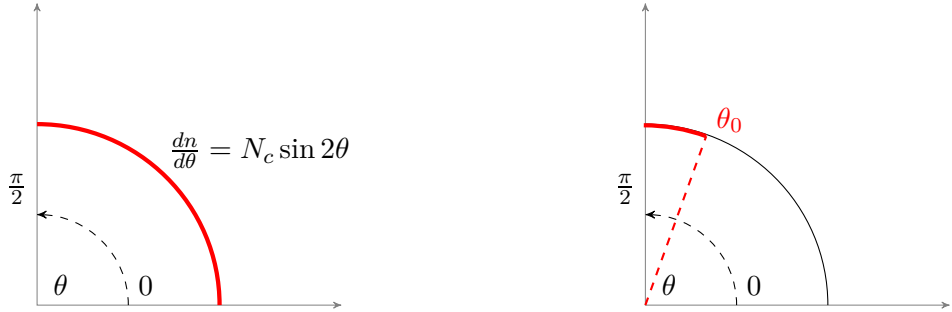

The analysis in section \ref{sec:supers} interprets the BPS limit of R-charged black holes \eqref{eq:10dBH} as the maximally entangled ensemble \eqref{eq:uniform} in the subspace of the Hilbert space describing at most $N_c$ giant gravitons. The naked singularity is spread over the entire transverse 5-sphere. This fact is reflected in the quantum mechanics by having a uniform probability for every available fermion excitation. However, the fermionic microscopic description and the holographic relation \eqref{eq:main-holo} suggest to consider distributions of giants that are {\it not} spread over the entire $\theta$ interval. These ensembles will be referred to as {\it localised} superstars. These are still located at the origin of AdS, but the sizes of the 3-cycles wrapped by giants in these ensembles lie in some interval $[\theta_0,\,\theta_1]\subset [0,\,\frac{\pi}{2}]$, as illustrated in figure \ref{fig:loc-bulk}. These are toy models for localised black holes in this sector of the theory, and indeed, they should be interpreted as BPS limits of non-extremal black holes localised in the transverse 5-sphere, which are not known in the literature\footnote{The author would like to thank Simon Ross for stressing this point.}.

The spectrum of the N free fermions can be equivalently encoded in terms of the number of columns $c_j$ carrying $j$ R-charge. The relation between this set and the YT row excitations $\{r_j\}$ is \cite{Suryanarayana:2004ig}
\begin{equation}
  c_N=r_1\,, \quad \text{and} \quad c_{N-i}=r_{i+1}-r_i \quad \quad i=1,2,\dots N-1
\end{equation}
By construction, these variables are not constrained by the Pauli exclusion principle. To design the ensembles one is interested in, one can introduce individual chemical potentials $\mu_i$ for each column $c_i$, so that
\begin{equation}
\langle c_i \rangle = \frac{e^{-\mu_i}}{1-e^{-\mu_i}}\,.
\label{eq:col-av}
\end{equation}
Tuning this set of chemical potentials, one can achieve any typical Young tableau\footnote{The validity of the ensemble description depends on the size of the ensemble fluctuations. In the following, it is assumed that the energy scales characterising these ensembles follow the results in \cite{D'Errico:2007jm,Balasubramanian:2005mg}.}. In particular, $\langle c_i\rangle =0$ requires $\mu_i\to \infty$, whereas $\langle c_i\rangle \sim \mathcal{O}(N)$ requires $\mu_i\sim \langle c_i\rangle^{-1}$. Introducing homogeneous chemical potentials $\mu$ for subsets of the R-charges $j$ will allow us to describe macroscopic energy scales that can have gravity duals. Next, we discuss some examples where these ideas are used.

\begin{figure}
	\centering
	\begin{tikzpicture}
	\draw [red, fill=red] (0,0) -- (2,0) -- (0,-3);
	\node [above, red] at (1,0) {$N_c$};
	\node [below left] at (0,-1.5) {$N-N_1$};
	\end{tikzpicture}
	\caption{Typical Young tableau consisting of $N_1$ fermions in the Fermi sea and $N-N_1$ fermions describing a localised superstar made of $N_c$ giant gravitons (red triangle).}
	\label{fig:loc-super-1}
\end{figure}
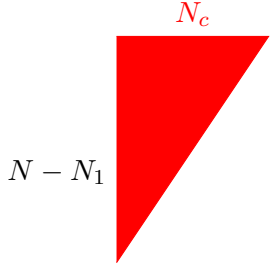

\paragraph{Single localised superstar.} Consider a superstar with $N_c$ giant gravitons made of $N-N_1$ fermions, while the remaining $N_1$ fermions remain in the Fermi sea, as in figure \ref{fig:loc-super-1}. The fermion excitation averages encoding such typical states require
\begin{equation}
  \langle r_i\rangle = 0\,, \quad \quad i=1,\dots, N_1 \quad \quad \langle r_i \rangle = \frac{N_c}{N-N_1} i\,, \quad \quad i=N_1+1,\dots N
\end{equation}
Its limit curve continuum limit is
\begin{equation}
y(x) = \left\{
\begin{array}{cc}
\left(1+ \frac{N_c}{N-N_1} \right) x\,, & x\in (N_1,N] \\
x\, & x \in [0,\,N_1]
\end{array}
\right. \,.
\end{equation}

In terms of average columns, this is equivalently described by
\begin{equation}
\begin{aligned}
\langle c_i\rangle &=0\,, \quad \quad i=N,N-1,\dots N-N_1\,, \\
\langle c_{N-N_1-j}\rangle &= \frac{N_c}{N-N_1}\,, \quad \quad j= 1,\dots N - N_1 - 1
\end{aligned}
\end{equation}
To describe the localised superstar ensemble, set $\mu_i\to \infty$ $(i=N,N-1,\dots N-N_1)$, introduce an homogeneous chemical potential for the remaining degrees of freedom, i.e. $\mu_{N-N_1-j}=\mu$ $(j= 1,\dots N - N_1 - 1)$ and take the $\mu\to 0$ limit on either a canonical ensemble enforcing the number of giants to be $N_c$ \cite{Balasubramanian:2005mg} or an ensemble where the number of giants is at least $N_c$ \cite{D'Errico:2007jm}. Defining the Hilbert space of such states as $\mathcal{H}_r$, the density matrix in the last ensemble reduces to
\begin{equation}
\rho_{\text{loc-superstar}} = \frac{1}{Z} \sum_{\vec{n}\in \mathcal{H}_r} |\Psi_{\vec{n}}\rangle\langle \Psi_{\vec{n}}|\,, \quad \quad Z = \binom{N-N_1+N_c}{N-N_1}
\label{eq:example-1}
\end{equation}

\begin{figure}
	\centering
	\begin{tikzpicture}
	\draw[fill=gray]  (0,0) circle [radius=2];
	\draw[fill=black]  (0,0) circle [radius=1.4];
	\end{tikzpicture}
	\caption{Phase space density corresponding to a Fermi sea of $N_1$ fermions, followed by a localised superstar.}
	\label{fig:disk-0}
\end{figure}
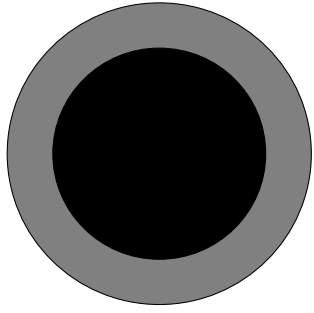

The gravity dual corresponds to the phase space density in figure \ref{fig:disk-0}. This describes $N_1$ fermions in the Fermi sea (black inner circle) together with a grayscale annulus with $N_c$ giants
\begin{equation}
u(0;r) = \left\{
\begin{array}{cc}
1\,, & r\in [0,\,r_1] \\
\frac{1}{1+\frac{N_c}{N-N_1}}\,, & r\in (r_1,\,r_2] \\
0\,, & r\in (r_2,\infty)
\end{array}
\right.
\end{equation}
with radia given by 
\begin{equation}
  \frac{r_1^2}{R_{\text{AdS}}^4} = \frac{N_1}{N}\,, \quad \quad \frac{r_2^2}{R_{\text{AdS}}^4} = 1+\frac{N_c}{N}\,.
\end{equation} 
This ensures the energy of these states equals $\Delta = \frac{1}{2}(N-N_1)N_c$, as corresponds to $N_c$ giants built out of $N-N_1$ fermions.

The energy scale $r_1$ corresponding to the highest energy among the $N_1$ fermions in the Fermi sea sets the geometric scale for the minimum 3-cycle size, through \eqref{eq:main-holo}, that giants are wrapping in this ensemble. In other words, the rotationally invariant phase space distribution of giants is
\begin{equation}
u_{\text{giant}}(\theta) = \frac{N_c}{N-N_1}\,, \quad \theta \in (\theta_0,\,\pi/2] \quad \text{with} \quad \cos\theta_0 = \sqrt{\frac{N_1}{N}} = \frac{r_1}{R_{\text{AdS}}^2}\,,
\end{equation}
vanishing otherwise. This interpretation can be confirmed by integrating this density over the relevant $\theta$ interval
\begin{equation}
N_c = \int dn = \frac{N_c}{1-N_1/N} \int^{\theta_0}_0 \sin 2\theta\,d\theta =  \frac{N_c}{1-N_1/N}\sin^2\theta_0\,.
\end{equation}
This reproduces the number of $N_c$ giants when $\cos\theta_0 = \sqrt{N_1/N}$.

\begin{figure}
	\centering
	\begin{tikzpicture}
	\draw [fill=white] (0,-3)  rectangle (3,0);
	\draw [blue, fill=blue] (0,-3) -- (0,-5) -- (3,-3);
	\node [above] at (1.5,0) {$N_c^\prime$};
	\node [below left] at (0,-1.5) {$N-N_A$};
	\node [below left] at (0,-4) {$N_A$};
	\end{tikzpicture}
	\caption{Young tableau consisting of $N_A$ fermions describing a superstar with $N_c^\prime$ giant gravitons, with the remaining $N-N_A$ fermions having an $N_c^\prime$ excitation.}
	\label{fig:loc-super-2}
\end{figure}
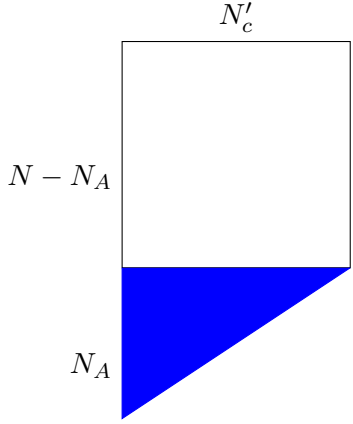

\paragraph{Single localised superstar and a soliton.} These tools can equally describe localised superstars made of giants wrapping intervals of smaller 3-cycles, such as the one illustrated in figure \ref{fig:loc-super-2}, whose typical YTs are characterised by
\begin{equation}
\langle r_i \rangle = N_c^\prime\,, \quad i=N_A+1,\dots N \quad \quad \quad \langle r_j \rangle =\frac{N_c^\prime}{N_A}\,j \quad j=1,2,\dots N_A
\end{equation}
where $\langle r_i \rangle$ describes the excitations of $N-N_A$ fermions, whereas $\langle r_j \rangle$ describes the blue triangle in figure \ref{fig:loc-super-2} responsible for its superstar interpretation. Its continuum limit is given by the limit curve
\begin{equation}
y(x) = \left\{
\begin{array}{cc}
\left( 1 + \frac{N_c^\prime}{N_A}\right) x\,, & x\in [0,N_A] \\
x + N_c^\prime\,, & x \in (N_A,\,N]
\end{array}
\right. \,.
\end{equation}
In terms of columns, this is equivalent to 
\begin{equation}
\begin{aligned}
  \langle c_i \rangle &= 0\,, \quad i=1,\dots ,N-N_A-1 \\
  \langle c_{N-j} \rangle &= \frac{N_c^\prime}{N_A}\,, \quad j=N_A,\dots 0
\end{aligned}	
\end{equation}

One can fix the individual chemical potentials to design an ensemble fulfilling these features. Defining the Hilbert space $\mathcal{H}_b$ of N fermions, where $N-N_A$ of them carry $N_c^\prime$ quanta, whereas the remaining $N_A$ fermions describe at least $N_c^\prime$ giants, the density matrix emerging in the vanishing chemical potential limit equals
\begin{equation}
\rho_{\text{loc-superstar}} = \frac{1}{Z} \sum_{\vec{n}\in \mathcal{H}_b} |\Psi_{\vec{n}}\rangle\langle \Psi_{\vec{n}}|\,, \quad \quad Z=\binom{N_A+N_c^\prime}{N_A}\,.
\end{equation}

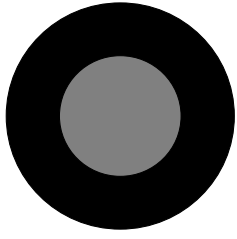
\begin{figure}
	\centering
	\begin{tikzpicture}
	\draw[fill=black]  (0,0) circle [radius=1.5];
	\draw[fill=gray]  (0,0) circle [radius=0.8];
	\end{tikzpicture}
	\caption{Phase space density corresponding to a localised superstar made of $N_A$ fermions building $N_c^\prime$ giants, together with $N-N_A$ excited fermions.}
	\label{fig:disk-3bis}
\end{figure}

The gravity dual corresponds to the phase space density in figure \ref{fig:disk-3bis} describing a grayscale distribution (the inner circle) in the range $0 \leq r \leq r_1$  together with a black annulus in the range $r_1 < r \leq r_2$ describing $N-N_A$ excited fermions, 
\begin{equation}
u(0;r) = \left\{
\begin{array}{cc}
\frac{1}{1+\frac{N^\prime_c}{N_A}}\, & r\in [0,\,r_1] \\
1\, & r\in (r_1,\,r_2] \\
0\, & r\in (r_2,\infty)
\end{array}
\right.
\end{equation}
with radia given by
\begin{equation}
\frac{r_1^2}{R_{\text{AdS}}^4} = \frac{N_A+ N_c^\prime}{N}\,, \quad \quad \frac{r_2^2}{R_{\text{AdS}}^4} = 1+\frac{N^\prime_c}{N}\,.
\end{equation} 
This ensures the average energy of these states equals $\langle \Delta \rangle = \frac{1}{2}N_AN^\prime_c + (N-N_A)N_c^\prime$, as corresponds to $N^\prime_c$ giants built out of $N_A$ fermions together with $N-N_A$ fermions excited by $N_c^\prime$ quanta.

The distribution of giants equals 
\begin{equation}
  u_{\text{giant}} = \frac{N^\prime_c}{N_A}\,, \quad \theta \in [\theta_1,\pi/2] \quad \text{with} \quad \cos\theta_1 = \sqrt{\frac{N_A}{N}} = 1- \frac{r_2^2-r_1^2}{R_{\text{AdS}}^4}\,,
\end{equation}
vanishing otherwise. Notice the upper size $\theta_1$ of the 3-cycles wrapped in this ensemble in consistent with the total number of giants being described since 
\begin{equation}
  N_c^\prime = \int dn = N^\prime_c\frac{N}{N_A} \int^{\frac{\pi}{2}}_{\theta_1} \sin2\theta\,d\theta= N_c^\prime\,\frac{N}{N_A} (1-\sin^2\theta_1),
\end{equation}
holds for this choice.

\begin{figure}
	\centering
	\begin{tikzpicture}
	\draw [fill=white] (0,-3)  rectangle (5,0);
	\draw [red, fill=red] (5,0) -- (7,0) -- (5,-3);
	\node [above] at (2.5,0) {M};
	\node [above, red] at (6,0) {$N_c^\prime$};
	\node [below left] at (0,-1.5) {$N-N_1$};
	\end{tikzpicture}
	\caption{Young tableau consisting of $N_1$ fermions in the Fermi sea and $N-N_1$ fermions with some superstar behaviour (red triangle) on top of some excitation $M$.}
	\label{fig:loc-super}
\end{figure}
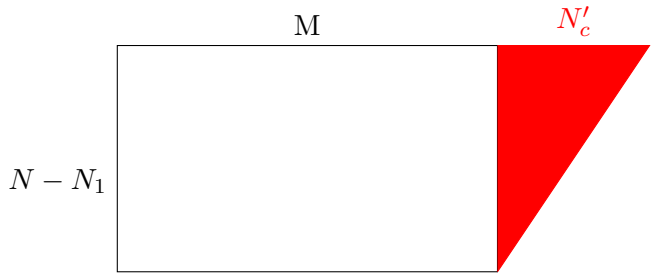

\paragraph{Single localised superstar gapped from the Fermi sea.} Consider the same split of fermions as in example 1, but adding $M$ extra quanta to the subset of $N-N_1$ fermions, as shown in figure \ref{fig:loc-super}. The fermion excitation averages encoding these typical states require
\begin{equation}
\langle r_i\rangle = 0\,, \quad \quad i=1,\dots, N_1 \quad \quad \langle r_i \rangle = M +\frac{N^\prime_c}{N-N_1} i\,, \quad \quad i=N_1+1,\dots N
\end{equation}
Its limit curve continuum limit
\begin{equation}
y(x) = \left\{
\begin{array}{cc}
M + x +\frac{N_c^\prime}{N-N_1} (x-N_1)\,, & x\in (N_1,N] \\
x\, & x \in [0,\,N_1]
\end{array}
\right. \,.
\end{equation}
has a discontinuity at $x=N_1$ encoding the extra $M$ quanta. This is reflected in the column averages
\begin{equation}
\begin{aligned}
  \langle c_i \rangle &= 0\,, \quad \quad i=N,N-1,\dots N-N_1+1\,,  \\
  \langle c_{N-N_1}\rangle &=M\,, \\
  \langle c_{N-N_1-j} \rangle &= \frac{N_c^\prime}{N-N_1}\,, \quad\quad j= 1,2, \dots N-N_1\,,
\end{aligned}
\end{equation}
containing a non-vanishing $\langle c_{N-N_1}\rangle =M$.

As discussed below equation \eqref{eq:col-av}, it is possible to tune the different chemical potentials $\mu_i$ to describe ensembles whose typical pure states look like the YT in figure \ref{fig:loc-super}. If $\mathcal{H}_r$ denotes the Hilbert space of states compatible with this ensemble and consider the limit maximising the entropy, one is left with 
\begin{equation}
\rho_{\text{loc-superstar}} = \frac{1}{Z} \sum_{\vec{n}\in \mathcal{H}_r} |\Psi_{\vec{n}}\rangle\langle \Psi_{\vec{n}}|\,, \quad \quad Z = \binom{N-N_1+N^\prime_c}{N-N_1}\,.
\end{equation}
Given the previous discussions on the superstar and localised superstars ensembles, this is natural since $N_1$ of the fermions are fixed in their Fermi sea, whereas the remaining $N-N_1$ fermions make up $N_c^\prime$ giant gravitons. What is different in this case is that the excitations of the $N-N_1$ fermions include an average fixed number $M$ of columns of size $N-N_1$. Hence, the sum over the excitation vectors $\vec{n}$ reflects that fact, but the combinatorics giving rise to the partition function normalisation do not change.

\begin{figure}
	\centering
	\begin{tikzpicture}
	\draw[fill=gray]  (0,0) circle [radius=2];
	\draw[fill=white]  (0,0) circle [radius=1.4];
	\draw[fill=black]  (0,0) circle [radius=0.8];
	\end{tikzpicture}
	\caption{Phase space density corresponding to a Fermi sea of $N_1$ fermions, followed by a gap and a localised superstar.}
	\label{fig:disk-4}
\end{figure}
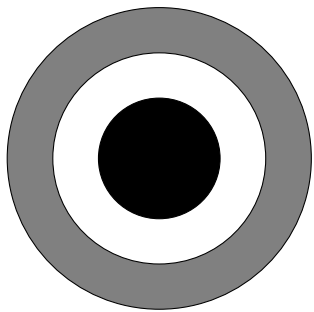

The gravity dual corresponds to the phase space density in figure \ref{fig:disk-4} describing a set of $N_1$ fermions in the Fermi sea (black inner circle of radius $r_1$), followed by a white annulus, describing the energy scale $M$, with outer radius $r_2$, together with a grayscale annulus  in the range $r_2 < r \leq r_3$ describing $N^\prime_c$ giant gravitons
\begin{equation}
u(0;r) = \left\{
\begin{array}{cc}
1\, & r\in [0,\,r_1] \\
0\, & r\in (r_1,\,r_2] \\
\frac{1}{1+N^\prime_c/(N-N_1)}\, & r\in (r_2,\,r_3] \\
0\, & r\in (r_3,\infty)
\end{array}
\right.
\end{equation}
The different radia are 
\begin{equation}
\frac{r_1^2}{R_{\text{AdS}}^4} = \frac{N_1}{N}\,, \quad \quad \frac{r_2^2}{R_{\text{AdS}}^4} = \frac{M+N_1}{N}\,, \quad\quad \frac{r_3^2}{R_{\text{AdS}}^4} = 1 + \frac{M+N_c^\prime}{N}\,.
\end{equation} 
This ensures the average energy of these states equals $\Delta = M(N-N_1) + \frac{1}{2}(N-N_1)N^\prime_c$, as corresponds to a superstar of $N^\prime_c$ giants built out of $N-N_1$ fermions together with $M$ giants carrying $N-N_1$ R-charge each.

The range in the size of the 3-cycles wrapped by the giants in this superstar is not different from that in example 1. Indeed, the current giants also wrap cycles in the range $\theta \in [0,\theta_0]$, with $\cos\theta_0 = \sqrt{N_1/N} = r_1/R_{\text{AdS}}^2$ still being determined by the energy scale $r_1$ of the highest fermion energy in the Fermi sea. The addition for this class of states is the existence of $M$ giants wrapping the precise 3-cycle with size determined by $\theta_0$. As before, this interpretation is confirmed by integrating the relevant distribution of giants
\begin{equation}
u_{\text{giant}} = \frac{N^\prime_c}{N-N_1}\,.
\end{equation}
over the relevant interval $[0,\theta_0]$
\begin{equation}
N^\prime_c = \int dn = \frac{N^\prime_c}{1-N_1/N} \int^{\theta_0}_0 \sin 2\theta\,d\theta =  \frac{N^\prime_c}{1-N_1/N}\sin^2\theta_0\,.
\end{equation}

Previous examples illustrate how to describe localised superstars in different regions of the transverse 5-sphere. Their gravity dual still involves a naked singularity at the origin of AdS, but the latter is only spread over the set of 3-cycles in the 5-sphere being wrapped by the distribution of giants. Which set is encoded by the localisation properties of the wave functions controlling the semiclassical limit of the reduced single particle density matrix, i.e. the gray area appearing in the LLM plane. 

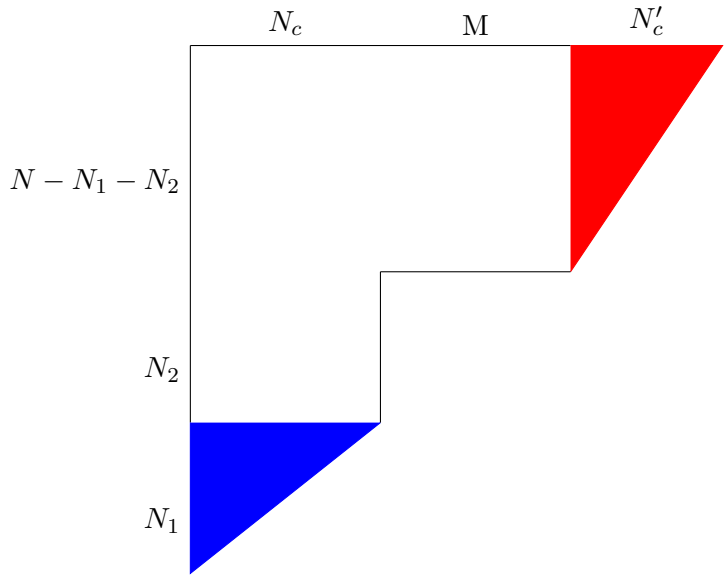
\begin{figure}
	\centering
	\begin{tikzpicture}
	\draw (0,0) to (5,0);
	\draw[red, fill=red] (5,0) -- (7,0) -- (5,-3);
	\draw (0,0) to (0,-5);
	\draw[blue, fill=blue] (0,-5) -- (0,-7) -- (2.5,-5);
	\draw (2.5,-3) to (2.5,-5);
	\node [above] at (1.25,0) {$N_c$};
	\node [above] at (3.75,0) {M};
	\node [above] at (6,0) {$N_c^\prime$};
	\node [below left] at (0,-1.5) {$N-N_1-N_2$};
	\node [below left] at (0,-4) {$N_2$};
	\node [below left] at (0,-6) {$N_1$};
	\draw (2.5,-3) to (5,-3);
	\end{tikzpicture}
	\caption{Young tableau consisting of $N_1$ fermions in a superstar ensemble with $N_c$ giant gravitons, $N_2$ fermions with $N_c$ excitation above Fermi sea and $N-N_1-N_2$ fermions with a superstar ensemble of $N_c^\prime$ giants on top of an $N_c + M$ excitation.}
	\label{fig:nonentangled-loc-super}
\end{figure}

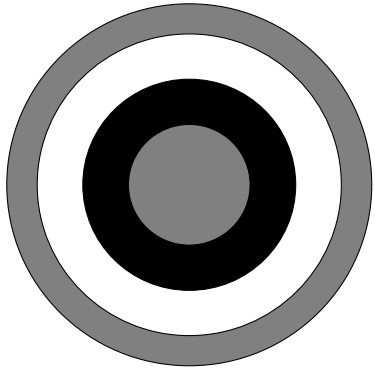
\begin{figure}
	\centering
	\begin{tikzpicture}
	\draw[fill=gray]  (0,0) circle [radius=2.4];
	\draw[fill=white]  (0,0) circle [radius=2.0];
	\draw[fill=black]  (0,0) circle [radius=1.4];
	\draw[fill=gray]  (0,0) circle [radius=0.8];
	\end{tikzpicture}
	\caption{Phase space density corresponding to a superstar made of $N_c$ giant gravitons, followed by $N_2$ fermions with $N_c$ excitations and a second superstar made of $N_c^\prime$ giants separated by a gap of energy $M$.}
	\label{fig:disk-5}
\end{figure}

\paragraph{Multiple localised superstars.} One can equally well describe states containing more than one localised superstar. Figure \ref{fig:nonentangled-loc-super} shows two localised superstars, indicated by the blue and red triangles, together with some energy gap in between. On the gravity side, this corresponds to the phase space density \ref{fig:disk-5}.

Typical states compatible with the YT \ref{fig:nonentangled-loc-super} carry excitation averages
\begin{equation}
\begin{aligned}
\langle r_i\rangle &= \frac{N_c}{N_1}\,i\,, \quad \quad i=1,\dots N_1 \\
\langle r_i\rangle &= N_c\,, \quad \quad i=N_A+1, \dots N_1+N_2 \\
\langle r_i\rangle &= N_c + M + \frac{N_c^\prime}{N-N_1-N_2}(i-N_1-N_2-1)\,, \quad \quad i=N_1+N_2+1,\dots N
\end{aligned}
\end{equation}
or equivalently, average column excitations
\begin{equation}
\begin{aligned}
\langle c_{N-j}\rangle &= \frac{N_c}{N_1}\,, \quad \quad j=0,1,\dots N_1-1 \\
\langle c_{N-j}\rangle &= 0\,, \quad \quad j=N_1,\dots N_1+N_2-1 \\
\langle c_{N-N_1-N_2}\rangle &= M\,, \\
\langle c_{N-j}\rangle &= \frac{N_c^\prime}{N-N_1-N_2}\,, \quad \quad j=N_1+N_2+1,\dots N
\end{aligned}
\end{equation}
The corresponding limit curve emerging in the continuum limit capturing these typical states is
\begin{equation}
y(x) = \left\{
\begin{array}{cc}
N_c+ M + x +\frac{N_c^\prime}{N-N_1-N_2} (x-N_1-N_2)\,, & x\in (N_1+N_2,N] \\
N_c + x\,, & x\in (N_1, N_1+N_2]\\
\left(1 + \frac{N_c}{N_1}\right)\,x\,, & x \in [0,\,N_1]
\end{array}
\right. \,.
\label{eq:mloc-sup}
\end{equation}
As before, this is discontinuous at $x=N_1+N_2$, capturing the average $\langle c_{N-N_1-N_2}\rangle = M$.

The number of such YTs is controlled by the partition function
\begin{equation}
Z = \binom{N_1+N_c}{N_1}\,\binom{N+N_c^\prime -N_1-N_2}{N-N_1-N_2}\,.
\end{equation}
The reason why this partition function involves a product is because for any subYT compatible with one of the triangles, all the subYT compatible with the second triangle should be accounted for. 

The gravity dual has a phase space density given in figure \ref{fig:disk-5}. It describes a localised superstar of $N_c$ giants made out of $N_1$ fermions (grayscale inner circle of radius $r_1$), together with a black annulus of outer radius $r_2$ describing $N_2$ fermions carrying an excitation of $N_c$ quanta, followed by a white annulus of outer radius $r_3$, describing the extra $M$ quanta that the remaining $N-N_1-N_2$ fermions carry and a final grayscale annulus, with outer radius $r_4$ describing the second $N_c^\prime$ giants out of $N-N_1-N_2$ fermions. Inspection of the limit curve \eqref{eq:mloc-sup} reveals the phase space density equals
\begin{equation}
u(0;r) = \left\{
\begin{array}{cc}
 \frac{1}{1+\frac{N_c}{N_1}}\,, & r \in [0,\,r_1) \\
 1\,, & r\in (r_1,r_2) \\
 0\,, & r\in (r_2,r_3) \\
 \frac{1}{1+\frac{N_c^\prime}{N-N_1-N_2}}\,, & r\in (r_3,r_4) \\
 0\,, & r> r_4
\end{array}
\right. \,.
\label{eq:mloc-phase}
\end{equation}
with radia given by
\begin{equation}
\begin{aligned}
\frac{r_1^2}{R_{\text{AdS}}^4} = \frac{N_1+N_c}{N}\,, \quad & \quad \frac{r_2^2}{R_{\text{AdS}}^4} = \frac{N_1+N_2+N_c}{N}\,, \\
\frac{r_3^2}{R_{\text{AdS}}^4} = \frac{N_1+N_2+N_c+M}{N}\,, \quad & \quad
\frac{r_4^2}{R_{\text{AdS}}^4} = 1 + \frac{M+N_c+N_c^\prime}{N}\,.
\end{aligned}
\end{equation} 
These geometric scales are compatible with an average energy
\begin{equation}
  \Delta = \frac{1}{2}\,N_1N_c + N_2N_c + (N-N_1-N_2)(N_c+M) + \frac{1}{2}(N-N_1-N_2)\,N_c^\prime\,,
\end{equation}
matching the interpretation of these states. In particular, the first localised superstar is responsible for the energy $\frac{1}{2}\,N_1N_c$, whereas the second carries energy $\frac{1}{2}(N-N_1-N_2)\,N_c^\prime$. Their respective phase space distributions equal
\begin{equation}
  u_{\text{loc-1}}= \frac{1}{1+\frac{N_c}{N_1}}\,, \quad \quad u_{\text{loc-2}}= \frac{1}{1+\frac{N_c^\prime}{N-N_1-N_2}}\,.
\end{equation}

Some of these scales are also responsible for the different localisation in the 5-sphere of both superstars. Indeed, from the phase space density \eqref{eq:mloc-phase}, we infer the density of giants in the first superstar (the blue triangle or the inner grayscale circle) and the second superstar (the red triangle and the grayscale annulus), are, respectively
\begin{equation}
  u_{\text{giant-1}} = \frac{N_c}{N_1}\,, \quad \quad u_{\text{giant-2}} = \frac{N^\prime_c}{N-N_1-N_2}\,. 
\end{equation}
Hence, the first superstar is composed of giants wrapping 3-cycles with size in the range $\theta\in [\theta_1,\pi/2]$ and $\cos\theta_1=\sqrt{N_1/N}$. The second superstar involves 3-cycles with size in the range $\theta\in [0,\theta_2,]$ and $\cos\theta_2 = \sqrt{(N_1+N_2)/N}$. Both these facts are compatible with
\begin{equation}
  N_c = N\int^{\frac{\pi}{2}}_{\theta_1}\, u_{\text{giant-1}}\,\sin2\theta\,d\theta \quad \text{and} \quad
  N^\prime_c = N\int^{\theta_2}_0\, u_{\text{giant-2}}\,\sin2\theta\,d\theta\,.
\end{equation}
The angular separation $\delta\theta = \theta_2-\theta_1$ is due to the existence of $N_2$ fermions, each carrying an excitation of $N_c$ quanta
\begin{equation}
  N_2 = \int^{r_2}_{r_1} \frac{rdrd\phi}{2\pi\hbar} = N(\cos^2\theta_2-\cos^2\theta_1)\,,
\end{equation}
where we used \eqref{eq:main-holo}\footnote{Remember there are $M$ giants localised at $\theta_2$ in this construction.}.

\section{Two boundary EPR=ER in LLM}
\label{sec:2-boundary}

The maximal Kruskal extension of large single R-charged AdS black holes \eqref{eq:5dBH} discussed in appendix \ref{kruskal} has a dual CFT description in terms of a maximally correlated state, the thermofield double state (TFD) \cite{Israel:1976ur,Maldacena:2001kr,Andrade:2013rra}, purifying the equilibrium mixed state at finite temperature and chemical potentials in \eqref{eq:dual-ensemble}
\begin{equation}
  |\text{TFD}(\tilde{\beta},\Phi)\rangle = \frac{1}{\sqrt{Z(\tilde{\beta},\mu)}}\sum_{\alpha\in \mathcal{H}_{\text{SYM,L}}\otimes \mathcal{H}_{\text{SYM,R}}} e^{-\tilde{\beta}(\Delta_\alpha - \Phi\,J_\alpha)/2} |\Delta_\alpha,J_\alpha\rangle_L |\Delta_\alpha,-J_\alpha\rangle_R
\label{eq:TFD-1}
\end{equation}
Quantum states $|\Psi\rangle_{\text{L}}|\Psi^\prime\rangle_{\text{R}}\equiv |\Psi\rangle_{\text{L}}\otimes |\Psi^\prime\rangle_{\text{R}}\in \mathcal{H}$ belong to the Hilbert space made of the tensor product $\mathcal{H}_{\text{SYM,L}}\otimes \mathcal{H}_{\text{SYM,R}}$ describing two isomorphic non-interacting $N=4$ SYM theories\footnote{The study of entanglement between two interacting $N=4$ SYM theories was first considered in \cite{Mollabashi:2014qfa}.}.

The BPS limit \eqref{BPSrho} reduces the maximally correlated state \eqref{eq:TFD-1} to
\begin{equation}
|\text{TFD}(\hat{\beta})\rangle = \frac{1}{\sqrt{Z(\hat{\beta})}}\sum_{\alpha\in\mathcal{H}_L\otimes\mathcal{H}_R} e^{-\hat{\beta}\Delta_\alpha/2} |\Delta_\alpha,\Delta_\alpha\rangle_L |\Delta_\alpha,-\Delta_\alpha\rangle_R\,.
\label{eq:TFD-2}
\end{equation}
where $\mathcal{H}_{\text{L}}$ and $\mathcal{H}_{\text{R}}$ are short notation for the half-BPS $\SO(4)$ invariant subspace $\mathcal{H}_{\text{LLM,L}}\subset \mathcal{H}_{\text{SYM,L}}$ and $\mathcal{H}_{\text{LLM,R}}\subset \mathcal{H}_{\text{SYM,R}}$, respectively. Notice how the BPS condition requires $\Delta_\alpha=-J_\alpha$ in $\mathcal{H}_{\text{R}}$, due to $|\Psi^\prime\rangle_{\text{R}}$ being the CPT transformed of $|\Psi\rangle_{\text{L}}$. This is compatible with supersymmetry since the supersymmetry charge $Q_\alpha = Q_{\text{L}\alpha}\otimes\mathbb{I}_{\text{R}} + \mathbb{I}_{\text{L}}\otimes Q_{\text{R}\alpha}$ satisfies $Q_\alpha|\Psi\rangle = 0$ when both $|\Psi\rangle_{\text{L}}$ and $|\Psi\rangle_{\text{R}}$ are half-supersymmetric. 

To quantify the quantum correlations between $\mathcal{H}_{\text{L}}$ and $\mathcal{H}_{\text{R}}$ in this kind of maximally correlated states, consider the $t=0$ 2-sided connected correlators
\begin{equation}
{\cal C}_{\rho} \equiv \text{tr}\left(\rho\,{\cal O}_{\text{L}}\otimes{\cal O}_{\text{R}}\right) - \left(\text{tr}\rho_{\text{L}}\,{\cal O}_{\text{L}}\right)\left(\text{tr}\rho_{\text{R}}\,{\cal O}_{\text{R}}\right)\,,
\label{eq:2side-corr}
\end{equation}
where $\rho$ is the density operator describing the state or the ensemble in $\mathcal{H}_{\text{L}}\otimes\mathcal{H}_{\text{R}}$, $\rho_{\text{L}}$ and  $\rho_{\text{R}}$ stand for its reduced density matrices and operators ${\cal O}_{\text{L}}$ and ${\cal O}_{\text{R}}$ are general gauge invariant operators acting on $\mathcal{H}_{\text{L}}$ and $\mathcal{H}_{\text{R}}$, respectively. 

In section \ref{sec:5d}, the length of the bridge in the near-extremal regime of these black holes \eqref{eq:bridge-div} was shown to increase as the non-extremal parameter $\mu$ decreases. This occurs even though the classical gravitational description does not develop a throat, but a naked singularity in the limit $\mu\to 0$. When interpreting the state \eqref{eq:TFD-2}, or its maximally entangled version, as two entangled distributions of giants sitting at the origin of two distinct AdS spaces, the non-vanishing correlations in the state \eqref{eq:TFD-2} ensure both spaces are connected quantum mechanically. It is the small entropy, see \eqref{eq:fentropy} or \eqref{eq:sstarentropy}, which bounds the amount of correlation \cite{cirac}, that is responsible for the lack of classical bulk connectivity, in agreement with the limiting bridge length behaviour. As soon as one includes quantum corrections to the classical supergravity description, this connectivity should be restored. 

Despite the expected lack of classical bridges in the BPS limit, the description of the naked singularity and its phase space interpretation in section \ref{sec:supers}, will allow us to identify the regions in spacetime where such quantum connectivity exists. These are precisely the gray areas in phase space describing the naked singularity in the semiclassical limit\footnote{When restricting the probe operators ${\cal O}_{\text{L}}$ and ${\cal O}_{\text{R}}$ to be half-BPS $\SO(4)$ symmetric, there exist conservation laws that can make correlations \eqref{eq:2side-corr} vanish. From now on, it is assumed that given some density $\rho$, the choice of probes ${\cal O}_{\text{L}}$ and ${\cal O}_{\text{R}}$ is such that \eqref{eq:2side-corr} does not vanish. Because of the arguments just presented, our interest in this work is in the localisation of the correlations in phase space and its bulk interpretation in terms of quantum connectivity between two otherwise disconnected spacetimes.}.

This main idea is spelt out in the coming subsections, building on the dictionary developed in section \ref{sec:review} and stressing the ability to design any desired correlation between both spacetimes given the analytic control over the microscopic degrees of freedom responsible for these correlations.

\subsection{Product states}

Product states $|\Psi\rangle_{\text{L}}\otimes |\Psi^\prime\rangle_{\text{R}} \in \mathcal{H}_{\text{L}}\otimes\mathcal{H}_{\text{R}}$ have no quantum correlations between $\mathcal{H}_{\text{L}}$ and $\mathcal{H}_{\text{R}}$ since their 2-sided connected correlators \eqref{eq:2side-corr} vanish. 

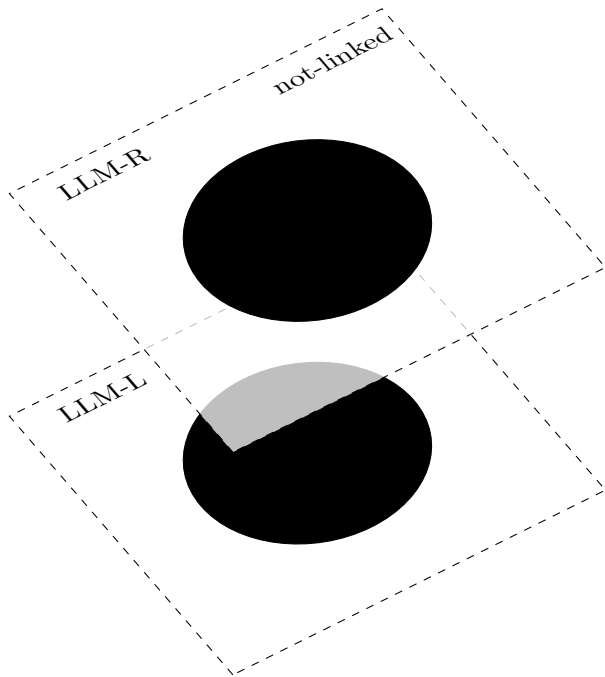
\begin{figure}
\centering	
	\begin{tikzpicture}[scale=.7,every node/.style={minimum size=1cm},on grid]
	
	\begin{scope}[
	yshift=-120,
	every node/.append style={yslant=\yslant,xslant=\xslant},
	yslant=\yslant,xslant=\xslant
	] 
	\draw[black, dashed, thin] (0,0) rectangle (7,7); 
	\draw[fill]  (3.5,3.5) circle [radius=2];
	\fill[black]
	(0.5,6.5) node[right, scale=.9] {LLM-L};
	
	\end{scope}
	
	\begin{scope}[
	yshift=0,
	every node/.append style={yslant=\yslant,xslant=\xslant},
	yslant=\yslant,xslant=\xslant
	]
	\fill[white,fill opacity=.75] (0,0) rectangle (7,7); 
	\draw[black, dashed, thin] (0,0) rectangle (7,7); 
	
	\draw[fill]  (3.5,3.5) circle [radius=2];
	\fill[black]
	(0.5,6.5) node[right, scale=.9] {LLM-R}
	(4.5,6.5) node[right, scale=.9] {not-linked};
	\end{scope} 
	
	\end{tikzpicture}
\caption{Two unentangled LLM droplets corresponding to smooth one sided geometries.}	
\label{fig1}
\end{figure}

If both $|\Psi\rangle_L$ and $|\Psi^\prime\rangle_R$ have smooth gravity duals, the $|\Psi\rangle_{\text{L}}\otimes |\Psi^\prime\rangle_{\text{R}}$ bulk description involves two disconnected LLM geometries. This situation is schematically represented in figure \ref{fig1} where one can see both LLM planes with their corresponding droplets\footnote{One should think of two independent disconnected geometries ending on their respective LLM planes, but one is only drawing the boundary condition encoded in the droplet that characterises the geometry uniquely in figure \ref{fig1}.}. Following the ideas in \cite{VanRaamsdonk:2010pw}, lack of correlation and entanglement is captured by the lack of connectivity between both spacetimes. In this situation, the statement is also true quantum mechanically.

\begin{figure}
	\centering	
	\begin{tikzpicture}[scale=.7,every node/.style={minimum size=1cm},on grid]
	
	\begin{scope}[
	yshift=-120,
	every node/.append style={yslant=\yslant,xslant=\xslant},
	yslant=\yslant,xslant=\xslant
	] 
	\draw[black, dashed, thin] (0,0) rectangle (7,7); 
	\draw[fill,gray]  (3.5,3.5) circle [radius=2];
	\fill[black]
	(0.5,6.5) node[right, scale=.9] {LLM-L};
	
	\end{scope}
	
	\begin{scope}[
	yshift=0,
	every node/.append style={yslant=\yslant,xslant=\xslant},
	yslant=\yslant,xslant=\xslant
	]
	\fill[white,fill opacity=.75] (0,0) rectangle (7,7); 
	\draw[black, dashed, thin] (0,0) rectangle (7,7); 
	
	\draw[fill,gray]  (3.5,3.5) circle [radius=2];
	\fill[black]
	(0.5,6.5) node[right, scale=.9] {LLM-R}
	(4.5,6.5) node[right, scale=.9] {not-linked};
	\end{scope} 
	
	\end{tikzpicture}
	\caption{Two unentangled LLM droplets corresponding to singular one sided geometries.}	
	\label{fig1-bis}
\end{figure}
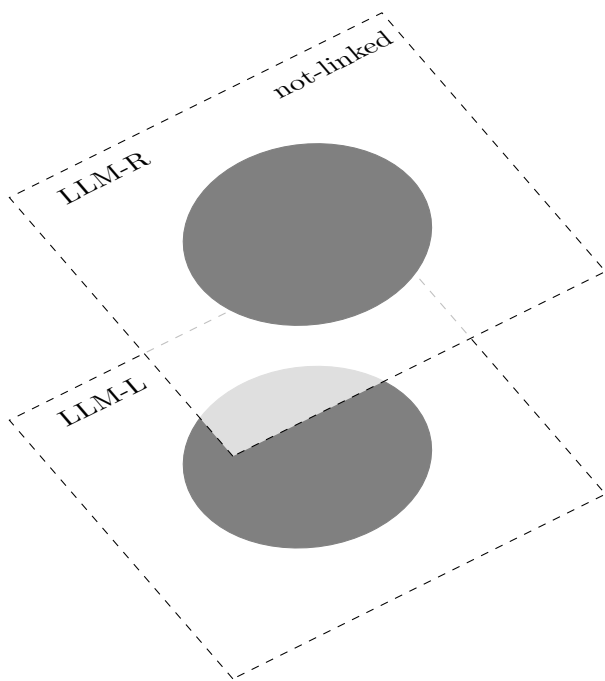

The same lack of connectivity holds for separable states, such as $\rho_{\text{superstar}}\otimes \rho_{\text{superstar}}$, as indicated in figure \ref{fig1-bis}. This is the analogue of an unentangled two black hole state in $\mathcal{H}_{\text{L}}\otimes\mathcal{H}_{\text{R}}$. Any observer having access to $\mathcal{H}_{\text{L}}\otimes \mathcal{H}_{\text{R}}$ would conclude there are no left-right correlations in this separable state. This is again an exact quantum mechanical statement. Consequently, two sided observers conclude there is no connectivity between the two singular gravity duals. Observers in $\mathcal{H}_{\text{L}}$ can never reach this conclusion due to the intrinsic quantum mechanical ambiguity in the reduced density matrix accessible to them, unless somehow there is exchange of classical information with observers in $\mathcal{H}_{\text{R}}$.

The bulk geometry dual to $\rho_{\text{superstar}}\otimes \rho_{\text{superstar}}$ corresponds to two disconnected superstar geometries. Hence both are singular in their deep interiors. Observers in $\mathcal{H}_{\text{L}}$ can validate the effective description provided by the superstar geometry. They may also interpret it as originating from the semiclassical description of a distribution of giant gravitons, but they can not make any precise statements regarding the existence of extra universes unless further information is provided to them. In fact, since LLM geometries are determined by the expectation value of the single particle phase space density, typical pure states describing at least $N_c$ giant gravitons $|\Psi_{\text{typical}}\rangle$ can not be distinguished from their ensemble averages when probed by low energy observables in classical gravity \cite{Balasubramanian:2005mg,Balasubramanian:2006jt,Balasubramanian:2007zt,Lashkari:2014pna}. Hence, the product state $|\Psi_{\text{typical}}\rangle \otimes |\Psi_{\text{typical}}\rangle$ is also effectively captured by figure \ref{fig1-bis} in classical gravity. It would indeed require the addition of higher order corrections in the bulk description to reproduce the different correlations between the N fermions responsible for the differences between $\rho_{\text{superstar}}$ and $|\Psi_{\text{typical}}\rangle\langle \Psi_{\text{typical}}|$.

\subsection{Maximally correlated states}

Maximally correlated states 
\begin{equation}
  |\Psi\rangle = \sum_{i=1}^K a_i\,|i\rangle_{\text{L}}\otimes |i\rangle_{\text{R}}
\label{eq:maxi}
\end{equation} 
have non-vanishing entanglement for $K>1$ between $\mathcal{H}_{\text{L}}$ and $\mathcal{H}_{\text{R}}$. The thermofield double state \eqref{eq:TFD-1} is one particular example where the vectors $\{|i\rangle_{\text{L}}\}$ and $\{|i\rangle_{\text{R}}\}$ stand for the energy eigenvectors of the non-interacting hamiltonian $H_{\text{L}}\otimes \mathbb{I}_{\text{R}} + \mathbb{I}_{\text{L}}\otimes H_{\text{R}}$. 

Given the holographic dictionary reviewed in section \ref{sec:review}, let us examine the correlation properties of these states and the relation between these and the existence of singularities from the perspective of single sided observers. Consider a maximally correlated state involving two orthogonal YT eigenstates $|\Psi_r\rangle$ $r=1,2$ 
\begin{equation}
|\Psi \rangle = \alpha\,|\Psi_1,\Psi_1\rangle + \sqrt{1-\alpha^2}\,|\Psi_2,\Psi_2\rangle\,, \quad \quad \alpha\in\mathbb{R}\,.
\label{toy}
\end{equation}
Both states describe $N-1$ fermions in the Fermi sea and differ by a single excitation $r_\gamma$ vs $r_{\tilde{\gamma}}$. Orthogonality guarantees both left and right reduced density matrices equal
\begin{equation}
\rho_{\text{L}} = \rho_{\text{R}} = \alpha^2|\Psi_1\rangle\langle \Psi_1| + (1-\alpha^2)|\Psi_2\rangle\langle \Psi_2|\,,
\label{toy-reduced}
\end{equation}
whereas their single particle versions are\footnote{These density matrices satisfy the standard normalisation $\text{Tr}\rho_{1\text{L}} = \text{Tr}\rho_{1\text{R}} =1$.}
\begin{equation}
  \rho_{1\text{L}} = \rho_{1\text{R}} = \frac{1}{N}\sum_{i=1}^{N-1} |i\rangle\langle i| + \frac{\alpha^2}{N}|\gamma\rangle\langle\gamma | +  \frac{1-\alpha^2}{N}|\tilde\gamma\rangle\langle\tilde\gamma | \,.
\end{equation}
Notice how the existence of extra correlations $(\alpha\neq 0,1)$\footnote{Due to the indistinguishibility of fermions, a single Young tableau encodes non-trivial correlations and entanglement. The increase in correlations alluded to refers to the nature of the maximally correlated state $|\Psi\rangle$ in equation \eqref{toy}.} between the excitations $|\gamma\rangle$ and $|\tilde{\gamma}\rangle$ is a source for grayness (singularity) in the gravity dual, according to our general discussion in \eqref{eq:singrho1}. The precise identification between these extra correlations and the origin of grayness can only be made by an observer having access to the entire quantum state since there is no quantum operation performed by a one sided observer that can help her determine the real source of the singularity. All she can infer, at most, is being in the ensemble \eqref{toy-reduced}, which already requires to have access to finite $N$ information.

Since the gravity dual is sensitive to the one particle density matrix, consider correlators \eqref{eq:2side-corr} of single particle operators ${\cal O}^{(1)}$. These equal
\begin{equation}
{\cal C}_{\Psi} =\alpha\,\sqrt{1-\alpha^2}\,\left({\cal O}_{L\tilde{\gamma}\gamma}\,{\cal O}_{R\tilde{\gamma}\gamma}
+ {\cal O}_{L\gamma\tilde{\gamma}} {\cal O}_{R\gamma\tilde{\gamma}}\right) 
+ \alpha^2\,(1-\alpha^2)({\cal O}_{R\gamma} - {\cal O}_{R\tilde{\gamma}})({\cal O}_{L\gamma} - {\cal O}_{L\tilde{\gamma}})
\label{toy-result}
\end{equation}
where different contributions were defined in these intermediate results
\begin{equation}
\begin{aligned}
\langle\Psi_1|{\cal O}^{(1)}_L |\Psi_1\rangle &= \frac{1}{N}\sum_{i=1}^{N-1}\langle i |{\cal O}^{(1)}_L |i\rangle + \frac{1}{N} \langle\gamma|{\cal O}^{(1)}_L |\gamma\rangle \equiv {\cal O}_{L,N-1} + {\cal O}_{L\gamma}\,, \\
\langle\Psi_2|{\cal O}^{(1)}_L |\Psi_2\rangle &= \frac{1}{N}\sum_{i=1}^{N-1}\langle i |{\cal O}^{(1)}_L |i\rangle + \frac{1}{N} \langle\tilde{\gamma}|{\cal O}^{(1)}_L |\tilde{\gamma}\rangle \equiv {\cal O}_{L,N-1} + {\cal O}_{L\tilde{\gamma}}\,,\\ 
\langle\Psi_2|{\cal O}^{(1)}_L |\Psi_1\rangle &= \frac{1}{N}\langle\tilde{\gamma}| {\cal O}^{(1)}_L |\gamma\rangle\equiv {\cal O}_{L\tilde{\gamma}\gamma}\,, \\
\langle\Psi_1|{\cal O}^{(1)}_L |\Psi_2\rangle &= \frac{1}{N}\langle\gamma| {\cal O}^{(1)}_L |\tilde{\gamma}\rangle\equiv {\cal O}_{L\gamma\tilde{\gamma}}\,,
\end{aligned}  
\end{equation}
with analogous expressions for the ${\cal O}_R^{(1)}$ matrix elements. As expected, this correlator only vanishes when the state \eqref{toy} is a product state\footnote{Remember it is assumed the probe operators have non-vanishing matrix elements for the quantum state under consideration.}, i.e. when the non-trivial correlation between the two excited fermions $|\gamma\rangle$ and $|\tilde{\gamma}\rangle$ is turned off. 

These claims follow from quantum mechanics. The further observation that can be added in our set-up is that, given a set of one-particle observables, their matrix elements support depend on the localisation properties of the quantum state wave functions in phase space. Hence,  their 2-sided correlations can also be localised. Consequently, if quantum correlations induce any kind of connectivity between left and right spacetimes \cite{VanRaamsdonk:2010pw} (assuming there is any gravity dual for them), the latter can only occur through the subregions of phase space where such correlations exist. 

The toy model just discussed \eqref{toy} has no gravity dual, but it helps to illustrate the main idea. The latter is developed below for maximally correlated states describing the purification of the superstar in \eqref{eq:TFD-2} and the localised superstar ensembles, both involving correlations between order $N$ of the fermions. To stress the ability to design any correlation between both spacetimes, the extension to maximal correlation between coherent states is also briefly discussed.

\paragraph{Superstar.} Consider the maximally correlated state whose reduced density matrix on either $\mathcal{H}_{\text{L}}$ or $\mathcal{H}_{\text{R}}$ equals the superstar density matrix \eqref{eq:uniform}
\begin{equation}
|\Psi \rangle_{\text{superstar}}= \frac{1}{\sqrt{Z}} \sum_{\vec{n}\in\mathcal{H}^\prime}  |\Psi_{\vec{n}}\rangle\otimes |\Psi_{\vec{n}}\rangle\,, \quad Z = \binom{N+N_c}{N}
\label{superstar}
\end{equation}
where $\mathcal{H}^\prime$ stands for the subspace of $N$ fermion states with at most $N_c$ giant gravitons. 

The quantum state \eqref{superstar} has non-vanishing correlations \eqref{eq:2side-corr}. For single particle operators ${\cal O}^{(1)}$, these equal
\begin{equation}
\begin{aligned}
{\cal C}_{\text{superstar}} &= \frac{N\,N_c}{(N+N_c)(N+N_c-1)}\sum_{k\neq j} \langle j|{\cal O}^{(1)}_L|k\rangle \langle j|{\cal O}^{(1)}_R|k\rangle \\
&+ \frac{N_c}{N}\,\frac{1}{(N+N_c)^2}\,\sum_i^{N+N_c} \langle i|{\cal O}^{(1)}_L|i\rangle\,\langle i|{\cal O}^{(1)}_R|i\rangle \\
& -  \frac{N_c}{N}\,\frac{1}{(N+N_c)^2}\frac{1}{N+N_c-1}\,
\sum_{i\neq j} \langle i|{\cal O}^{(1)}_L|i\rangle\,\langle j|{\cal O}^{(1)}_R|j\rangle\,.
\end{aligned}
\label{eq:2sidedcorr}
\end{equation}
But the amount of correlation between $\mathcal{H}_{\text{L}}$ and $\mathcal{H}_{\text{R}}$ is bounded by the mutual information \cite{cirac}
\begin{equation}
I(\text{L};\text{R}) \geq \frac{\left(\langle \Psi|{\mathcal O}_{\text{L}}{\mathcal O}_{\text{R}}|\Psi\rangle - \langle\Psi| {\mathcal O}_{\text{L}}|\Psi\rangle \langle\Psi| {\mathcal O}_{\text{R}}|\Psi\rangle\right)^2}{2\|{\mathcal O}_{\text{L}}\|^2\|{\mathcal O}_{\text{R}}\|^2}\,,
\label{eq:mutualbound}
\end{equation}	
where $\|{\mathcal O}_{\text{L}}\|$ and $\|{\mathcal O}_{\text{L}}\|$ are the largest eigenvalues for these single particle operators. Since $I(\text{L};\text{R}) = 2S_{\text{superstar}} \propto N\ll N^2$ for the state \eqref{superstar}, these correlations are suppressed in the large N limit to be accessible to classical bulk physics. Formally, if we were to reproduce these correlations using some bulk geodesic distance \cite{Louko:2000tp,Balasubramanian:1999zv}, the latter would include an additional divergent $\log N$ piece, indicating they are infinitely far apart in the $N\to \infty$ where the classical bulk description emerges.

This conclusion is consistent with the near-extremal bridge analysis in \eqref{eq:bridge-div}. In that regime, supergravity can still be trusted and the bridge length increases as the system becomes more extremal, i.e. as the non-extremal degrees of freedom become more diluted. Hence, one is already observing that quantum correlations between the non-extremal degrees of freedom responsible for the entropy scaling like $N^2$ and for the presence of a bridge, are decreasing. In the presence of such excitations, the entropy and correlations due to the quantum source of the singularity are subleading, but in the BPS limit \eqref{BPSrho}, they are everything that is left. Our arguments above indicate that the remaining BPS correlations are not strong enough to support a classical bridge, but they still describe quantum connectivity.

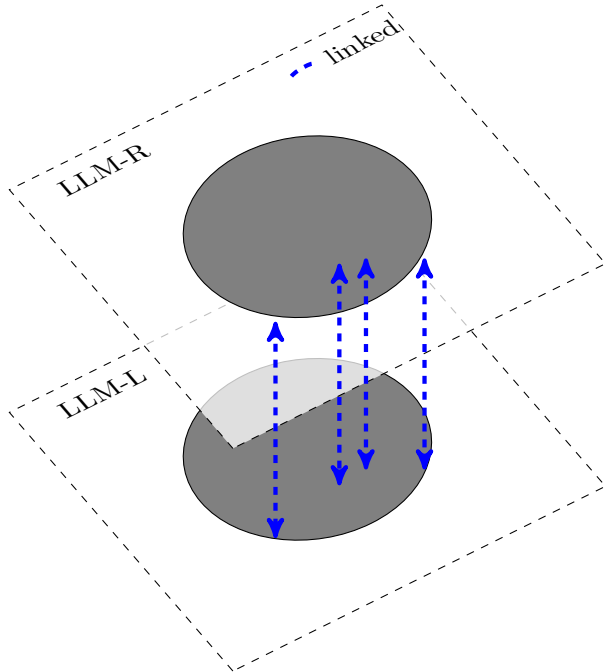
\begin{figure}	
	\centering
	\begin{tikzpicture}[scale=0.7,every node/.style={minimum size=1cm},on grid]
	
	\begin{scope}[
	yshift=-120,
	every node/.append style={yslant=\yslant,xslant=\xslant},
	yslant=\yslant,xslant=\xslant
	] 
	\draw[black, dashed, thin] (0,0) rectangle (7,7); 
	\draw[fill=gray]  (3.5,3.5) circle [radius=2];
	\fill[black]
	(0.5,6.5) node[right, scale=.9] {LLM-L};
	\end{scope}
	
	\begin{scope}[
	yshift=0,
	every node/.append style={yslant=\yslant,xslant=\xslant},
	yslant=\yslant,xslant=\xslant
	]
	\fill[white,fill opacity=.75] (0,0) rectangle (7,7); 
	\draw[black, dashed, thin] (0,0) rectangle (7,7); 
	
	\draw[dashed, ultra thick, blue]
	(5,6.5) to[out=30,in=30] (5.4,6.5); 
	
	\draw[fill=gray]  (3.5,3.5) circle [radius=2];
	\fill[black]
	(0.5,6.5) node[right, scale=.9] {LLM-R}
	(5.5,6.5) node[right, scale=.9] {linked};
	\end{scope} 
	
	\draw[<->,dashed, ultra thick, blue](3.6,3.6) to (3.6,-0.4);
	\draw[<->,dashed, ultra thick, blue](.8,2.4) to (.8,-1.7);
	\draw[<->,dashed, ultra thick, blue](2.0,3.5) to (2.0,-0.7);
	\draw[<->,dashed, ultra thick, blue](2.5,3.6) to (2.5,-0.4);
	
	\end{tikzpicture}
	\caption{Schematic representation of the gravity dual for two maximally entangled superstars. The existence of quantum correlations and entanglement is simulated by linking both LLM geometries along the droplets.}
	\label{fig2}	
\end{figure}

Since single sided observers are only sensitive to the superstar density matrix \eqref{eq:uniform}, their gravity duals are the superstar geometries themselves. These considerations suggest there is no classical bulk distinction between the gravity dual of the separable state $\rho_{\text{superstar}}\otimes \rho_{\text{superstar}}$ and the maximally correlated state \eqref{superstar}. This lack of classical gravity distinguishability should not be surprising given the need for quantising the classical moduli space of LLM geometries in order to reproduce the dual fermion Hilbert space \cite{Grant:2005qc,Maoz:2005nk}, following the covariant quantisation methods introduced in \cite{Crnkovic:1986ex,Zuckerman:1989cx}. Quantum mechanically, both states are different, though this conclusion is only accessible to a two sided observer who can measure the quantum correlations \eqref{eq:2sidedcorr}. 

Our conclusion makes any direct classical test of the EPR=ER conjecture not possible. But our discussion on the origin of the LLM singularities due to the existence of fermion correlations (see \eqref{eq:singrho1}) offers a different perspective. Despite not being able to encode the precise correlations in the classical geometry, the compact support of the bulk singularity in the LLM phase space\footnote{The interior of the droplet with radius $r_{\text{sup}}^2$ in \eqref{eq:r-super}.} is due to the semiclassical localisation properties of the quantum wave functions describing the ensemble accessible to a single sided observer. Hence, the only available connectivity between the two geometries experienced by both single sided observers is through the grayscale droplet \cite{VanRaamsdonk:2010pw}. 

In some tautological sense, this is an explicit construction of "quantum bridges" \cite{Maldacena:2013xja}, since the quantum state \eqref{superstar} is an EPR-like state in $\mathcal{H}_{\text{L}}\otimes\mathcal{H}_{\text{R}}$.  The magic of the holographic dictionary in the half-BPS sector of N=4 SYM is the existence of an infra-red bulk picture where some coarse-grained description of the phase space of the same quantum mechanical system emerges, allowing a potential bulk connectivity interpretation of the same quantities. 

To keep track of these facts, figure \ref{fig2} schematically represents the gravity dual to the state \eqref{superstar} in terms of two linked LLM grayscale droplets. There is no precise geometric definition as to what this linking means, but it does not involve regions of phase space outside of the droplet in the semiclassical regime. Notice that, in some heuristic sense, the bit thread picture advocated in \cite{Freedman:2016zud} is being realised here through the gluing of the LLM droplets by linking the correlated microscopic degrees of freedom (the eigenvalues of the adjoint matrices $X,\,X^\dagger$ in $N=4$ SYM).

\begin{figure}
	\centering
	\begin{tikzpicture}[xscale=8, yscale=8]
	\draw[<->, help lines] (0.5,1.5) -- (0.5,1)  -- (1,1);
	\node[below left] at (0.8,1) {$r_{\text{L}}=0$};
	\draw[red, ultra thick] (0.8,1) to [out=90 ,in=0] (0.5,1.3);
	\draw[->, dashed] (0.65,1) to [out=90 ,in=0] (0.5,1.15);
	\node [above right] at (0.51,1) {$\theta_{\text{L}}$};
	\node[above left] at (0.5,1.15) {$\frac{\pi}{2}$};
	\node[above right] at (0.65,1) {$0$};
	\draw[<->, help lines] (1.5,1.5) -- (1.5,1)  -- (2,1);
	\node[below left] at (1.8,1) {$r_{\text{R}}=0$};
	\draw[->, dashed] (1.65,1) to [out=90 ,in=0] (1.5,1.15);
	\node [above right] at (1.51,1) {$\theta_{\text{R}}$};
	\draw[blue, ultra thick] (1.8,1) to [out=90 ,in=0] (1.5,1.3);
	\node[above left] at (1.5,1.15) {$\frac{\pi}{2}$};
	\node[above right] at (1.65,1) {$0$};
	\draw[<->, blue, very thick] (0.65,1.15) to [out=-60, in=90] (1.65,1.15);
	\draw[<->, red, very thick] (0.65,1.15) to [out=45, in=210] (1.65,1.15);
	\end{tikzpicture}
	\caption{Two entangled distributions of giant gravitons located at $r_{\text{L}}=0$ and $r_{\text{R}}=0$, spread over two different spheres along $\theta_{\text{L}}$ and $\theta_{\text{R}}$ connected through the correlations encoded in the phase space formulation of their joint quantum mechanical description.}
	\label{fig:loc-bulk-2}
\end{figure}
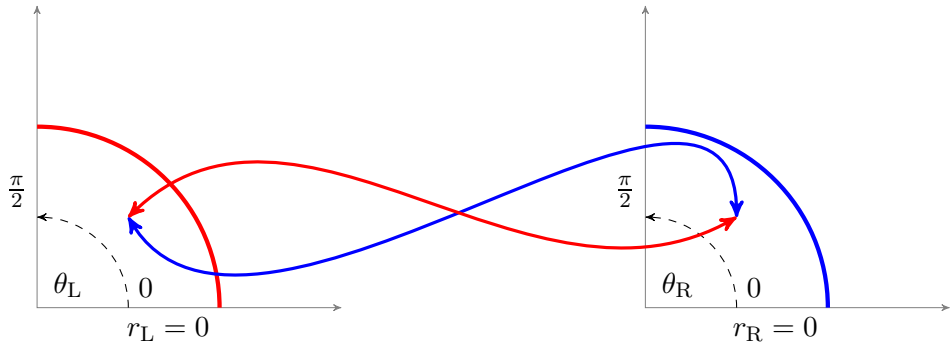

It is perhaps worth stressing this point using the geometric description in terms of the BPS limits of the non-extremal black holes \eqref{eq:5dBH}. The left observer sees a distribution of giants along $\theta_{\text{L}}$ at the origin of its AdS universe $r_{\text{L}}=0$, whereas the right observer sees the same distribution along $\theta_{\text{R}}$ at the origin of its AdS universe $r_{\text{R}}=0$. Since moving inside each droplet is equivalent to moving along the respective 5-sphere at the respective AdS origin, the linking between droplets in figure \ref{fig2} describes the connectivity between the origins of two AdS spacetimes along their entangled giant gravitons distributed along both transverse 5-spheres (see figure \ref{fig:loc-bulk-2}).

\paragraph{Localised superstar.} To stress the relevance of the localisation of quantum correlations in phase space for the connectivity between two LLM geometries, consider the localised superstar discussed in \eqref{eq:example-1} in subsection \ref{sec:localised}. The maximally correlated state purifying this density operator in $\mathcal{H}_{\text{L}}\otimes\mathcal{H}_{\text{R}}$ equals
\begin{equation}
|\Psi \rangle_{\text{localised}} = \frac{1}{\sqrt{Z_{\text{r}}}} \sum_{\vec{n}\in \mathcal{H}_r} |\Psi_{\vec{n}}\rangle \otimes |\Psi_{\vec{n}}\rangle\,, \quad \text{with} \quad  Z_{\text{r}}= \binom{N_2+N_c}{N_2}\,,
\label{eq:annulus}
\end{equation}
where $\vec{n}$ labels the set of $N$ fermion states where $N_1$ of them remain in the Fermi sea, whereas $N_2=N-N_1$ of them explore all excitations compatible with having at least $N_c$ giant gravitons.

When quantum correlations \eqref{eq:2side-corr} are computed in this state \eqref{eq:annulus}, only the excitations of the $N_2$ fermions will contribute. Labelling these by $a,\,b$, 2-sided correlators for single particle operators equal
\begin{equation}
\begin{aligned}
{\cal C}_{\text{localised}} &= \frac{N_2\,N_c}{N^2\,(N_2+N_c)^2}\,\sum_{a} \langle a| {\cal O}^{(1)}_L |a\rangle \langle a| {\cal O}^{(1)}_R |a\rangle \\
&+ \frac{N_2\,N_c}{(N_2+N_c)(N_2+N_c-1)} \sum_{a\neq b} \langle a| {\cal O}^{(1)}_L |b\rangle \langle a| {\cal O}^{(1)}_R |b\rangle \\
&- \frac{N_2\,N_c}{N^2(N_2+N_c)^2(N_2+N_c-1)}\,\sum_{a\neq b}\langle a| {\cal O}^{(1)}_L |a\rangle \langle b| {\cal O}^{(1)}_R |b\rangle\,.
\end{aligned}
\end{equation}

\begin{figure}
\centering
	\begin{tikzpicture}[scale=0.7,every node/.style={minimum size=1cm},on grid]
	
	\begin{scope}[
	yshift=-120,
	every node/.append style={yslant=\yslant,xslant=\xslant},
	yslant=\yslant,xslant=\xslant
	] 
	\draw[black, dashed, thin] (0,0) rectangle (7,7); 
	\draw[fill=gray]  (3.5,3.5) circle [radius=2];
	\draw[fill] (3.5,3.5) circle [radius=1];
	\fill[black]
	(0.5,6.5) node[right, scale=.9] {LLM-L};
	
	\end{scope}
	
	\begin{scope}[
	yshift=0,
	every node/.append style={yslant=\yslant,xslant=\xslant},
	yslant=\yslant,xslant=\xslant
	]
	\fill[white,fill opacity=.75] (0,0) rectangle (7,7); 
	\draw[black, dashed, thin] (0,0) rectangle (7,7); 
	
	\draw[dashed, ultra thick, blue]
	(5,6.5) to[out=30,in=30] (5.4,6.5); 
	
	\draw[fill=gray]  (3.5,3.5) circle [radius=2];
	\draw[fill] (3.5,3.5) circle [radius=1];
	\fill[black]
	(0.5,6.5) node[right, scale=.9] {LLM-R}
	(5.5,6.5) node[right, scale=.9] {linked};
	\end{scope} 
	
	\draw[<->,dashed, ultra thick, blue](3.6,3.6) to (3.6,-0.4);
	\draw[<->,dashed, ultra thick, blue](.8,2.4) to (.8,-1.7);
	\draw[<->,dashed, ultra thick, blue](3,3.6) to (3,-0.4);
	
	\end{tikzpicture}
	
\caption{Schematic representation of the gravity dual for two maximally entangled modified superstars. The existence of quantum correlations and entanglement is captured by the gluing of both LLM geometries along the gray annulus.}
\label{fig3}
\end{figure}
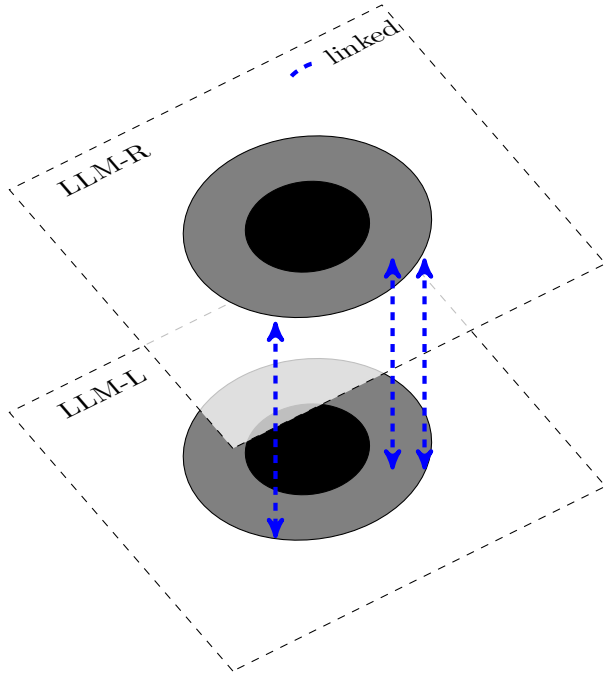

\begin{figure}
	\centering
	\begin{tikzpicture}[xscale=8, yscale=8]
	\draw[<->, help lines] (0.5,1.5) -- (0.5,1)  -- (1,1);
	\node[below left] at (0.8,1) {$r_{\text{L}}=0$};
	\draw (0.8,1) to [out=90 ,in=0] (0.5,1.3);
	\draw[red, ultra thick] (0.6,1.28) arc (70:90:0.30);
	\draw[red, dashed, thick] (0.5,1) to (0.6,1.28);
	\node [red, above right] at (0.6,1.27) {$\theta_0$};
	\draw[->, dashed] (0.65,1) to [out=90 ,in=0] (0.5,1.15);
	\node [above right] at (0.51,1) {$\theta_{\text{L}}$};
	\node[above left] at (0.5,1.15) {$\frac{\pi}{2}$};
	\node[above right] at (0.65,1) {$0$};
	
	\draw[<->, help lines] (1.5,1.5) -- (1.5,1)  -- (2,1);
	\node[below left] at (1.8,1) {$r_{\text{R}}=0$};
	\draw[->, dashed] (1.65,1) to [out=90 ,in=0] (1.5,1.15);
	\node [above right] at (1.51,1) {$\theta_{\text{R}}$};
	\node[above left] at (1.5,1.15) {$\frac{\pi}{2}$};
	\node[above right] at (1.65,1) {$0$};
	\draw (1.8,1) to [out=90 ,in=0] (1.5,1.3);
	\draw[blue, ultra thick] (1.6,1.28) arc (70:90:0.30);
	\draw[blue, dashed, thick] (1.5,1) to (1.6,1.28);
	\node [blue, above right] at (1.6,1.27) {$\theta_0$};
	
	\draw[<->, blue, very thick] (0.52,1.2) to [out=-60, in=90] (1.55,1.2);
	\draw[<->, red, very thick] (0.52,1.2) to [out=45, in=210] (1.55,1.2);
	\end{tikzpicture}
	\caption{Two entangled distributions of giant gravitons located at $r_{\text{L}}=0$ and $r_{\text{R}}=0$, spread over the localised regions $[\theta_0,\pi/2]$ in two different spheres along $\theta_{\text{L}}$ and $\theta_{\text{R}}$ connected through the correlations encoded in the phase space formulation of their joint quantum mechanical description.}
	\label{fig:loc-bulk-3}
\end{figure}
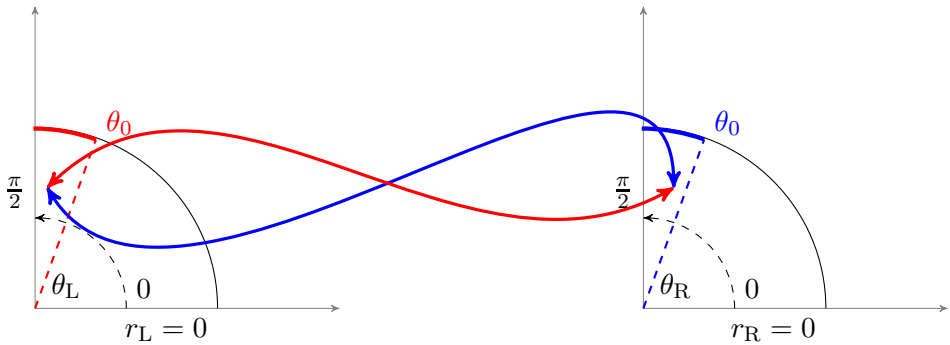

The absence of left-right correlations between the $N_1$ fermions is reflected in the gravity dual by leaving their black inner disks unlinked, as it occurred for product states. Otherwise, both gray annulus are linked, as indicated in figure \ref{fig3}. Hence, the connectivity between both spacetimes can only happen through quantum bridges localised in the annulus, whereas subleading higher particle correlators would require to replace the LLM phase space with the entire N-particle phase space description. This localised connectivity is shown in figure \ref{fig:loc-bulk-3} in terms of localised giant gravitons over the two transverse 5-spheres.

\paragraph{Correlation design.} To stress the ability to design quantum correlations localised in specific regions of phase space and, consequently, to design some potentially arbitrary, within the approximations, connectivity between both LLM geometries, the discussion around state \eqref{toy} can be extended to coherent states \cite{Hillery:1983ms}
\begin{equation}
  |\alpha_n \rangle = e^{\alpha_n\,a^\dagger - \alpha_n^\star\,a}\,|0\rangle = e^{-|\alpha_n|^2/2}\,\sum_{s=0}^\infty \frac{\alpha_n^s}{(s!)^{1/2}}|s\rangle\,.
\end{equation}
These are coherent superpositions of standard creation $(a^\dagger)$ and anihilation operators $(a)$, satisfying $[a,\,a^\dagger]=1$, $a|0\rangle = 0$, $a|n\rangle = \sqrt{n}\,|n-1\rangle$ and $a^\dagger|n\rangle = \sqrt{n+1} |n+1\rangle$, whose wave functions are localised within an $\hbar$ area around the phase space point $\alpha_n \equiv \frac{x_n + ip_n}{\sqrt{2\hbar}}$.

Even though coherent states are not orthogonal
\begin{equation}
\langle \alpha_1|\alpha_2\rangle = e^{-(x_1-x_2)^2/(4\hbar)}\,e^{-(p_1-p_2)^2/(4\hbar)}\,e^{i(x_1p_2-p_1x_2)/(2\hbar)}\,,
\end{equation}
they effectively behave as such for semiclassical separations
\begin{equation}
\langle \alpha_1|\alpha_2\rangle \to 0 \quad \text{when} \quad x_1-x_2,\,p_1-p_2 \sim N\,\sqrt{\hbar}\,, \quad \hbar \to 0
\label{semiclass}
\end{equation}

Their localisation in phase space allows to design specific correlations between phase space regions. To illustrate this idea, consider the state\footnote{Strictly speaking, one should consider a large number of fermions excited in nearby phase space cells so that the resulting state admits a classical description. This technical point is ignored below for the sake of reducing the technicalities while stressing the physical point sharply.}
\begin{equation}
\begin{aligned}
	|\Psi\rangle &= \sqrt{p}\,|\Psi_1,\Psi_1\rangle + \sqrt{1-p}\,|\Psi_2,\Psi_2\rangle\,, \\
    |\Psi_1\rangle &= A\,\left(|\alpha_1,\alpha_2\rangle - |\alpha_2,\alpha_1\rangle\right)\,, \\
    |\Psi_2\rangle &= B\,\left(|\alpha_1,\alpha_3\rangle - |\alpha_3,\alpha_1\rangle\right)\,.
\end{aligned}
\end{equation}
This is a maximally correlated state between states built from the Slater determinant of two coherent states (either $|\alpha_1\rangle$ and $|\alpha_2\rangle$, or $|\alpha_1\rangle$ and $|\alpha_3\rangle$). Constants $A$ and $B$ are fixed by proper normalisation
\begin{equation}
\langle\Psi_1|\Psi_1\rangle = 1 \quad \Rightarrow \quad |A|^2 = \frac{1}{2\left(1-|\langle \alpha_1|\alpha_2\rangle|^2\right)}\,,
\end{equation}
with an analogous expression for $|B|^2$. 

The state $|\Psi_1\rangle$ describes a state of two fermions localised at $\alpha_1$ and $\alpha_2$. The state $|\Psi_2\rangle$ does so at $\alpha_1$ and $\alpha_3$. Both states are non-orthogonal since
\begin{equation}
\langle\Psi_2|\Psi_1\rangle = 2A\,B^\star \left(\langle \alpha_3|\alpha_2\rangle - \langle \alpha_1|\alpha_2\rangle\,\langle \alpha_3|\alpha_1\rangle\right)\,.
\end{equation}
But working in the semiclassical regime \eqref{semiclass}, they effectively are. Hence, the reduced density matrices in the semiclassical approximation\footnote{For a discussion on whether the entanglement entropy in holographic CFTs can be given by the expectation value of a linear operator, see \cite{Almheiri:2016blp}. For general holographic expectations for linear superpositions of states, see the discussion in section 3 in \cite{Papadodimas:2015jra}.}
\begin{equation}
  \rho_L=\rho_R \approx p |\Psi_1\rangle\langle \Psi_1| + (1-p)\,|\Psi_2\rangle\langle \Psi_2|
\end{equation}
equal the ones computed in our toy model \eqref{toy-reduced}. The only difference is the localisation properties of their wave functions. 

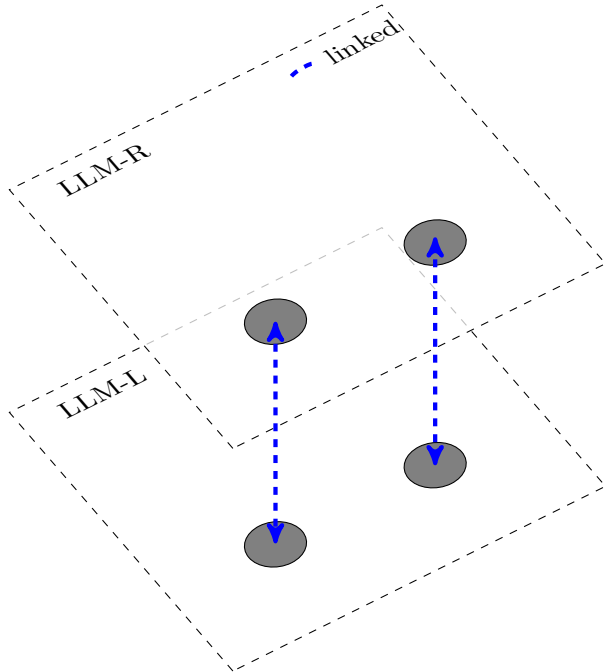
\begin{figure}
\centering
	\begin{tikzpicture}[scale=0.7,every node/.style={minimum size=1cm},on grid]
	
	\begin{scope}[
	yshift=-120,
	every node/.append style={yslant=\yslant,xslant=\xslant},
	yslant=\yslant,xslant=\xslant
	] 
	\draw[black, dashed, thin] (0,0) rectangle (7,7); 
	\draw[fill=gray]  
	(5,2) circle (.5) 
	(2,2) circle (.5); 
	
	\fill[black]
	(0.5,6.5) node[right, scale=.9] {LLM-L};
	\end{scope}
	
	\begin{scope}[
	yshift=0,
	every node/.append style={yslant=\yslant,xslant=\xslant},
	yslant=\yslant,xslant=\xslant
	]
	\fill[white,fill opacity=.75] (0,0) rectangle (7,7); 
	\draw[black, dashed, thin] (0,0) rectangle (7,7); 
	
	\draw[dashed, ultra thick, blue]
	(5,6.5) to[out=30,in=30] (5.4,6.5); 
	
	\draw [fill=gray]
	(5,2) circle (.5) 
	(2,2) circle (.5); 
	
	\fill[black]
	(0.5,6.5) node[right, scale=.9] {LLM-R}
	(5.5,6.5) node[right, scale=.9] {linked};
	
	\end{scope} 
	
	\draw[<->,dashed,ultra thick,blue](3.8,4) to (3.8,-0.32);
	\draw[<->,dashed,ultra thick,blue](.8,2.4) to (.8,-1.8);
	
	\end{tikzpicture}
\caption{Schematic representation of the gravity dual for two entangled fermions in coherent states. These allow to localise in phase space the left-right quantum correlations through the linking of their corresponding supports.}
\label{fig4}
\end{figure}	

Exact quantum correlators depend on off-diagonal matrix elements in the coherent state basis. Their contribution is subleading in the semiclassical limit. Hence, correlators behave like in \eqref{toy-result}, with wave functions sharply picked in phase space. Following previous arguments, the gluing between spacetimes should only occur in Planck area cells centered around $\alpha_2$ and $\alpha_3$ in phase space, as illustrated in figure \ref{fig4}. For classical gravity dual considerations, one should consider order $N$ coherent states to have macroscopic droplet entangling regions in phase space. In fact, since the superstar entropy \eqref{eq:sstarentropy} can be reproduced by thinking of it in terms of a gas of coherent states in a compact region of phase space \cite{Balasubramanian:2007zt}, it already provides an example for such a macroscopic coherent state configuration.

\subsection{Comments on traversability}
\label{sec:2trav}

The discussion in this section follows \cite{Gao:2016bin,Maldacena:2017axo} and it focuses on the possibility of using the quantum correlations and associated entanglement described earlier as a resource to connect both spacetimes.

It is apparent from the Penrose diagram \ref{fig:Penrose} of the eternal AdS black hole, that no information can be sent between both boundaries because of the lack of causal connection. In the quantum theory, this is because there is no interaction between the degrees of freedom in both Hilbert spaces. More explicitly, consider a quantum state $|\Psi\rangle \in \mathcal{H}_{\text{L}}\otimes \mathcal{H}_{\text{R}}$ and turn on some perturbation $\left(e^{i\delta_{\text{L}}\hat{\mathcal{R}}_{\text{L}}}\right)$ at $t=0$. One can probe whether this perturbation reaches $\mathcal{H}_{\text{R}}$ by computing whether the $\hat{\mathcal{R}}_{\text{R}}$ expectation value gets modifed by the action of the perturbation \cite{Maldacena:2017axo}
\begin{equation}
  \langle\Psi |\,e^{-i\delta_{\text{L}}\hat{\mathcal{R}}_{\text{L}}}\,\hat{\mathcal{R}}_{\text{R}}(t)\,e^{i\delta_{\text{L}}\hat{\mathcal{R}}_{\text{L}}}
  |\Psi\rangle = \langle\Psi |\,e^{-i\delta_{\text{L}}\hat{\mathcal{R}}_{\text{L}}}\,U^\dagger(t)\hat{\mathcal{R}}_{\text{R}}\,U(t)\,e^{i\delta_{\text{L}}
  	\hat{\mathcal{R}}_{\text{L}}}|\Psi\rangle\,,
\label{eq:mprobe}
\end{equation}
under the evolution operator $U(t)=e^{-itH}$. In the absence of interaction, $H=H_{\text{L}}\otimes\mathbb{I}_{\text{R}} + \mathbb{I}_{\text{L}}\otimes H_{\text{R}}$, the evolution operator $\hat{\mathcal{R}}_{\text{R}}(t)$ only acts on $\mathcal{H}_{\text{R}}$. Consequently, it commutes with the unitary perturbation $e^{i\delta_{\text{L}}\hat{\mathcal{R}}_{\text{L}}}$ and has no influence on the observable in $\mathcal{H}_{\text{R}}$. Hence, the wormhole is non-traversable.

Analogously, if both boundary theories are not classically coupled, i.e. if observers in different universes can not share the outcomes of their local measurements, the standard teleportation protocol \cite{teleportation} fails, despite the existence of quantum correlations when the state $|\Psi\rangle$ is entangled. 

Reference \cite{Gao:2016bin} realised that turning on some double trace relevant interaction between the two CFTs in a thermofield double state can generate a one-loop stress tensor violating the averaged null energy condition (ANEC). This is a necessary condition for wormholes to become traversable \cite{Morris:1988tu,Visser:2003yf,Hochberg:1998ii}. Physically, the negative energy density defocuses gravity allowing the wormhole, i.e. the quantum bridge, to open up. Technically, the backreaction of the one-loop bulk stress tensor pushes the horizon enough so that if the perturbation is sent far enough in the past from $\mathcal{H}_{\text{L}}$, it can be measured by an observer in $\mathcal{H}_{\text{R}}$ without diving into the black hole interior, making the wormhole traversable. This phenomenon is nicely captured by the probe \eqref{eq:mprobe} \cite{Maldacena:2017axo} 
\begin{equation}
\begin{aligned}
  \langle \Psi | e^{-i\delta_{\text{L}}\hat{\mathcal{R}}_{\text{L}}(-t)}\, \mathcal{R}_{\text{R}}(t)\,e^{i\delta_{\text{L}}\hat{\mathcal{R}}_{\text{L}}(-t)}|\Psi\rangle 
  & \approx \langle\Psi| \hat{\mathcal{R}}_{\text{R}} |\Psi \rangle \\
  &-g\,\delta_{\text{L}}\,\langle \Psi |\left[\hat{\mathcal{R}}_{\text{L}}(-t)\,,\mathcal{O}_{\text{L}}\right]\left[\mathcal{O}_{\text{R}}\,,\hat{\mathcal{R}}_{\text{R}}(t)\right]|\Psi\rangle  + \dots
\end{aligned}
\label{eq:trav-effect}
\end{equation}
where $\mathcal{R}_{\text{R}}(t)=e^{-igV}\,\hat{\mathcal{R}}_{\text{R}}\,e^{igV}$ depends on the interaction $e^{igV}= e^{ig\mathcal{O}_L(0)\mathcal{O}_R(0)}$ between $\mathcal{H}_{\text{L}}$ and $\mathcal{H}_{\text{R}}$  turned on at $t_L=t_R=0$  and it was assumed that both $g$ and $\delta_{\text{L}}$ were small to keep the leading contribution to the expectation value\footnote{Some readers may wonder what the relation is between the protocol used to send a message, such as \eqref{eq:trav-effect} and the standard one used in teleportation \cite{teleportation}. It was stressed in \cite{Maldacena:2017axo} that turning on the interaction and performing a non-projective measurement in $\mathcal{H}_L$ was similar to the standard teleportation protocol. This discussion was extended to include the dynamics of scrambling in \cite{Susskind:2017nto}, where the authors also estimate how complex it is to achieve all these operations.}.

These arguments are rather general, especially in the quantum mechanics side with no holographic interpretation attached to them. If one applies them within the half-BPS sector of N=4 SYM, they will {\it not} give rise to a traversable wormhole\footnote{What is meant here is that the quantum bridge associated with the quantum mechanics of the N free fermions will not see such effect within the half-BPS sector.} because of the lack of dynamics in this sector, i.e. because the effect requires non-trivial commutators in \eqref{eq:trav-effect}. But, turning on $g_{\text{YM}}$, allowing to explore the full N=4 SYM Hilbert space $\mathcal{H}_{\text{SYM}}$ dynamics should change the conclusion for adequate choices of the operators $\mathcal{O}_{\text{L}}$ and $\mathcal{O}_{\text{R}}$.

In particular, it should be possible to generalise the double trace interactions introduced in \cite{Gao:2016bin}, to take into account the microscopic description of the system in terms of excitations of giant gravitons,  to discuss traversable wormholes in the near-extremal single R-charged AdS black holes discussed in section \ref{sec:5d}.

\section{Single boundary EPR=ER in LLM}
\label{sboundary}

The notion of entanglement depends on the subsystem factorisation of the Hilbert space \cite{Zanardi:2004}. The two boundary discussion in section \ref{sec:2-boundary} has a manifest factorisation allowing to define a precise notion of entanglement between $\mathcal{H}_{\text{L}}$ and $\mathcal{H}_{\text{R}}$. But the ideas introduced in \cite{Maldacena:2013xja} are even more exciting in single boundary set-ups, where one could consider entangled black holes \cite{Susskind:2014yaa}, for example.

Such notion requires some factorisation of the Hilbert space (to properly define entanglement) and some (effective) bulk locality to distinguish the black hole locations. If such formulation would exist, even if approximately, it would not only allow to test the EPR=ER conjecture itself \cite{Maldacena:2013xja}, but together with \cite{Gao:2016bin}, it would very strongly suggest that turning on non-local boundary interactions between the degrees of freedom responsible for both entangled black holes, would make the associated wormhole traversable, if the interactions satisfy some conditions.

In this section, a very modest step is taken in this direction in the limited, but controllable, corner of the half-BPS $\SO(4)$ invariant sector of N=4 SYM. There are no black holes in this sector, but there are superstars and in section \ref{sec:localised}, it was explained how to describe the notion of a localised superstar. What is missing is the existence of some Hilbert space factorisation allowing to unambiguously quantify the entanglement between two such localised superstars. This technical problem in our set-up is a general difficulty when encoding quantum information in gauge and gravity theories \cite{Donnelly:2016auv,Donnelly:2017jcd}.

The effective field theory techniques sharpened in \cite{Berenstein:2017rrx} will allow us to circumvent this difficulty. These will be used below to introduce some notion of entanglement between different localised gases of pointlike gravitons and to explore connections to teleportation protocols, connectivity in space and traversability in this single boundary set-up. 

The type of non-locality required in our half-BPS discussions is in R-charge Hilbert space, since it is the entanglement associated with the  factorisation of the latter that is responsible for the non-trivial quantum correlations among the fermions\footnote{The notion of R-charge entanglement entropy in the half-BPS sector of N=4 SYM has been recently stressed in the literature \cite{Berenstein:2016pcx,Berenstein:2017abm,Lin:2017dnz,Berenstein:2017rrx}. For earlier discussions on entanglement entropy and internal space in AdS/CFT, see \cite{Andrade:2013rra,Belin:2013uta,Karch:2014pma}.}.

The phase space formulation of quantum mechanics treats space and momentum democratically. Since the LLM plane provides a semiclassical description of a single fermion phase space, it allows us to explore a different factorisation of the Hilbert space. One related to the emergent one dimension where the fermion trapping potential acts. The latter is natural from the condensed matter description of a system of N free fermions. In the second part of this section, together with appendix \ref{cond-mat}, we briefly review the calculation of real space R-charge entanglement performed in the condensed matter literature\footnote{This formalism has also been used recently in \cite{Hartnoll:2015fca, Hartnoll:2016mdv}.}.

\subsection{Entangled gas of gravitons}
\label{sec:egrav}

To circumvent the difficulty of having a properly factorised Hilbert space, one can consider the effective factorisation in the N fermion quantum mechanics, recently advocated in \cite{Berenstein:2016pcx} and subsequently pursued in \cite{Berenstein:2017abm,Lin:2017dnz,Berenstein:2017rrx}\footnote{There exists earlier work by de Mello Koch and collaborators \cite{Koch:2008ah,Koch:2016jnm}, where even tough this explicit factorisation terminology was not stressed, the localisation properties for the type of excitations considered here was already discussed.}.

\begin{figure}
	\centering
	\begin{tikzpicture}
	\draw (0,0) to (5,0);
	\draw (0,0) to (0,-5);
	\draw (2.5,-3) to (2.5,-5);
	\draw (0,-5) to (2.5,-5);
	\draw (2.5,-3) to (5,-3);
	\draw (5,0) -- (5,-3); 
	\node [above] at (2.5,0) {$\text{M}_1$};
	\node [above] at (3.7,-3) {$\text{M}_2$};
	\node [below right] at (5,-1.5) {$\text{L}_1$};
	\node [below right] at (2.5,-4) {$\text{L}_2$};
	\end{tikzpicture}
	\caption{Young tableau corresponding to the reference state $|\Box\rangle$.}
	\label{fig:entsuper-0}
\end{figure}
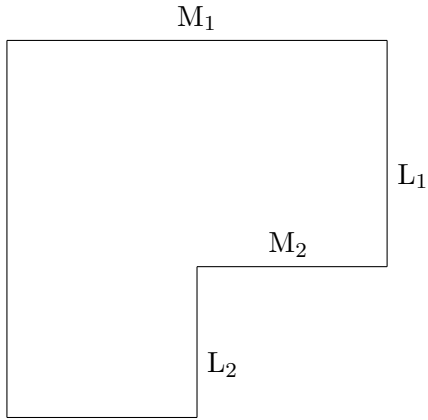

One of the ideas introduced in \cite{Berenstein:2017abm} and developed more precisely in \cite{Berenstein:2017rrx} using the language of code subspaces \cite{Almheiri:2014lwa,Harlow:2016vwg} is to consider an approximate Hilbert space factorisation describing small excitations around some {\it reference} state. Consider a reference state $|\Box\rangle$ having a gravity dual, such as the one in figure \ref{fig:entsuper-0}. The code subspace Hilbert space is defined as the approximate Hilbert space 
\begin{equation}
  \mathcal{H}_{\text{code\,}|\Box\rangle} = \mathcal{H}_{\text{M}_1+1,1}\otimes \mathcal{H}_{\text{M}_2+1,\text{N}_1+1}\otimes \mathcal{H}_{1,\text{L}_2+1}\otimes \mathcal{H}_{\text{M}_1,\text{L}_1}\otimes \mathcal{H}_{\text{M}_2,\text{L}_2}
\label{eq:factor}
\end{equation}
where the pairs $(x,y)$ in $\mathcal{H}_{(x,y)}$ label the row and the column position where quantum excitations (boxes) are added using the convention that $x$ grows to the right and $y$ grows downwards. 

As explained in \cite{Berenstein:2017rrx}, the validity of the effective field theory constructed out of reference states with $\text{M}_i,\,\text{L}_i$ of order $N$ requires the size of the small Young tableaux describing their fluctuations not to be larger than squares of size $\sqrt{N}$. Hence, the energy carried by these excitations is order $\mathcal{O}(\text{N})$, at most, and should be interpreted as describing pointlike gravitons (or a gas of them) rather than giant gravitons. 

\begin{figure}
	\centering
	\begin{tikzpicture}
	\draw (0,0) to (5,0);
	\draw (0,0) to (0,-5);
	\draw (2.5,-3) to (2.5,-5);
	\draw (1,-5) -- (2.5,-5);
	\draw (2.5,-3) to (5,-3);
	\draw[red, thick] (5,-1) rectangle (6,0);
	\draw[blue, thick] (0,-6) rectangle (1,-5);
	\draw (5,-1) -- (5,-3);
	\draw[<->, blue, very thick] (0.45,-5.55) to [out=90, in=0] (5.5,-0.5);
	\node[blue, below left] at (0,-5.2) {$\sim \sqrt{\text{N}}$};
	\draw[<->, red, very thick] (0.45,-5.55) to [out=0, in=180] (5.5,-0.5);
	\node[red, below right] at (6.1,-0.2) {$\sim \sqrt{\text{N}}$};
	\end{tikzpicture}
	\caption{Young tableau consisting of a reference state $|\Box\rangle$ together with (at most) $\sqrt{\text{N}}\times \sqrt{\text{N}}$ entangled graviton excitations.}
	\label{fig:entgraviton}
\end{figure}
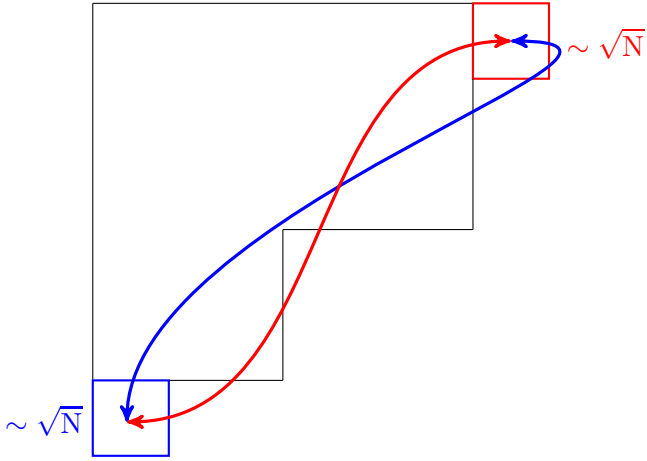

To describe quantum states resembling entangled gases of gravitons, one can entangle excitations between  $\mathcal{H}_{\text{M}_1+1,1}$ in the top right corner and $\mathcal{H}_{1,\text{L}_2+1}$ in the bottom left corner, while keeping some macroscopic classical scale, i.e. leaving the details of the reference state in the middle region of the YT in figure \ref{fig:entsuper-0} untouched. For the purposes of this discussion and also to match the notation in section \ref{sec:review}, these factors are relabelled as $\mathcal{H}_{\text{M}_1+1,1}=\mathcal{H}_{\text{red}}$, $\mathcal{H}_{1,\text{L}_2+1}=\mathcal{H}_{\text{blue}}$, while jointly referring to all other factors as $\mathcal{H}_{\text{C}}$. This is illustrated in figure \ref{fig:entgraviton}.

The existence of such effective factorisation allows to define entanglement between $\mathcal{H}_{\text{red}}$ and $\mathcal{H}_{\text{blue}}$ at once. To describe a pair of entangled gases of gravitons, consider a maximally correlated state
\begin{equation}
	|\Psi\rangle = \sum_{\text{E}} a_{\text{E}}\,|\text{E}_{\text{red}},\,\text{E}_{\text{blue}},\,\Box\rangle\,,
	\label{eq:pure}
\end{equation}
where the sum over excitations is bounded from above by the cut-off defining $\mathcal{H}_{\text{code\,}|\Box\rangle} $, as explained in \cite{Berenstein:2017rrx}. The coefficients $a_{\text{E}}$ can be chosen so that the reduced density matrix
\begin{equation}
	\rho_{\text{red},\Box} = \sum_{\text{E}} |a_{\text{E}}|^2\,|\text{E}_{\text{red}},\,\Box\rangle \langle \text{E}_{\text{red}},\,\Box |\,,
\label{eq:redstar}
\end{equation}
maximises the entropy while keeping the average energy in $\mathcal{H}_{\text{red}}$ fixed. Assuming $|\mathcal{H}_{\text{red}}|=|\mathcal{H}_{\text{blue}}|$ for simplicity, there is an equivalent expression for $\rho_{\text{blue},\Box}$ with the same average energy in $\mathcal{H}_{\text{blue}}$. Hence, the ensemble \eqref{eq:redstar} is designed to reproduce the coarse-grained properties satisfied by a localised gas of gravitons, as soon as either $\mathcal{H}_{\text{red}}$ or $\mathcal{H}_{\text{blue}}$ are integrated out. In particular, its von Neumann entanglement entropy equals the ensemble thermodynamic entropy.

\paragraph{Comments on the bulk interpretation.}  The $\mathcal{O}(N)$ scaling of the maximal energy carried by these excitations \eqref{eq:pure} prevents them from having a classical supergravity description. But the connection between the fermion phase space and the bulk LLM plane, suggests to interpret this construction in terms of quantum mechanics over the classical geometry dual to the reference state $|\Box\rangle$. 

\begin{figure}
	\centering
	\begin{tikzpicture}
	\draw[fill=black]  (0,0) circle [radius=2];
	\draw[fill=white]  (0,0) circle [radius=1.6];
	\draw[fill=black]  (0,0) circle [radius=1.2];
	\draw[fill=white]  (0,0) circle [radius=0.9];
	\draw[fill=black]  (0,0) circle [radius=0.6];
	\end{tikzpicture}
	\caption{Single particle phase space density corresponding to the reference state $|\Box \rangle$ in figure \ref{fig:entsuper-0}.}
	\label{fig:refstate}
\end{figure}
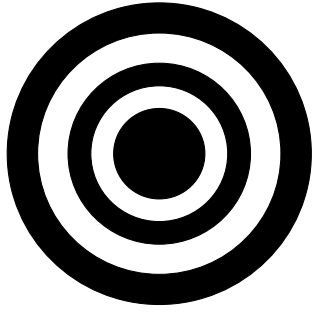

The maximal entropy ensemble \eqref{eq:redstar} provides a one parameter reduced density matrix labelled by the energy. When \eqref{eq:pure} carries no excitations, the state equals the reference state $|\Box\rangle$. By assumption, this has a gravity dual whose one particle phase space density is illustrated in figure \ref{fig:refstate}. Using the holographic dictionary discussed in section \ref{sec:review}, the relation between the YT data in figure \ref{fig:entsuper-0} and the radia describing this phase space is given by
\begin{equation}
\begin{aligned}
  \frac{r_1^2}{R_{\text{AdS}}^4} = \frac{N_1}{N}\,, \quad & \quad \frac{r_2^2}{R_{\text{AdS}}^4} = \frac{M_1-M_2+N_1}{N}\,, \\
  \frac{r_3^2}{R_{\text{AdS}}^4} = \frac{M_1-M_2+ N_1 + L_2}{N}\,, \quad & \quad \frac{r_4^2}{R_{\text{AdS}}^4} = \frac{M_1+N_1+L_2}{N}\,, \\
  \frac{r_5^2}{R_{\text{AdS}}^4} &= 1 + \frac{M_1}{N}\,,
\end{aligned}
\end{equation}
where $N_1$ is the number of fermions remaining in the Fermi sea and $r_5>r_4> \dots >r_1$.

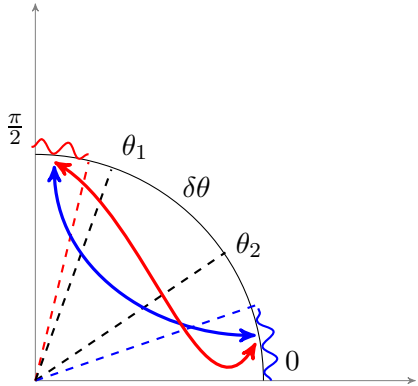
\begin{figure}
	\centering
	\begin{tikzpicture}[xscale=10, yscale=10]
	\draw[<->, help lines] (0.5,1.5) -- (0.5,1)  -- (1,1);
	\draw (0.8,1) to [out=90 ,in=0] (0.5,1.3);
	\node [above right] at (0.68,1.23) {$\delta\theta$};
	\draw[blue, thick, snake it] (0.81,1) arc (0:20:0.28);
	\draw[dashed, thick] (0.5,1) to (0.75,1.17);
	\draw[blue, dashed, thick] (0.5,1) to (0.79,1.1);
	\node [above right] at (0.75,1.15) {$\theta_2$};
	\draw[red, thick, snake it] (0.57,1.3) arc (75:90:0.29);
	\draw[red, dashed, thick] (0.5,1) to (0.57,1.29);
	\draw[dashed, thick] (0.5,1) to (0.6,1.28);
	\node[above right] at (0.6,1.28) {$\theta_1$};
	\node[above left] at (0.5,1.3) {$\frac{\pi}{2}$};
	\node[above right] at (0.815,1) {$0$};
	\draw[<->, blue, very thick] (0.79,1.06) to [out=180, in=270] (0.525,1.285);
	\draw[<->, red, very thick] (0.79,1.05) to [out=240, in=330] (0.525,1.29);
	\end{tikzpicture}
	\caption{Bulk effective representation of connectivity in terms of the $\theta$ angle measuring the 3-cycles in the 5-sphere.}
	\label{fig:bulk}
\end{figure}

Since the energy of the fermions in the Fermi sea describes the size of 3-cycles in the interval $[\theta_1,\pi/2]$, whereas the energy of the most excited fermions in the outer annulus in figure \ref{fig:refstate} corresponds to the size of 3-cycles in the interval $[0,\theta_2]$,
\begin{equation}
\cos\theta_1 = \sqrt{\frac{N_1}{N}}\,, \quad \cos\theta_2 = \sqrt{\frac{N_1 + L_2}{N}}\,, \quad \delta\theta = \theta_1-\theta_2
\end{equation}
there exists some region in the 5-sphere of size $\delta\theta$, describing the intermediate scales either in the YT \ref{fig:entsuper-0} or in the phase space density \ref{fig:refstate} which is intrinsic to the reference state $|\Box\rangle$. This corresponds to figure \ref{fig:bulk} {\it without} any of the blue-red entangled excitations.

As the energy of the ensemble \eqref{eq:redstar} increases, maximally correlated blue and red excitations (the coloured rectangles in figure \ref{fig:entgraviton}) will increase the entanglement between the subsystems $\mathcal{H}_{\text{red}}$ and $\mathcal{H}_{\text{blue}}$.
By design, these correlations are between degrees of freedom that are approximately localised in two distinct regions of the 5-sphere, as illustrated in figure \ref{fig:bulk} by the coloured wavy excitations\footnote{Figure \ref{fig:bulk} is not scaled properly. All the effects being described are suppressed in the classical limit, but they were magnified in figure \ref{fig:bulk} to reinforce the potential finite N interpretation of the construction.}. Hence, by construction, the state \eqref{eq:pure} is an EPR-like state with some approximate bulk localisation properties, resembling a pair of entangled gases of gravitons, which are geometrically separated along the 5-sphere by, at least, $L\,\delta \theta$, due to the existence of a classical scale defining the reference state $|\Box\rangle$, as illustrated in figures \ref{fig:entgraviton} and \ref{fig:bulk}.

\paragraph{Comments on connectivity and teleportation.} Contrary to our two boundary discussion, setting the excitations in \eqref{eq:pure} to vanish, does not cancel the correlations in the reference state $|\Box\rangle$. Indeed, the analogue of the 2-sided correlator \eqref{eq:2side-corr} is, modulo conservation law constraints, non-vanishing 
\begin{equation}
\langle\Box |\mathcal{O}_{\text{red}}\otimes \mathcal{O}_{\text{blue}} | \Box \rangle_c \neq 0 
\end{equation}
due to the existing correlations (and entanglement) in the reference state. This is relevant because one expects the entanglement spectrum in this state to be responsible for the connectivity of the LLM plane itself, as argued and supported by explicit calculations in \cite{Lin:2017dnz}\footnote{See \cite{Gentle:2015jma,Kim:2016dzw} for studies of Ryu-Takayangi surfaces in the M-theory versions of LLM and \cite{Balasubramanian:2017hgy} for the existence of shadows in LLM.}.  According to the holography dictionary reviewed in section \ref{sec:review}, this corresponds to connectivity in the $\theta$ direction of the transverse 5-sphere located at the origin of AdS$_5$. 

Clearly, if one turns on a perturbation in $\mathcal{H}_{\text{red}}$ at finite (small) $g_{\text{YM}}$, it can reach and affect observables in $\mathcal{H}_{\text{blue}}$, due to the non-trivial dynamics in $N=4$ SYM. Using the bulk geometry dual to the reference state, the time it takes this perturbation to reach its receiver must satisfy the bound
\begin{equation}
\delta t \geq R_{\text{AdS}}\,\delta \theta\,,
\label{eq:cbound}
\end{equation}
where $\delta\theta$ stands for the angle separation due to the quantum energy gap provided by the reference state $|\Box\rangle$, as illustrated in figure \ref{fig:bulk}. 

Maximally correlated states as \eqref{eq:pure} have new sources of correlation and one can ask whether these extra correlations can modify the bulk connectivity properties, following the EPR=ER \cite{Maldacena:2013xja} and traversable wormhole \cite{Gao:2016bin} ideas.

These extra correlations provide a new resource capable, among others, of teleportation \cite{teleportation}. The holographic bulk description of the reference state provides a classical communication channel between local observers in $\mathcal{H}_{\text{red}}$ and $\mathcal{H}_{\text{blue}}$ located in different regions of the transverse 5-sphere at the origin of AdS. Hence, teleportation can be achieved by any local interaction between them. 

Consider the task of sending a message from $\mathcal{H}_{\text{blue}}$ to $\mathcal{H}_{\text{red}}$, as probed by \eqref{eq:trav-effect}. The perturbation can reach its destination using the correlations responsible for the existence of a gravity dual in the region of the 5-sphere of size $\delta\theta$ in figure \eqref{fig:bulk}. Hence, it will again preserve the causality bound \eqref{eq:cbound}. 

The work in \cite{Gao:2016bin} suggests a second possibility : to turn on a non-local interaction between $\mathcal{H}_{\text{blue}}$ and $\mathcal{H}_{\text{red}}$
\begin{equation}
  V = \lambda\,\mathcal{O}_{\text{red}}\otimes\mathcal{O}_{\text{blue}}
\label{eq:nonlocal}
\end{equation}
 If the latter violates the ANEC, it can open the quantum bridge allowing traversability between $\mathcal{H}_{\text{blue}}$ and $\mathcal{H}_{\text{red}}$, without crossing the bulk region described by $\delta\theta$. 

As discussed in section \ref{sec:2trav}, for this mechanism to exist one needs to explore the dynamics of the full N=4 SYM due to the lack of dynamics within the half-BPS sector. It is important to stress the notion of non-locality in \eqref{eq:nonlocal} is based on the effective factorisation \eqref{eq:factor}. Hence, it refers to non-locality in R-charge, or energy, due to the BPS condition. Equivalently, it refers to non-locality in the radial LLM droplet direction, i.e. in the bulk transverse 5-sphere, due to the holographic dictionary in this sector of the theory.

\subsection{Free fermion QM perspective}
\label{r-charge}

Once the system of N free fermions in a 1d harmonic potential is derived from the original N=4 SYM degrees of freedom \cite{Corley:2001zk,Berenstein:2004kk}, the 1d trapping potential introduces a manifest locality allowing to decompose the Hilbert space into $\mathcal{H}_A$ and its complementary $\mathcal{H}_{A^c}$, where $A$ stands for some interval in the real line where the 1d potential acts. Due to the interpretation of the LLM plane as the phase space of a single fermion, such locality is manifest in the bulk, at least in its deep interior.

This notion of "real space" entanglement in R-charge, meaning the Fourier transformed of the R-charge momentum entanglement, is the one used in condensed matter physics when computing the same quantity for free cold atoms trapped in a 1d harmonic potential. This is an exciting area of research in the field of optically trapped ultra-cold atom gases given the experimental access claimed both in equilibrium \cite{nature-ee} and out-of-equilibrium \cite{science-ee}. See \cite{ultracold} for a review.

\paragraph{Condensed matter results.} There are several methods in the condensed matter literature to compute this entanglement. These are reviewed in appendix \ref{cond-mat}, where the notation below is more thoroughly defined. This body of work shows
that for Slater determinants, the Renyi entropies of the reduced density matrix $\rho_A$ equal 
\begin{equation}
S_q(A) = \sum_{i=1}^N e_q(a_i) \quad \text{with} \quad e_q(x) \equiv \frac{1}{1-q}\log[x^q + (1-x)^q]\,,
\end{equation}
where $a_i$ stands for the i-th eigenvalue of the overlap $A_{mn}$ or correlation $\mathcal{C}_A(x,y)$ matrices
\begin{equation}
\begin{aligned}
A_{nm} &= \int_A dz\, \phi_n^\star(z)\,\phi_m(z) \quad \quad n,m = 1,\dots ,N \\
\mathcal{C}_A(x,y) &= I_A(x)\langle c^\dagger(x)\,c(y)\rangle I_A(y)\,,
\end{aligned}
\end{equation}
which share the same spectrum \cite{Calabrese-3,Calabrese-4}. 

When the pure state is the Fermi sea of N fermions, it is further known (see \cite{Calabrese-2} for example) that the Renyi entropies in the large N limit are dominated by the variance in the number of fermions in the region $A$ according to
\begin{equation}\label{Nrenyi}
\frac{S_q(A)}{V_A^{(2)}} = \frac{(1+q^{-1}) \pi^2}{6} + \dots \quad \text{with} \quad V_A^{(2)} \equiv \left\langle N_A^2 \right\rangle - \left\langle N_A \right\rangle^2\,.
\end{equation}

Either using the connection to random matrix theory \cite{2014PhRvL.112y4101M} or some effective 2d CFT approach involving a massless Dirac fermion propagatin in a non-trivial background determined by the Fermi momentum \cite{Calabrese-EFT}, the entanglement entropy $q=1$ in the Fermi sea has been computed for different regions $A$ 
\begin{equation}
\begin{aligned}
S_q(x) & = \frac{q+1}{12q}\log[2N(1-x^2/R_{\text{AdS}}^4)^{3/2}]\,, \quad \text{when} \quad A=[-\infty,x] \\
S_1(x_1,\,x_2) & = S_1(x_1) + S_1(x_2) + \frac{1}{6}\log\left|\frac{a-b + ic}{a+b+ic}\right|^2\,, \quad \text{when} \quad A=[x_1,\,x_2]
\end{aligned}
\label{calabrese-results}
\end{equation}
where $a=\sqrt{1-x_1^2/R_{\text{AdS}}^4}$, $b=\sqrt{1-x_2^2/R_{\text{AdS}}^4}$ and $c=(x_1-x_2)/R_{\text{AdS}}^2$. 

At finite $N$, the entanglement entropy is finite. This was interpreted in \cite{Hartnoll:2015fca,Hartnoll:2016mdv} as an indication of the fine grained structure of spacetime in the context of 1+1 dimensional string theory.

The extension of these calculations to two non-overlapping intervals requires a 4-pt function of twist operators in the upper half plane in the effective 2d CFT introduced in appendix \ref{cond-mat}. In the particular case of two non-overlapping and semi-infinite intervals, i.e. $A = (-\infty,\,x_1]$ and $B=[x_2,\,\infty)$, one can use the 2-pt function for twist operators in \cite{Casini-huerta}. From the latter, one can recover a positive mutual information
\begin{equation}
I(A;B) = -\frac{1}{6}\log\frac{\left(\sqrt{1-x_1^2/R_{\text{AdS}}^4} - \sqrt{1-x_2^2/R_{\text{AdS}}^4}\right)^2 + (x_1-x_2)^2/R_{\text{AdS}}^4}{\left(\sqrt{1-x_1^2/R_{\text{AdS}}^4} + \sqrt{1-x_2^2/R_{\text{AdS}}^4}\right)^2 + (x_1-x_2)^2/R_{\text{AdS}}^4} > 0\,.
\end{equation}

\paragraph{Holographic comments.} The Ryu-Takayanagi formula \cite{Ryu:2006bv} for the
holographic description of entanglement entropy in real space has provided important clues about the nature and the emergence of spacetime. It is natural to wonder whether entanglement entropy in the R-charge Hilbert space provides a similar understanding for the compact transverse space. The work in \cite{Berenstein:2016pcx,Lin:2017dnz} supports this expectation. 

The quantum mechanical fermion picture and its semiclassical phase space realisation in the LLM plane provide an alternative factorisation of the Hilbert space to study entanglement in R-charge "real space", the conjugate to R-charge momentum. The condensed matter results teach us two points. First, the existence of an effective 2d CFT with central charge $c_{\text{eff}}=1$ confirms any dual gravity description must involve a highly curved spacetime. Second, the amount of entanglement entropy in some region A equals the variance in the number of particles in that region. This number involves the charge \eqref{LLM-charge}. For example, for a symmetric interval $x\in [-\ell,\,\ell]$ in the Fermi sea, the number of particles is computed by integrating the phase space density over a surface $\Sigma$ anchored in the region of phase space allowed by $A$, that is, for all momenta allowed in the droplet region defined by the vaccuum 
\begin{equation}
x\in [-\ell,\,\ell] \quad \Rightarrow \quad p \in [-\sqrt{R_{\text{AdS}}^4- x^2},\,\sqrt{R_{\text{AdS}}^4- x^2}]\,.
\end{equation}
To compute the variance in this quantity in the bulk, one would still need to use the quantisation results in \cite{Maoz:2005nk}.

\section{Conclusions and discussion}
\label{sec:discussion}

In this work, the original idea relating correlations and entanglement with the connectivity of spacetime \cite{VanRaamsdonk:2010pw} has been revised in the half-BPS sector of $N=4$ SYM. Supersymmetry reduces certain aspects of the quantum dynamics to a system of N free fermions in a 1d harmonic oscillator. The semiclassical limit of the phase space density of a single fermion controls the boundary conditions uniquely determining the solution to the classical type IIB supergravity equations of motion describing the gravity dual \cite{Lin:2004nb}. These known facts correlate the localisation properties of the quantum mechanical wave functions describing quantum states with some geometrical features in the gravity dual. 

In particular, the existence of uncertainty in the quantum state of any fermion gives rise to a bulk singularity, as discussed at the end of section \ref{sec:dictionary} and more extensively in section \ref{sec:supers}. This observation has implications when discussing two entangled, but not interacting, half-BPS sectors in two $N=4$ SYM theories. Given our control on the quantum dynamics, one can design any correlation between fermions localised at the origin of two, otherwise disconnected, AdS gravity duals. The localisation of the wave functions establishes that any classical connectivity between the two geometries can only occur through the regions in phase space where correlations exist. These are precisely the regions in the LLM plane where a bulk singularity develops (as discussed in section \ref{sec:supers}). 

The microscopic interpretation of these singularities as the backreaction of distributions of giant gravitons and dual giants allows to interpret this connectivity in terms of connectivity of the different sets of 3-spheres pinching at the apices of the two cones describing the singularity on each side of the gravity dual.  Depending on the R-charge carried by the correlated giant and dual gravitons, bulk singularities and connectivity occur in different regions of the 5-sphere. This allowed us to introduce the notion of localised superstar in section \ref{sec:localised}, whose non-extremal gravity duals are believed to exist, but are not known in the literature.

Because of the amount of supersymmetry, the entropy of microstates carrying conformal dimension $\Delta \sim N^2$ scales linearly with $N$. This means these BPS configurations have no macroscopic horizons and give rise to naked singularities sourced by the distribution of giant gravitons. This poses natural questions regarding the possible EPR=ER interpretation \cite{Maldacena:2013xja} of these results. To clarify this, the geometry of the near-extremal limit of 5d R-charged black holes was performed in section \ref{sec:5d}, following its maximal Kruskal extension detailed in appendix \ref{kruskal}. The length of the bridge between two very separated 5d R-charged black holes grows indefinitely in the near-extremal limit, while its cross-section tends to zero. Equivalently, the bulk description of the entangled system gets pinched in an explicit realisation of van Raamsdonk ideas \cite{VanRaamsdonk:2010pw}. 

Furthermore, our microscopic interpretation relates the infinite length of the bridge with the existence of quantum correlations which are not strong enough to give rise to a classical bridge. Indeed, the entropy and mutual information scale linearly with $N$. These bound the amount of correlation through \eqref{eq:mutualbound}. Hence, these are quantum bridges. Notice that despite the absence of an infinite throat, as it occurs in the smooth extremal limit of black holes, the conclusion regarding the infinite bulk separation remains. As a further check on the physics of this near extremal limit, the shock wave analysis describing the backreaction of some perturbation sent from one of the boundaries when reaching the horizon was performed in appendix \ref{shock} and the same features as in \cite{Leichenauer:2014nxa} were found.

The EPR=ER ideas should also hold for a single asymptotic boundary, i.e. for a single half-BPS sector in the context of this manuscript. Generalisations of the ensembles described in section \ref{sec:localised} would describe quantum states giving rise to bulk singularities localised in different regions of the transverse 5-sphere at the origin of AdS. Supersymmetry guarantees the absence of force between these singularities. Hence, if one interprets these as the BPS limit of their near extremal versions, they provide a good laboratory to study entangled black holes. The conceptual difficulty is to properly define the notion of entanglement which requires a proper factorisation of the Hilbert space. To avoid this, the effective field theory approach developed in \cite{Berenstein:2017rrx} was followed in section \ref{sec:egrav}, giving rise to entangled gases of pointlike gravitons in different regions of the 5-sphere. Since the construction depends on some reference state having a gravity dual, this was interpreted in the bulk as quantum mechanics on the 5-sphere. This provides an explicit realisation for a quantum teleportation protocol in which the dual geometry of the reference state provides a classical channel between the two subsystems located at different regions of the 5-sphere. Null geodesic propagation on the 5-sphere gives rise to causality bounds guaranteeing the traversability of the quantum bridge between both subsystems is consistent with causality. 

In section \ref{r-charge}, it was finally stressed that the free fermion picture allows to compute the notion of entanglement in real space, as it is done in the condensed matter literature, where real space stands here for the effective one dimension in which the fermions are subject to the harmonic potential. This is a highly non-local notion from the original $N=4$ SYM gauge theory perspective, where the natural notion of entanglement is in R-charge momentum space. Interestingly, the bulk description provided by the LLM plane treats both the R-charge momentum and its conjugate real dimension on an equal footing, since it is the phase space of a single fermion that emerges. Computing this entanglement in global AdS confirms the quantum nature of the bridges discussed in earlier sections.

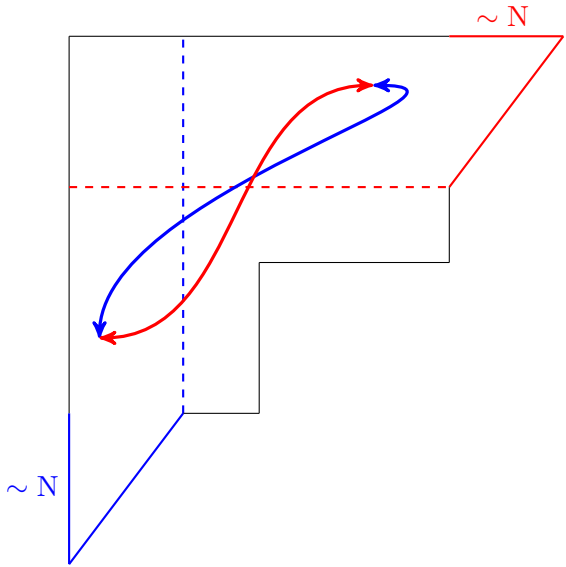
\begin{figure}
	\centering
	\begin{tikzpicture}
	\draw (0,0) to (5,0);
	\draw[red, thick] (5,0) -- (6.5,0);
	\draw (0,0) to (0,-5);
	\draw[blue, thick] (0,-5) to (0,-7);
	\draw (2.5,-3) to (2.5,-5);
	\draw[blue, dashed, thick] (1.5,-5) to (1.5,0);
	\draw (1.5,-5) -- (2.5,-5);
	\draw (2.5,-3) to (5,-3);
	\draw[red, dashed, thick] (0,-2) to (5,-2); 
	\draw (5,-2) -- (5,-3);
	\draw[red, thick] (5,-2) -- (6.5,0);
	\draw[blue, thick] (0,-7) -- (1.5,-5);
	\draw[<->, blue, very thick] (0.4,-4) to [out=90, in=0] (4,-0.65);
	\draw[<->, red, very thick] (0.4,-4) to [out=0, in=180] (4,-0.65);
	\node[blue, below left] at (0,-5.7) {$\sim \text{N}$};
	\node[red, above] at (5.7,0) {$\sim \text{N}$};
	\end{tikzpicture}
	\caption{Schematic representation of a quantum state describing a pair of entangled superstars (triangular excitations at the edges) separated by a large energy scale.}
	\label{fig:entsuper}
\end{figure}

\paragraph{Comments on entangled localised superstars.} Even though the EFT description in subsection \ref{sec:egrav} breaks down before reaching the $\mathcal{O}(N^2)$ energy required to describe a superstar ensemble, some of the features discussed for the entangled gas of pointlike gravitons are expected to hold for a pair of localised entangled superstars, a situation schematically illustrated in figure \ref{fig:entsuper}.

\begin{figure}
	\centering
	\begin{tikzpicture}[xscale=10, yscale=10]
	\draw[<->, help lines] (0.5,1.5) -- (0.5,1)  -- (1,1);
	\draw (0.8,1) to [out=90 ,in=0] (0.5,1.3);
	\node [above right] at (0.68,1.23) {$\delta\theta$};
	\draw[blue, ultra thick] (0.8,1) arc (0:29:0.33);
	\draw[blue, dashed, thick] (0.5,1) to (0.755,1.15);
	\node [blue, above right] at (0.80,1.08) {$\delta\theta_{\text{b}}$};
	\draw[red, ultra thick] (0.6,1.28) arc (70:90:0.30);
	\draw[red, dashed, thick] (0.5,1) to (0.6,1.28);
	\node [red, above right] at (0.53,1.31) {$\delta\theta_{\text{r}}$};
	\draw[<->, blue, very thick] (0.785,1.08) to [out=180, in=270] (0.53,1.29);
	\draw[<->, red, very thick] (0.785,1.07) to [out=240, in=330] (0.535,1.29);
	\end{tikzpicture}
	\caption{Bulk effective representation of connectivity in terms of the $\theta$ angle measuring the 3-cycles in the 5-sphere.}
	\label{fig:bulk-1}
\end{figure}
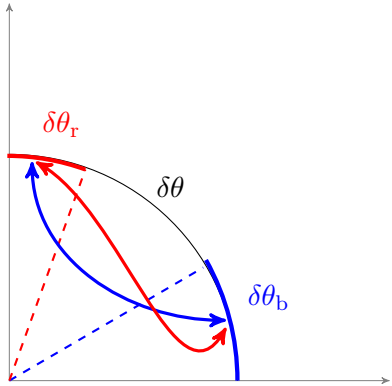

One expects entangled superstars to allow for some effective description in terms of approximately localised superstars in different regions of the 5-sphere, as illustrated in figure \ref{fig:bulk-1}, with extra quantum correlations between them, in analogy with \eqref{eq:pure}. This is what the EFT used in section \ref{sec:egrav} achieved for pointlike gravitons. What is missing is a factorisation of the Hilbert space in R-charge momentum space allowing us to distinguish the degrees of freedom of both superstars, i.e. the red and blue quanta of the schematic YT in figure \ref{fig:entsuper}, together with those responsible for the connectivity of space between the localised superstars, i.e. the black quanta in figure \ref{fig:entsuper}. Remember that even though the energies of the entangled subsystems would be of order $\mathcal{O}(N^2)$, the size of the relevant Hilbert spaces is not large enough to allow a classical geometric description. Hence, the connectivity induced by the extra quantum correlations would still give rise to quantum bridges, as in the two boundary discussion in section \ref{sec:2-boundary}.

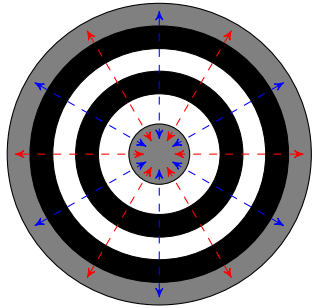
\begin{figure}
	\centering
	\begin{tikzpicture}
	\draw[fill=gray]  (0,0) circle [radius=2];
	\draw[fill=black] (0,0) circle [radius=1.7]; 
	\draw[fill=white]  (0,0) circle [radius=1.4];
	\draw[fill=black]  (0,0) circle [radius=1.1];
	\draw[fill=white]  (0,0) circle [radius=0.8];
	\draw[fill=gray]  (0,0) circle [radius=0.4];
	\draw[<->, blue, dashed] (0,0.2) to (0,1.9);
	\draw[<->, blue, dashed] (0,-0.2) to (0,-1.9); 
	\draw[<->, red, dashed] (0.1,0.17) to (0.95,1.64); 
	\draw[<->, blue, dashed] (0.17,0.1) to (1.64,0.95);
	\draw[<->, red, dashed] (-0.1,0.17) to (-0.95,1.64); 
	\draw[<->, blue, dashed] (-0.17,0.1) to (-1.64,0.95); 
	\draw[<->, blue, dashed] (-0.17,-0.1) to (-1.64,-0.95);
	\draw[<->, red, dashed] (-0.1,-0.17) to (-0.95,-1.64);  
	\draw[<->, red, dashed] (0.1,-0.17) to (0.95,-1.64); 
	\draw[<->, blue, dashed] (0.17,-0.1) to (1.64,-0.95); 
	\draw[<->, red, dashed] (0.2,0) to (1.9,0); 
	\draw[<->, red, dashed] (-0.2,0) to (-1.9,0); 
	\end{tikzpicture}
	\caption{Expected phase space density describing a pair of entangled superstars, with dashed arrows stressing the existence of additional correlations between regions of phase space that are properly encoded in the N-particle phase space.}
	\label{fig:disk-2}
\end{figure}

A bulk observer having access to the full Hilbert space should describe the 5-sphere at the origin of AdS$_5$ as illustrated in figure \ref{fig:bulk-1}. The expected phase space density should look like \ref{fig:disk-2}, where the blue and red dashed lines are schematically representing the existence of extra correlations between the grayscale distributions supported where both superstars are located. To reproduce such correlations would require the use of the full quantum mechanics.

If there exists some quantum mechanical decomposition of the Hilbert space allowing to separate each superstar as a proper quantum subsystem, it would then be natural to trace over the degrees of freedom non-accessible to an observer tied to such subsystem and ask what spacetime geometry would holographically reproduce her local physics. Given the lack of dynamics in the half-BPS sector of N=4 SYM, it is natural to expect such observer would describe a single localised superstar in the presence of some additional horizon/singularity preventing her to unambiguously infer the existence of a second superstar. This expectation is reminiscent of Rindler physics \cite{Bisognano:1975ih,Unruh:1976db}, where an observer with constant proper acceleration has no access to the full Minkowski spacetime. To account for this fact, her description involves a bulk horizon. Whether this factorisation exists and if so, whether it involves non-trivial UV-IR mixing is an important question.

The interpretation of entangled superstars in terms of connectivity is analogous to the one advocated for the entanglement of gases of pointlike gravitons in section \ref{sec:egrav}. The existent entanglement among the black quanta (the intermediate energy modes) in figure \ref{fig:entsuper} should be responsible for the connectivity between both localised superstars \cite{Lin:2017dnz} using the N=4 SYM hamiltonian and it should provide with some causality bound analogous to the one in \eqref{eq:cbound} controlling both the reception of messages and the teleportation of quantum states between observers operating in their respective localised superstars.

If there exist non-local interactions between the degrees of freedom describing both localised superstars allowing to violate the ANEC, this could open the quantum bridge making it traversable, following the ideas in \cite{Gao:2016bin,Maldacena:2017axo}. Such traversable bridge would provide an alternative bulk connection between the localised superstars. Even though one may expect such connectivity not to violate the causality bound \eqref{eq:cbound}, this may depend on the type of interaction turned on. Furthermore, it is known that wormholes that are permanently opened contain closed timelike curves under some conditions \cite{Frolov:1990si}\footnote{The author would like to thank Roberto Emparan for stressing this point.}. It is important to generically clarify this point, independently of the set-up described in this work.

It would be important to achieve any progress in the description of entangled macroscopic black holes in single boundary holographic set-ups where our understanding on the microscopic degrees of freedom is comparable to the one in this corner of N=4 SYM.

\subsection*{Acknowledgements}
The author would like to thank Sanjaye Ramgoolam, Robert de Mello Koch and especially, Vijay Balasubramanian and Vishnu Jejjala  for encouragement to write these vague thoughts. The author would like to thank the many audiences and institutions where this material was informally presented during the last two years. The author would like to thank all the support received from the Perimeter Institute during his sabbatical (January-June 2017) and the atmosphere at AIMS South Africa during the last stages of this investigation. This work is supported by the Science and Technology Facilities Council (STFC) [grant number ST/L000458/1]. This research was supported in part by Perimeter Institute for Theoretical Physics. Research at Perimeter Institute is supported by the Government of Canada through the Department of Innovation, Science and Economic Development and by the Province of Ontario through the Ministry of Research, Innovation and Science.


\newpage

\appendix

 \section{5d single R-charged AdS black hole : Kruskal extension}
\label{kruskal}

To construct the Kruskal extension of the 5d single R-charged AdS black hole
\begin{equation}
	\begin{aligned}
		ds^2 &= H^{-2/3}\,f \left(-dt^2 + \frac{H}{f^2}dr^2\right) + H^{1/3}\,r^2\,d\Omega_3^2\,, \\
		H&=1+\frac{q}{r^2}\,, \quad f=1-\frac{\mu}{r^2} + \frac{r^2}{R_{\text{AdS}}^2}\,H\,,
	\end{aligned}
\end{equation}
we follow the procedure outlined in \cite{Graves:1960zz}. First, the tortoise coordinate is introduced 
\begin{equation}
dr_\star \equiv \frac{\sqrt{H}}{f}\,dr = \frac{\sqrt{r^2+q}\,rR_{\text{AdS}}^2}{(r^2-r_+^2)(r^2-r_-^2)}\, dr\,,
\end{equation}
with $2r^2_\pm = -(q+R_{\text{AdS}}^2) \pm \left[(q+R_{\text{AdS}}^2)^2+4\mu\,R_{\text{AdS}}^2\right]^{1/2}$. Decomposing into simple fractions and changing the integration variable $r^2=x$, the defining integral
\begin{equation}
  r_\star =\frac{R_{\text{AdS}}^2}{2} \frac{1}{r_+^2-r_-^2} \int \sqrt{x+q}\,\left(\frac{1}{x-r_+^2} - \frac{1}{x-r_-^2}\right)\,dx \equiv I_1 + I_2
\label{eq:tortoise}
\end{equation}
can be performed using the table integral
\begin{equation}
\int \frac{\sqrt{ax+b}}{x}\,dx = \left\{
\begin{array}{ll}
2\left(\sqrt{ax+b} - \sqrt{b}\coth^{-1} \frac{\sqrt{ax+b}}{\sqrt{b}}\right) & b>0\,,\,\, ax > 0 \\
2\left(\sqrt{ax+b} - \sqrt{b}\tanh^{-1} \frac{\sqrt{ax+b}}{\sqrt{b}}\right) & b>0\,,\,\, ax < 0 \\
2\left(\sqrt{ax+b} - \sqrt{-b}\arctan^{-1} \frac{\sqrt{ax+b}}{\sqrt{-b}}\right) & b<0\,, \\
\end{array}\right.
\end{equation}
where an arbitrary constant was not included. Since $r_-^2 < 0$, the piece $I_2$ in \eqref{eq:tortoise} always corresponds to the third branch with $b=q-|r_-|^2<0$. The piece $I_1$ corresponds to the first branch for $r^2>r_+^2$ and to the second branch for $r^2< r_+^2$, with $b=q+r_+^2$. All cases involve $a=1$. 

Second, we look for a coordinate transformation $u_s=u_s(r_\star,t)$ and $v_s=v_s(r_\star,t)$ satisfying
\begin{equation}
\phi(r_\star)(-dt^2 + dr_\star^2) = w^2(u_s,v_s)(du_s^2-dv_s^2)\,, \quad \text{with} \quad \phi (r_\star) = f(r)\,H^{-2/3}(r)
\end{equation}
and absorbing the zero at $f(r_+)$ for any instant of time. The existence of the map requires 
\begin{equation}
\begin{aligned}
w^2\left((\partial_{r_\star} u_s)^2 -(\partial_{r_\star} v_s)^2\right) &= \phi\,, \\
w^2\left((\partial_t u_s)^2 - (\partial_t v_s)^2\right) &= -\phi\,, \\
\partial_{r_\star}u_s\,\partial_t u_s &= \partial_{r_\star} v_s \partial_t v_s\,.
\end{aligned}
\end{equation}
The general solution involves 
\begin{equation}
  u_s(r_\star,t) = h(v) + g(u)\,, \quad \quad v_s(r_\star,t) = h(v) - g(u)\,,
\end{equation}
with $v=t+r_\star$ and $u=t-r_\star$, while the conformal factor $w(u_s,v_s)$ satisfies
\begin{equation}
w^2 = -\frac{\phi(r_\star)}{4h^\prime(v)\,g^\prime(u)}\,,
\end{equation}
where primes stand for derivatives with respect to the relevant lightlike coordinates $u$ or $v$.
Choosing the waves $h(v)$ and $g(u)$ as 
\begin{equation}
h(v) = \frac{1}{2}\,e^{\gamma\,v}\,, \quad g(u) = \frac{1}{2}\,e^{-\gamma\,u}
\end{equation}
the conformal factor becomes time independent 
\begin{equation}
w^2(u_s,v_s) = \phi(r_\star)\,\gamma^{-2}\,e^{-2\gamma r_\star}\,.
\end{equation}
Removing the zero of $\phi(r)$ at $r=r_+$, fixes $\gamma$ to equal the surface gravity $\kappa=\frac{2\pi}{\beta}=2\pi\,T$
\begin{equation}
  \gamma =\kappa = \frac{r_+^2 + |r_-|^2}{R_{\text{AdS}}^2\,\sqrt{q+r_+^2}} = \frac{q+R_{\text{AdS}}^2+2r_+^2}{R_{\text{AdS}}^2\,\sqrt{q+r_+^2}}\,.
\label{eq:zerocancel}
\end{equation}
To derive this result, one must analyse the behaviour of the tortoise coordinate \eqref{eq:tortoise} close to the horizon. Consider $r^2=r_+^2+y$ with $0< y\ll r_+^2$ (a similar analysis for $y<0$ gives rise to the same conclusion). The dominant expansion
\begin{equation}
\begin{aligned}
r_\star(y) &\approx \frac{R_{\text{AdS}}^2}{r_+^2 + |r_-|^2}\sqrt{|r_-|^2-q}\arctan \sqrt{\frac{r_+^2+q}{|r_-|^2-q}} - \frac{R_{\text{AdS}}^2\sqrt{q+r_+^2}}{2(r_+^2+|r_-|^2)}\log \frac{4(q+r_+^2)}{y} \\
& \equiv D - B \log \frac{4(q+r_+^2)}{y}\,,
\end{aligned}
\label{eq:tortoise-exp}
\end{equation}
involves a divergent logarithm of $y$ in the limit $y\to 0$. Using the expansions
\begin{equation}
\begin{aligned}
  e^{-2\kappa r_\star(y)}&\approx e^{-2\kappa D}\,\frac{[4(q+r_+^2)]^{2B\gamma}}{y^{2B\gamma}}\,, \\
  \phi(y)&\approx \frac{r_+^2-r_-^2}{R_{\text{AdS}}^2r_+^2H(r_+^{2/3})}\,y
\end{aligned}
\label{eq:small-y}
\end{equation}
one reaches the conclusion that absence of zeroes requires $2B\gamma = 1$, from which \eqref{eq:zerocancel} follows.

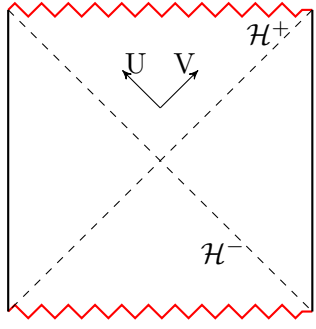
\begin{figure}
	\centering
	
	\begin{tikzpicture}
	
	\draw[thick,red,zigzag] (-\L,\L) coordinate(stl) -- (\L,\L) coordinate (str);
	\draw[thick,red,zigzag] (-\L,-\L) coordinate(stl) -- (\L,-\L) coordinate (str);
	\draw[thick,black] (-\L,\L) -- (-\L,-\L);
	\draw[thick,black] (\L,\L) -- (\L,-\L);
	\draw[dashed,black] (\L,\L) -- (-\L,-\L);
	\draw[dashed,black] (-\L,\L) -- (\L,-\L);
	
	\draw[->,black] (0,0.7) -- (0.5,1.2);
	\draw[->,black] (0,0.7) -- (-0.5,1.2);
	
	\draw[black] (0.2*\L,-0.6*\L) node[right] (scrip) {$\mathcal{H}^-$}
	(0.5*\L,0.85*\L) node[right] (scrip) {$\mathcal{H}^+$};
	\draw[black] (0.6,1.3) node[left] (scrip) {V} (-0.6,1.3) node[right] (scrip) {U};

	\end{tikzpicture}
	\caption{Schematic maximal Kruskal extension of the 5d black hole \eqref{eq:5dBH}.}
	\label{fig:penrose-0}	
\end{figure}

Having identified the smooth coordinates, one can introduce standard Kruskal coordinates $U$ and $V$ covering the maximal extension of the 5d BH \eqref{eq:5dBH}, as illustrated in the Penrose diagram \ref{fig:penrose-0}. We use the conventions where $U>0$ in the left exterior, whereas $V>0$ in the right exterior. In the latter, the Kruskal coordinates satisfy
\begin{equation}
  U = -(u_s-v_s) = -e^{-\kappa\,u}\,, \quad\quad V=u_s+v_s= e^{\kappa\,v}\,.
\end{equation}

\section{Shock-wave analysis}
\label{shock}

The backreaction of a perturbation reaching a black hole horizon, due to the blue-shift it experiences, was originally studied in \cite{Dray:1984ha}. This work was applied in a holographic context to estimate the scrambling time in a bulk calculation in \cite{Shenker:2013pqa}. In this appendix, we follow the ideas in \cite{Shenker:2013pqa} and the tools developed in \cite{Leichenauer:2014nxa}, applied to the specific black holes \eqref{eq:5dBH}\footnote{For further work on the subject involving localised shock waves, see \cite{Roberts:2014isa}. For an specific analysis on rotating and charged BTZ black holes, see \cite{Reynolds:2016pmi}.}.

\begin{figure}
	\centering
	
	\begin{tikzpicture}
	
	\draw[thick,red,zigzag] (-\L,\L) coordinate(stl) -- (1.1,\L) coordinate (str);
	\draw[thick,black] (-\L,\L) -- (-\L,-1.1);
	\draw[dashed,blue] (-\L,-1.1) -- (1.1,\L);
	\draw[black] (-\L,\L) -- (-0.45,0.45);
	
	\draw[thick,red,zigzag] (1.6,\L) coordinate(stl) -- (2.3,\L) coordinate (str);
	\draw[dashed,blue] (-1.7,-1.3) -- (1.6,\L);
	\draw[thick,black] (-1.7,-1.3) -- (-1.7,-\L);
	\draw[thick,red,zigzag] (-1.7,-\L) coordinate(stl) -- (2.3,-\L) coordinate (str);
	\draw[thick,black] (2.3,\L) -- (2.3,-\L);
	\draw[black] (-1.7,-\L) -- (2.3,\L);
	\draw[black] (-0.2,0.2)-- (2.3,-\L);
	
	\draw[blue] (0.2,-0.6*\L) node[right] (scrip) {R}
	(-1.7,0.8) node[right] (scrip) {L};
	\draw[blue] (-2.3,-1.3) node[right] (scrip) {$t_0$};
	
	\draw[magenta,->] (2.5,0) -- (3.5,0);
	
	\draw[thick,red,zigzag] (3.7,\L) coordinate(stl) -- (8,\L) coordinate (str);
	\draw[thick,red,zigzag] (4.7,-\L) coordinate(stl) -- (9,-\L) coordinate (str);
	\draw[thick,black] (3.7,\L) -- (4.7,-\L);
	\draw[thick,black] (8,\L) -- (9,-\L);
	\draw[dashed,blue] (4.7,-\L) -- (8,\L);
	\draw[black] (3.7,\L) -- (6.1,-0.31);
	\draw[black] (6.6,0.3) -- (9,-\L);
	
	\draw[blue] (6.25,-0.1) node[right] (scrip) {$\alpha$};
	
	\end{tikzpicture}
	\caption{The shock-wave geometry corresponds to the gluing of two-half spacetimes along the lightlike perturbation trajectory in the limit $t_0\to \infty$, $\delta M\to 0$ keeping $\alpha\propto e^{\kappa t_0}\delta M$ fixed, giving rise to a shift $V_L=V_R + \alpha$ in the Kruskal coordinate by $\alpha$.}
	
	\label{fig:shock-wave}	
\end{figure}
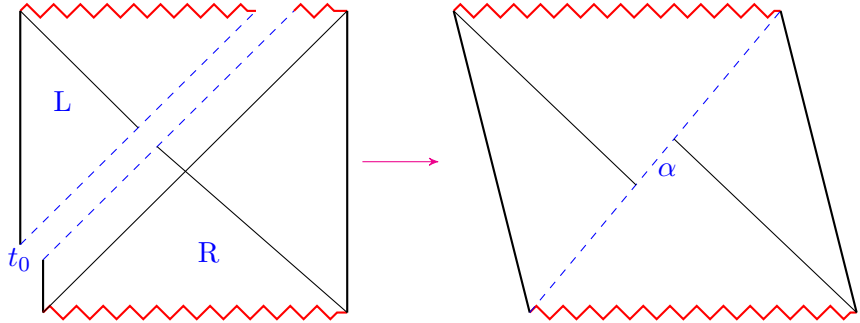

Consider turning on a perturbation of arbitrarily low energy and no charge from the left boundary, at very early times $t_0$ and travelling along a constant null $U$ trajectory, as indicated in the left diagram in figure \ref{fig:shock-wave}. As first shown in \cite{Dray:1984ha}, the non-trivial effect of this perturbation on the background in the limit $\delta M\to 0$ keeping $\delta M\,e^{\kappa t_0}$ fixed is to describe a shock-wave propagating along the horizon $\mathcal{H}^+$. This is equivalent to gluing two black holes with a shift in the $V$ coordinate, as illustrated in the second diagram in figure \ref{fig:shock-wave}. 

Label the coordinates and parameters of the two black holes by L and R. Using time translations, one can choose $t_L=t_R=t_0$. The gluing of the two BH spacetimes is achieved using the continuity in the radius of the 3-sphere and conservation of energy $(M_R = M_L + \delta M)$, and charge $(\delta Q=0)$. The last condition reduces to $(2q+\mu)\delta q=-q\delta\mu$. This fixes
\begin{equation}
  \delta M = \frac{\omega_3}{8\pi G_5}\,\delta\mu\left(\frac{3}{2}- \frac{1}{2+\frac{\mu}{q}}\right)\,,
\end{equation}
which is equivalent to a horizon size increase $\delta R$, i.e. $r_{+R}^2 = r_{+L}^2 + \delta R$,
\begin{equation}
  \delta R = \frac{\delta\mu}{2r_{+L}^2+q+R_{\text{AdS}}^2}\left(L^2 + \frac{r_+^2}{2+\frac{\mu}{q}}\right)\,.
\end{equation}

Since the perturbation follows null geodesics of constant $U$, these are described by
\begin{equation}
U_{L}= e^{\kappa_L (t_0-r_{\star L}(\infty))}\,, \quad \quad U_{R}= e^{\kappa_R (t_0-r_{\star R}(\infty))}\,,
\end{equation}
where there is no sign because the perturbation is turned on the left exterior region\footnote{The constants $r_{\star L}(\infty)$ and $r_{\star R}(\infty)$ are not physically relevant. They could have been removed by fixing the integration constants in\eqref{eq:tortoise}.}. Since the description of both geodesics in the two pieces of spacetime satisfy
\begin{equation}
U_{L}V_L = - e^{2\kappa_L r_{\star L}(r_L)}\,, \quad U_{R}V_R = - e^{2\kappa_R r_{\star R}(r_R)}\,,
\end{equation}
the quotient between these two relates $V_L$ with $V_R$. In the limit $\delta R\to 0$ keeping $\delta R\,e^{\kappa t_0}$ fixed, the trajectories are always very close to the horizon. Hence, we can use \eqref{eq:tortoise-exp} for both spacetimes. The only non-trivial surviving effect gives rise to the shift
\begin{equation}
V_L= V_R + \frac{\delta R\,e^{\kappa t_0}}{4(q+r_{+R}^2)}\,e^{2\kappa D_R}\,e^{-\kappa r_\star(\infty)} \equiv v_R + \alpha\,.
\end{equation}
This is the standard shift in the shock-wave geometry \cite{Dray:1984ha}.

The scrambling time $t_\star$ is the time scale associated with $\alpha\sim 1$ \cite{Shenker:2013pqa}
\begin{equation}
t_\star \approx \frac{\beta}{2\pi}\,\log \frac{4(q+r_{+R}^2)}{\delta R}\,,
\end{equation}
where we only kept the logarithmic terms dependent on the perturbation. For very massive black holes, $\mu\gg q\sim R_{\text{AdS}}^2$, this reduces to 
\begin{equation}
  t_\star \approx \frac{\beta}{2\pi}\log \frac{8M}{\delta M}\,,
\end{equation}
whereas for near-extremal black holes, $\mu\ll q \sim R_{\text{AdS}}^2$,
\begin{equation}
t_\star \approx \frac{\beta}{2\pi}\log \frac{4(1+w)M}{\delta M}\,,
\end{equation}
where $w=q/R_{\text{AdS}}^2$. Both are consistent with the more general result discussed in \cite{Leichenauer:2014nxa}.

\section{Thermodynamic stability of near-extremal R-charged black holes}
\label{sec:stability}

The purpose of this appendix is to analyse the thermodynamic stability of single R-charged black holes \eqref{eq:5dBH} in the near extremal regime $\mu\ll q$. There is a large literature on the thermodynamics of black holes in AdS, starting with \cite{Hawking:1982dh}, and in particular, for R-charged AdS black holes, see for example \cite{Hawking:1999dp,Batrachenko:2004fd,Son:2006em,Cvetic:1999ne,Gubser:2000ec,Gubser:2000mm,Chamblin:1999hg,Chamblin:1999tk}. The material discussed below is not new, but makes this work more self-contained.

Consider the Gibbs' potential
\begin{equation}
G = M - TS + \Phi\,Q\,.
\end{equation}
Stable thermodynamic equilibrium requires $\left.\delta G\right|_{T,\Phi}=0$ and $\left.\delta^2 G\right|_{T,\Phi}>0$. The first condition is responsible for the first law \eqref{eq:1stlaw}, i.e. $TdS = dM - \Phi dQ$. It follows
\begin{equation}
\left(\frac{\partial G}{\partial S}\right)_{T,\Phi}=  \left(\frac{\partial M}{\partial S}\right)_{T,\Phi} - T\,, \quad \left(\frac{\partial G}{\partial Q}\right)_{T,\Phi}=  \left(\frac{\partial M}{\partial Q}\right)_{T,\Phi} + \Phi\,.
\end{equation}
Hence, the second condition is equivalent to 
\begin{equation}
  \left.\delta^2 G\right|_{T,\Phi}>0 \quad \Rightarrow \quad \text{det}\left(\frac{\partial^2M}{\partial x_i\partial x_j}\right)_{T,\Phi} > 0\,.
\label{eq:stability-condition}
\end{equation}
This requires the positivity of the determinant of the Hessian of the mass $M(S,Q)$ as a function of $x_i,x_j=S,Q$.

To perform this analysis, the maps $\mu=\mu(S,J)$ and $q=q(S,J)$ are required. Since the entropy is linear in the non-extremal parameter in the regime $\frac{\mu}{q}\ll 1$ (see \eqref{eq:near-extreme}), there is no loss of generality in considering the expansions
\begin{equation}
\begin{split}
q &= q_0 + q_1\,S + q_2\,S^2 + \dots \\
\mu &= \mu_0\,S + \mu_1\,S^2 + \dots
\end{split}
\end{equation}
where all coefficients are functions of $Q$. From the mass and electric charge formulas \cite{Behrndt:1998jd,Batrachenko:2004fd} 
\begin{equation}
\begin{split}
  M &=\frac{\omega_3}{8\pi\,G_5}\left(\frac{3}{2}\mu + q\right) =\frac{\omega_3}{8\pi\,G_5}\left(q_0 + S(q_1+\frac{3}{2}\mu_0) + S^2(q_2+\frac{3}{2}\mu_1)+\dots \right)\,, \\
  Q &= \frac{\omega_3}{8\pi\,G_5}\,\sqrt{q(q+\mu)} = \frac{\omega_3}{8\pi\,G_5}\left(q_0 + S(q_1 + \frac{\mu_0}{2}) + S^2(q_2+ \frac{\mu_1}{2} - \frac{\mu_0^2}{8q_0}) + \dots \right)
\end{split}
\label{eq:inv-charge}
\end{equation}
it follows
\begin{equation}
\begin{split}
\left(\frac{\partial^2 M}{\partial S^2}\right)_{T,\Phi}  &= \frac{\omega_3}{4\pi G_5}\left(q_2+\frac{3}{2}\mu_1\right) \sim \mathcal{O}(1) \\
\left(\frac{\partial^2 M}{\partial Q^2}\right)_{T,\Phi}  &= \frac{8\pi G_5}{\omega_3}\,S\,\frac{\partial^2\mu_0}{\partial q_0^2} + \mathcal{O}(S^2) \sim \mathcal{O}(S) \\
\left(\frac{\partial^2 M}{\partial Q\partial S}\right)_{T,\Phi}  &= \frac{\partial (q_1 + 3\mu_0/2)}{\partial q_0} + \mathcal{O}(S) \sim \mathcal{O}(1)\,, 
\end{split}
\end{equation}
where it was assumed that all leading contributions were non-vanishing and $\partial_Q$ was replaced by a rescaled version of $\partial_{q_0}$.
Hence, the condition \eqref{eq:stability-condition} will be violated if $\frac{\partial (q_1 + 3\mu_0/2)}{\partial q_0} \neq 0$.

Inverting \eqref{eq:inv-charge}, it follows
\begin{equation}
\begin{split}
 \mu_0 &= \frac{4G_5}{\omega_3}\,\frac{q_0+R_{\text{AdS}}^2}{R_{\text{AdS}}^2\,\sqrt{q_0}}\,, \quad \quad \mu_1 = -\frac{16G_5^2}{\omega_3^2}\,\frac{(q_0-R_{\text{AdS}}^2)^2}{q_0^2\,R_{\text{AdS}}^4}\,, \\
  q_0 & = \frac{8\pi G_5}{\omega_3}\,Q\,, \quad q_1 = -\frac{\mu_0}{2}\,, \quad q_2 = \frac{2G_5^2}{q_0^2\,R_{\text{AdS}}^4\,\omega_3^2}\,(4(q_0-R_{\text{AdS}}^2)^2+(q_0+R_{\text{AdS}}^2)^2)\,. \\
\end{split}
\end{equation}
Since $q_1 + \frac{3}{2}\mu_0 = \mu_0$, the thermodynamic stability is controlled by
\begin{equation}
  0 < \text{det}\left(\frac{\partial^2M}{\partial x_i\partial x_j}\right)_{T,\Phi} = - \left(\frac{\partial\mu_0}{\partial q_0}\right)^2 + \mathcal{O}(S) = -\frac{4G^2_5}{\omega^2_3\,R_{\text{AdS}}^4\,q_0^{3}} (R_{\text{AdS}}^2-q)^2 + \mathcal{O}(S)
\end{equation}
Hence, single R-charged black holes are thermodynamically unstable in the regime $R_{\text{AdS}}^2 \sim q \gg \mu$. This means the lesson extracted from \eqref{eq:bridge-div}, describing the lengthening on the bridge as $\frac{\mu}{q}$ decreases is not reliable. Despite this fact, this behaviour agrees with the one encountered in the BPS limit, as discussed in section \ref{sec:2-boundary}.

\section{Matching the superstar geometry to a singular LLM geometry}
\label{sec:matching}

The purpose of this appendix is to find an explicit map between the coordinates used for non-extremal black holes \eqref{eq:10dBH}
\begin{equation}
\begin{aligned}
ds^2 &=\sqrt{\gamma}\left[-H^{-1}\,f\,dt^2 + \frac{dr^2}{f} + r^2\,d\Omega_3^2 + R_{\text{AdS}}^2\,d\theta^2\right] + \frac{R_{\text{AdS}}^2}{\sqrt{\gamma}}\sin^2\theta\,d\tilde{\Omega}_3^2 \\
&+ \frac{H}{\sqrt{\gamma}}\cos^2\theta \left(R_{\text{AdS}}\,d\phi + A\right)^2\,,
\end{aligned}
\end{equation}
with $\gamma = 1+ \frac{q}{r^2}\sin^2\theta$, and $H(r),\,A(r)$ given in \eqref{eq:5dBH}, and the coordinates for rotationally invariant LLM configurations 
\begin{equation}
\begin{aligned}
ds^2 &= - \frac{y}{\sqrt{\frac{1}{4}-z^2}}(dt_{\text{LLM}} + V_\varphi d\varphi_{\text{LLM}})^2 + \frac{\sqrt{\frac{1}{4}-z^2}}{y} (dy^2 + dr_{\text{LLM}}^2 + r_{\text{LLM}}^2\,d\varphi_{\text{LLM}}^2) \\
&+ y\, \sqrt{\frac{\frac{1}{2}+z}{\frac{1}{2}-z}} d\Omega_3^2
+ y\, \sqrt{\frac{\frac{1}{2}-z}{\frac{1}{2}+z}} d \tilde \Omega_3^2\,, \quad i=1,2
\end{aligned}
\end{equation}
in the BPS extremal limit $\mu=0$. In fact, the analysis below shows the map does not exist when $\mu\neq 0$, as stressed in \cite{Lin:2004nb}.

Matching the size of the 3-spheres, one derives 
\begin{equation}
y(r,\theta) = rR_{\text{AdS}}\sin\theta\,, \quad \quad z(y,r_{\text{LLM}})= \frac{1}{2}\frac{r^2 + \sin^2\theta (q-R_{\text{AdS}}^2)}{r^2 + \sin^2\theta (q+R_{\text{AdS}}^2)}\,.
\label{eq:scalar-super}
\end{equation}
In the remaining 4d metric, focus first on the 2d metric spanned by $\{y,\,r_{\text{LLM}}\}$ and look for a change of coordinates $r_{\text{LLM}}=h(r,\theta)$. Absence of cross-terms $dr\,d\theta$ gives rise to the constraint
\begin{equation}
\partial_r h\,\partial_\theta h = -R_{\text{AdS}}^2\,r\,\sin\theta\cos\theta\,.
\label{eq:constraint-a}
\end{equation}
Matching $d\theta^2$ and $dr^2$ gives rise to 
\begin{equation}
\frac{R_{\text{AdS}}^2 r^2 \cos^2\theta + (\partial_\theta h)^2}{r^2 + \sin^2\theta (q+R_{\text{AdS}}^2)} = R_{\text{AdS}}^2\,, \quad \quad
\frac{R_{\text{AdS}}^2\sin^2\theta + (\partial_r h)^2}{r^2 + \sin^2\theta (q+R_{\text{AdS}}^2)} = \frac{1}{f}\,.
\label{eq:constraint-b}
\end{equation}
Solving the first equation for $\partial_\theta h$ and integrating, one obtains
\begin{equation}
(\partial_\theta h)^2 = R_{\text{AdS}}^2\sin^2\theta (r^2+q + R_{\text{AdS}}^2) \quad \Rightarrow \quad
h(r,\theta) = R_{\text{AdS}} \cos\theta \sqrt{r^2+q + R_{\text{AdS}}^2} + K(r)\,,
\end{equation}
where the sign was fixed to ensure $r_{\text{LLM}}\geq 0$. Plugging this integration in \eqref{eq:constraint-a}, one derives
\begin{equation}
\frac{dK}{dr} = 0 \quad \Rightarrow \quad K(r)= K_0\,.
\end{equation}
To reproduce the global AdS map \eqref{vac-map} in the limit $q=0$, one must set $K_0=0$. This analysis already determines $r_{\text{LLM}}$ to be
\begin{equation}
  r_{\text{LLM}} = R_{\text{AdS}}\cos\theta\,\sqrt{r^2 + q + R_{\text{AdS}}^2}\,,
\end{equation}
leaving the second equation in \eqref{eq:constraint-b} as an integrability condition
\begin{equation}
\frac{R_{\text{AdS}}^2\sin^2\theta + (\partial_r h)^2}{r^2 + \sin^2\theta (q+R_{\text{AdS}}^2)} = \frac{1}{1 + r^2/R_{\text{AdS}}^2 + q/R_{\text{AdS}}^2} = \frac{1}{f} \quad \Leftrightarrow \quad \mu=0\,.
\end{equation}
Hence, the map to LLM only exists when the configuration is supersymmetric, as stressed in \cite{Lin:2004nb}.

In the BPS limit, $\tilde{q}=q$ the gauge field reduces to
\begin{equation}
  A = \left(H^{-1}-1\right)\,dt\,.
\end{equation}
One is left to match the 2d dimensional submanifold spanned by $t,\phi$.  The large gauge transformation
\begin{equation}
  \phi = \varphi_{\text{LLM}} + t_{\text{LLM}}\,, \quad t = R_{\text{AdS}}\,t_{\text{LLM}}\,,
\label{eq:large-gauge}
\end{equation}
maps both metrics if the LLM gauge field equals
\begin{equation}
V_\varphi = - \frac{R_{\text{AdS}}^2\cos^2\theta}{r^2+\sin^2\theta (q+R_{\text{AdS}}^2)}\,.
\label{eq:gauge-supers}
\end{equation}

The only step remaining is to show \eqref{eq:scalar-super} and \eqref{eq:gauge-supers} satisfy the LLM equations of motion. 
According to \eqref{eq:grayscalar}, the singular droplet configuration describing the superstar geometry must equal 
\begin{equation}
  z = \left(\frac{1}{2}-z_{\text{sup}}\right) f_-(r_{\text{sup}}) + \frac{1}{2}\left(\frac{1}{2}+z_{\text{sup}}\right)\,.
\label{eq:d1}
\end{equation}
where $r^2_{\text{sup}}= R_{\text{AdS}}^2(q+R_{\text{AdS}}^2)$, is the size of the singular droplet as discussed below \eqref{eq:z-superstar}. Using the identity
\begin{equation}
  f_-(r_{\text{sup}})= \frac{1}{2}\frac{r^2 - (q+R_{\text{AdS}}^2)\sin^2\theta}{r^2+(q+R_{\text{AdS}}^2)\sin^2\theta}\,,
\end{equation}
\eqref{eq:d1} equals \eqref{eq:scalar-super}.

A similar discussion to the one leading to \eqref{eq:grayscalar}, allows to infer the LLM vector field for a singular droplet configuration 
\begin{equation}
  V_\varphi = \left(\frac{1}{2}-z_{\text{sup}}\right)\left(\frac{1}{2}- f_+(r_{\text{sup}})\right)\,.
\label{eq:gaugefield}
\end{equation}
Using the identity
\begin{equation}
  f_+(r_{\text{sup}})= \frac{1}{2}\frac{r^2 + (q+R_{\text{AdS}}^2)(1+\cos^2\theta)}{r^2+(q+R_{\text{AdS}}^2)\sin^2\theta}\,,
\end{equation}
\eqref{eq:gaugefield} equals \eqref{eq:gauge-supers}, completing the match between both metrics.

\section{Free fermion entanglement entropy in condensed matter}
\label{cond-mat}

Any system of free fermions is described by a quadratic hamiltonian
\begin{equation}
  \text{H} = -\sum_{m,n} t_{m,n}c_n^\dagger\,c_m\,.
\end{equation}
In the language of second quantization, $m,n$ label sites in the lattice, whereas $c_n$, $c^\dagger_n$ are annihilation and creation fermion operators at site $n$. Its eigenstates are Slater determinants $|\Psi\rangle$. 

The set of two-point functions
\begin{equation}
  C_{nm} = \langle\Psi |c_n^\dagger\,c_m |\Psi\rangle\,,
\end{equation}
defines an hermitean matrix $C$. Consider a subsystem $A$ of $M$ sites labelled by $i,j$. By definition, the reduced density matrix $\rho_A$ satisfies
\begin{equation}
  C_{ij} = \text{tr}\left(\rho_A\,c_i^\dagger\,c_j\right)\,.
\end{equation}
Since the theory is free, Wick's theorem allows to write $\rho_A$ in terms of the so called entanglement hamiltonian $K_{ij}$ \cite{chung-peschel,henley,Peschel-1,Peschel-2}
\begin{equation}
  \rho_A = {\cal K}\,e^{-{\cal H}}\,, \quad \text{with} \quad {\cal H}=\sum_{i,j} K_{ij} c_i^\dagger\,c_j
\end{equation}
The eigenvalues of $K_{ij}$ can be related to those of the original two-point function using the relation
\begin{equation}
  K = \log \frac{1-C_A}{C_A}\,,
\end{equation}
where $C_A$ and $K$ stand for the matrices with entries $C_{ij}$ and $K_{ij}$, respectively.

Furthermore, it was shown in \cite{Calabrese-3,Calabrese-4} that the spectrum of the continuous version of $C_{ij}$
\begin{equation}
  \mathcal{C}_A(x,y) = I_A(x)\langle c^\dagger(x)\,c(y)\rangle I_A(y)\,,
\end{equation}
where $I_A$ stands for the projector into the subsystem $A$, equals the spectrum of the overlap matrix
\begin{equation}
  A_{nm} = \int_A dz\, \phi_n^\star(z)\,\phi_m(z) \quad \quad n,m = 1,\dots ,N
\end{equation}
which is defined in terms of the single particle energy eigenfunctions
\begin{equation}
\phi_n(x) = \left[\frac{\alpha}{\sqrt{\pi}\,2^n\,n!}\right]^{1/2}\,e^{-\alpha^2 x^2/2}\,H_n(\alpha\,x)\equiv a_n\, e^{-\alpha^2 x^2/2}\,H_n(\alpha\,x)\quad \text{with}\quad \alpha = \sqrt{\frac{m\omega}{\hbar}}\,.
\end{equation}

In terms of these eigenvalues $a_i$, the Renyi entropies equal \cite{Calabrese-3,Calabrese-4}
\begin{equation}
  S_q(A) = \sum_{i=1}^N e_q(a_i)\,, \quad \text{where} \quad e_q(x) \equiv \frac{1}{1-q}\log\left[x^q + (1-x)^q\right]\,.
\end{equation}
In particular, the entanglement entropy $S_1$ reduces to
\begin{equation}
  S_1(A) = \sum_{i=1}^N H(a_i)\,, \quad \text{where} \quad H(x) = -x\log x - (1-x)\log (1-x)\,.
\end{equation}
This is the sum of Shannon's entropies for a binary distribution associated to each eigenvalue.

This formulation allows a numerical analysis of the Renyi entropies for a system of N free fermions in the ground state \cite{eisler,vicari}. There also exist some exact analytic results for the Renyi entropies in the ground state \cite{Klich-Levitov, Song-1, Song-2}
\begin{equation}
\begin{aligned}
  S_q(A) &= \sum_{k=1}^\infty s_k^{(q)}\,V_A^{(2k)}\,, \\
  s_k^{(q)} &= (-1)^k\,(2\pi)^{2k}\frac{2\zeta[-2k,(1+q)/2]}{(q-1)q^{2k}k!}\,, \\
  V_A^{(m)} &= \left(-i\partial_{\lambda}\right)^m\log \left.\langle e^{i\lambda N_A}\rangle\right|_{\lambda=0}
\end{aligned}
\label{eq:exact-renyi}
\end{equation}
where $\zeta$ is the generalised Riemann zeta function in terms of the cumulants $V_A^{(2k)}$ of the number of particles $N_A$ in the region $A$.

\paragraph{Random matrix theory approach.} As mentioned in section \ref{sec:review}, the squared of the wave function for the ground state of the N fermions in a 1d harmonic oscillator potential
\begin{equation}
	|\Psi_{\text{vac}}(\vec{x})|^2 = \frac{1}{Z_N}\,e^{-\alpha^2\sum_{i=1}^N x_i^2}\,\prod_{j\neq k} (x_j-x_k)^2
\end{equation}
corresponds to the joint probability distribution of the $\{x_i=\lambda_i\}$ eigenvalues in a random gaussian unitary matrix model ensemble \cite{mehta}. Using this connection, the large $N$ methods developed in random matrix theory have been used to evaluate $\langle N_A\rangle$, the averaged number of particles for the symmetric closed interval $[-\ell,\,\ell]$ around the origin \cite{mehta} and its variance \cite{2014PhRvL.112y4101M}\footnote{See \cite{Calabrese-2} for generalisations of these results, \cite{2016arXiv160103178M} for studies of $\beta$-ensembles and \cite{2016arXiv160904366D} for d-dimensional trapped potentials at finite temperature.}
\begin{equation}
\begin{aligned}
	\langle N_A \rangle &= \frac{N}{\pi} \left(\frac{\alpha\ell}{\sqrt{N}}\sqrt{2-\frac{\alpha^2\ell^2}{N}} +2 \arcsin \frac{\alpha \ell}{\sqrt{2N}}\right) \\
	V_A^{(2)} &= \left\{ \begin{array}{cc}
	\frac{1}{\pi^2}\log \left[N\frac{\alpha\ell}{\sqrt{N}}\left(2-\frac{\alpha^2\ell^2}{N}\right)^{3/2}\right]\,,  & N^{-1}\ll \frac{\alpha\ell}{\sqrt{N}} < \sqrt{2}  \\
	\tilde{V}_2(s)\,, &  \frac{\alpha\ell}{\sqrt{N}} = \sqrt{2} + \frac{s}{2}N^{-2/3} \\
	e^{-2N\phi(\alpha\ell/\sqrt{N})}\,, & \frac{\alpha\ell}{\sqrt{N}} > \sqrt{2}
	\end{array}\right. 
\end{aligned}	
\end{equation}
where $\frac{\alpha\ell}{\sqrt{N}}\lessgtr \sqrt{2}$ means $\left|\frac{\alpha\ell}{\sqrt{N}} -\sqrt{2}\right|\ll N^{-2/3}$. The function $\tilde{V}_2(s)$ has an analytical expression in terms of Airy functions \cite{2014PhRvL.112y4101M}, but its asymptotics are
\begin{equation}
	\tilde{V}_2(s) \sim \left\{\begin{array}{cc}
		\frac{3}{2\pi^2}\log |s|\,, & s\to -\infty \\
		e^{-4s^{3/2}/3}\,, & s\to \infty
	\end{array} \right. 
\end{equation}
Finally,
\begin{equation}
	\phi(s) = \frac{s}{2}\sqrt{s^2-2} + \log \left(s-\sqrt{s ^2-2}\right)/\sqrt{2}\,.
\end{equation}

\paragraph{Effective CFT formulation.} It was pointed out in \cite{Calabrese-EFT} that despite the inhomogeneity in the system of the $N$ fermions, there must exist some intermediate scale $\ell$ in the range $\rho^{-1} \ll \ell \ll \frac{\rho}{|\partial_x\rho|}$, where the system is effectively homogeneous, with a local Fermi momentum $k_F(x) = \pi\,\rho(x)$. 

For a system of N free fermions with vanishing potential in $\hbar=1$ units and labelling $y$ as euclidean time, this expectation can be derived by approximating the one particle function as  \cite{Calabrese-EFT}
\begin{equation}
\begin{aligned}
\langle c^\dagger (x,y) c(0,0)\rangle &= \int_{-k_F}^{k_F} \frac{dk}{2\pi}\,e^{-i[kx + i\varepsilon(k)\frac{y}{\hbar}]} \\
&\simeq \int_{-\infty}^{k_F} \frac{dk}{2\pi}\,e^{-i[kx + i(k-k_F)v_F\,y]} + \int^{\infty}_{-k_F} \frac{dk}{2\pi}\,e^{-i[kx - i(k+k_F)v_F\,y]} \\
&= \frac{i}{2\pi}\left[\frac{e^{-ik_F x}}{x+iv_Fy} -\frac{e^{ik_F x}}{x-iv_Fy} \right]\,.
\end{aligned}
\end{equation}
where $x,\,v_F y \gg k_F^{-1}$ was used in the second line and $\varepsilon(k)$ was expanded around the Fermi level, i.e. $v_F = \left.\frac{d\varepsilon(k)}{dk}\right|_{k_F}$. Thus, the propagator of a translationally invariant massless Dirac fermion controls the entanglement entropy in this regime of scales. The same conclusion holds, locally, when the fermions are subject to any potential $V(x)$, in particular $V(x) = \frac{1}{2}x^2$, in units $m=\omega=1$.

In the case of a 1d trapped potential, the semicircle Wigner distribution is non-zero for $x\in [-L,L]$ where $L=\sqrt{2\mu}=\sqrt{2N}$. The above local description was extended to the domain $[-L,\,L]\times \mathbb{R}$ by realising the effective massless Dirac fermion propagates in a non-trivial background determined by the local Fermi momentum $k_F(x)$ \cite{Calabrese-EFT}. Since the propagator of a right fermion in the most general 2d background $ds^2 = e^{2\sigma} dzd\bar{z}$ is
\begin{equation}
\langle \psi^\dagger(z+\delta z)\psi (z)\rangle = \frac{1}{e^\sigma\,\delta z}\,,
\end{equation}
to prove the existence of such a global effective description is equivalent to finding a map $(x,y)\to z(x,y)$ such that
\begin{equation}
e^{\sigma(x,y)}\delta z(x,y) = \delta x + iv_F(x)\delta y\,.
\end{equation}
Since $\delta z = \partial_x z\,\delta x + \partial_y z\,\delta y$, this condition is equivalent to
\begin{equation}
  e^\sigma \partial_x z=1\,, \quad \quad e^\sigma\partial_y z = iv_F(x)\,.
\end{equation}
Using the integrability condition $\partial^2_{xy}z = \partial^2_{yx}z$ leads to
\begin{equation}
(iv_F\partial_x - \partial_y)\sigma = i\partial_x v_F \quad \Rightarrow \quad e^\sigma=v_F \quad \text{with} \quad \partial_x z = \frac{1}{v_F(x)}\,,\,\,\,\partial_y z = i\,,
\end{equation}
which can be integrated to give
\begin{equation}
z(x,y) = \int^x \frac{1}{v_F(x^\prime)} dx^\prime + iy\,, \quad e^\sigma = v_F(x)\,,
\end{equation}
where $v_F(x)$ is determined from the ground state condition $k_F^2/2 - \mu + V(x)=0$.

This effective CFT approximation holds away from the edges of the Wigner semicircle distribution since
\begin{equation}
  \rho^{-1} \ll \ell \ll \frac{\rho}{|\partial_x\rho|} \quad \Leftrightarrow \quad N^{-1/2}\sim \rho^{-1} \ll \ell \ll L \sim N^{1/2}
\end{equation}
and gives rise to 
\begin{equation}
z(x,y) = \arcsin \frac{x}{L} + iy\,, \quad \quad e^\sigma = v_F = \sqrt{L^2-x^2}\,.
\end{equation}
Notice the coordinate $z\in [-\pi/2,\,\pi/2]\times \mathbb{R}$ lives on an infinite strip.

Having established this connection, the 2d CFT tools to compute entanglement entropy (see \cite{Calabrese:2009qy} for a review) can be applied to the problem of computing entanglement entropy in the ground state for the non-relativistic fermions. As an example, consider the region $A=[-\infty,x]$. The Renyi entropy equals
\begin{equation}
S_n(A) \simeq \frac{1}{1-n}\log \epsilon^{\Delta_n} \langle\mathcal{T}_n(x,0)\rangle\,,
\end{equation}
where $\mathcal{T}_n$ corresponds to a twist operator with conformal dimension $\Delta_n = \frac{n^2-1}{12n}$ since the central charge $c$ for a massless Dirac fermion is $c=1$. $\epsilon$ stands for a UV cut-off in the CFT literature. In the current discussion it corresponds to $\epsilon = \epsilon_0/k_F(x)$, i.e. the natural cut-off associated with the effective field theory approximation.

To compute the correlator, first consider a Weyl rescaling to map $e^{2\sigma}dz\,d\bar{z}\to dz\,d\bar{z}$. The expectation value transforms as $\langle\mathcal{T}_n\rangle \to e^{\sigma\Delta_n}\,\langle\mathcal{T}_n\rangle$. Second, map the strip to the upper half plane using $g(z)= e^{i(z+\pi/2)}$, under which the correlator transforms as
\begin{equation*}
\langle\mathcal{T}_n(z,\bar{z})\rangle = \left|\frac{dg}{dz}\right|^{\Delta_n}\,\langle\mathcal{T}_n(g(z),g(\bar{z})\rangle\,.
\end{equation*}   
Third, use the correlator on an upper half plane $(\text{Im}(g(z)))^{-\Delta_n}$. Altogether, 
\begin{equation}
\langle\mathcal{T}_n(z,\bar{z})\rangle = \left(e^{\sigma}\, \left|\frac{dg}{dz}\right|^{-1}\,\text{Im}(g(z))\right)^{-\Delta_n}\,.
\end{equation}
This gives rise to a Renyi entropy
\begin{equation}
S_n(A) = \frac{n+1}{12n}\log \left[k_F(x)\,e^{\sigma}\, \left|\frac{dg}{dz}\right|^{-1}\,\text{Im}(g(z))\right] = \frac{n+1}{12n}\log[2N(1-x^2/L^2)^{3/2}]\,,
\end{equation}
where some subleading pieces were dropped. This equals the expression computable by random matrix technology away from the edges and the first result in \eqref{calabrese-results}\footnote{The results in \eqref{calabrese-results} are written in coordinates adjusted to the LLM original description \cite{Lin:2004nb}. Hence, distances $x, x_1$ and $x_2$ in \eqref{calabrese-results} have length squared units. To ease the comparison, notice that $x=\alpha x_{\text{LLM}}$, $\hbar_{\text{LLM}}=(\hbar/(m\omega))^2$ and the phase space densities are related by $\rho_{\text{LLM}}(x_{\text{LLM}})=\alpha\,\rho(x)$.}. The second is computed in \cite{Calabrese-EFT}.


{}

\end{document}